\documentclass{article}

\usepackage{amsmath}
\usepackage{tikz}
\usepackage{url}
\usepackage{amsfonts,amssymb, booktabs} 
\usetikzlibrary{arrows}
\usepackage{mathtools}
\usepackage{longtable} 
\usepackage{float}
\usepackage{placeins}
\usepackage{graphicx}
\usepackage{comment}

\usepackage{afterpage}
\usepackage{emptypage}

\usepackage{algorithm}
\usepackage{algorithmic}
\usepackage{subcaption}
\usepackage{chngcntr}

\usepackage[utf8]{inputenc}

\title{Impact of Non-Informative Censoring on Propensity Score Based Estimation of Marginal Hazard Ratios }

\usepackage{authblk}

\author[]{Guilherme W. F. Barros\thanks{Corresponding author: guilherme.barros@umu.se} }
\author[]{ Jenny H\"aggstr\"om}
\affil[]{Department of Statistics, Ume{\aa} School of Business, Economics and Statistics, Ume{\aa} University, SE-901 87 Ume{\aa}, Sweden}

\usepackage{csquotes}
\usepackage{booktabs}
\usepackage{multirow}
\usepackage[margin=2cm]{geometry}
\usepackage{longtable}
\usepackage{comment}
\usepackage{hyperref}
\usepackage[
backend=biber,
style=authortitle,
citestyle=authoryear,
maxbibnames=100,
natbib,
firstinits,
dashed = FALSE
]{biblatex}
\renewbibmacro{in:}{}
\DeclareFieldFormat[article, inbook, incollection, inproceedings, misc, thesis, unpublished]
{title}{#1}
\DeclareFieldFormat[article]
{pages}{#1}
\DeclareBibliographyCategory{cited}
\AtEveryCitekey{\addtocategory{cited}{\thefield{entrykey}}}

\newcommand{\commenttext}[1]{}

\begin{document}
\renewcommand{\arraystretch}{1.2}
\maketitle

\begin{abstract}
\noindent
In medical and epidemiological studies, one of the most common settings is studying the effect of a treatment on a time-to-event outcome, where the time-to-event might be censored before end of study.  A common parameter of interest in such a setting is the marginal hazard ratio (MHR). When a study is based on observational data propensity score (PS) based methods are often used in an attempt to make the treatment groups comparable despite having a non-randomized treatment. Previous studies have shown censoring to be a factor that induces bias when using PS based estimators. In this paper we study the magnitude of the bias under different rates of non-informative censoring when estimating MHR using PS weighting or PS matching. A bias correction involving the conditional probability of event is suggested and compared to conventional PS based methods.

\end{abstract}


\section{Introduction}
\label{section:Introduction}

The marginal (i.e., population-averaged over time) hazard ratio (MHR) is commonly estimated in randomized controlled trials (RCTs) investigating the effect of a binary treatment on a time-to-event outcome. Fitting a Cox proportional hazards (CPH) model \citep{DC:72} including only the treatment variable (i.e., not adjusting for any baseline covariates) to a full counterfactual sample results in an unbiased MHR estimate if there is no censored data \citep{wyss}. A counterfactual sample is the ideal RCT: a sample consisting of two groups identical in all respects except that one group is treated and the other group is untreated. Although an RCT might not result in a full counterfactual sample it is expected to be free from baseline confounding, i.e., randomizing the treatment ensures that the groups are comparable and that the effect of treatment on outcome can be estimated without bias. 

Having observational data, i.e., non-randomized treatment assignment, means that confounding must be addressed  to make the groups similar enough in preparation to analyzing  the relationship between treatment and outcome.  In a now highly-cited paper \citep{austin-3} it was demonstrated how propensity score (PS) methods, such as inverse probability of treatment weighting (IPTW) and propensity score matching (PSM), in combination with the CPH model could be used to estimate MHR in observational studies. 

Concerns regarding the interpretation of MHR have previously been raised \citep{hernan,Aalen-MHR,ryalen} and it was recently shown that, although MHR can be consistently estimated in an uncensored setting, estimates of MHR will be biased if the outcome is censored before the end of follow-up \citep{wyss, fireman}. This is a problem both in observational studies and RCTs, even in the case of non-informative censoring where the mechanism for the individual's censoring time is independent of the time-to-event. To correct for this bias, \cite{wyss} suggested upweighting "the late underrepresented risk sets" but, as far as the authors are aware, there has been no concrete suggestions on how to construct such weights and no studies to date which investigate how this bias behaves under various censoring rates. Notably, findings related to PS based estimation of MHR come from settings with no censoring \citep{austin-mis-full, austin-variance} or few censoring scenarios \citep{hajage_censoring_bias, wyss, fireman}. 

The objective of this paper is to study finite sample properties of PS based MHR estimators  under different rates of non-informative censoring. In an attempt to bias correct the conventional PS based estimators a modification, involving the conditional probability of event, is suggested. The article is organized as follows: In Section \ref{section:Methodology}, relevant notation and concepts regarding MHR estimation and PS methods are introduced. Section \ref{section:simulation} outlines the Monte Carlo simulation study, designed to investigate the impact of non-informative censoring on MHR estimation. In Section \ref{section:results} results obtained from the simulation study are presented. Section \ref{section:discussion} concludes the paper with a short discussion of the results and some possible areas for future research.

\section{Methods}
\label{section:Methodology}

Suppose that we have a sample of size $N$ and let $\{T_i, D_i, Z_i, \mathbf{X}_i; i = 1, 2, ..., N \}$ denote i.i.d data for the $N$ subjects. $T_i = \min\{Y_i, C_i\}$, the minimum of the true time-to-event $Y_i$ and the censoring time $C_i$, $D_i = 1_{Y_i\leq C_i}$ is the event indicator, $Z_i$ the treatment status ($Z_i = 1$ if treated; $Z_i = 0$ if untreated) and $\mathbf{X}_i$ a $K$-vector of observed baseline covariates. The censoring is assumed to be non-informative \citep{censoring_types}.

\subsection{MHR estimation}
In an RCT MHR would commonly be estimated by a CPH model including the treatment status as the only covariate,
\begin{equation}
\label{equation:cph}
\lambda(t|Z) = \lambda_0 (t)\text{exp}(\alpha Z),
\end{equation}

\noindent
where $\lambda_0(t)$ is the hazard function when $Z = 0$ and $\mathbf{X}$ is marginalized, and $\text{exp}(\alpha)$ is the hazard ratio between $Z = 1$ and $Z = 0$ when $\mathbf{X}$ is marginalized, i.e., MHR. If we instead model the hazard as
\begin{equation}
\label{equation:cph_chr}
\lambda(t|Z, \mathbf{X}) = \lambda_0^* (t)\text{exp}(\alpha^* Z + \beta_1X_1 + \cdots + \beta_KX_K),
\end{equation}

\noindent
then $\lambda_0^*(t)$ is the hazard function when $Z = X_1 = \cdots = X_K = 0$, and $\text{exp}(\alpha^*)$ is the hazard ratio between $Z = 1$ and $Z = 0$ when values of $X_1, \ldots, X_K$ are held constant, i.e., the conditional hazard ratio (CHR). 

The maximum partial likelihood estimator $\hat{\alpha}$ represents an average over time of the log MHRs between $Z = 1$ and $Z = 0$ (a time-averaged treatment effect) and in RCTs, where we assume $\mbox{Pr}(Z = 1|X_1, \ldots, X_K) = \mbox{Pr}(Z = 1)$, usually $|\alpha| < |\alpha^*|$ and $\alpha$ will tend to some limit between 0 and $\alpha^*$ as $\sum_k^K |\beta_k| \rightarrow \infty$ \citep{lin13}. The mechanism behind $\alpha$ diverging from $\alpha^*$ toward 0 is due to differential depletion of susceptibles, i.e., over time those most susceptible to suffer the outcome drop out of the risk set heterogeneously across treatment groups \citep{hernan, wyss}.

\begin{figure}[htp]
\centering

\begin{subfigure}{0.49\columnwidth}
\centering
\includegraphics[width=\textwidth]{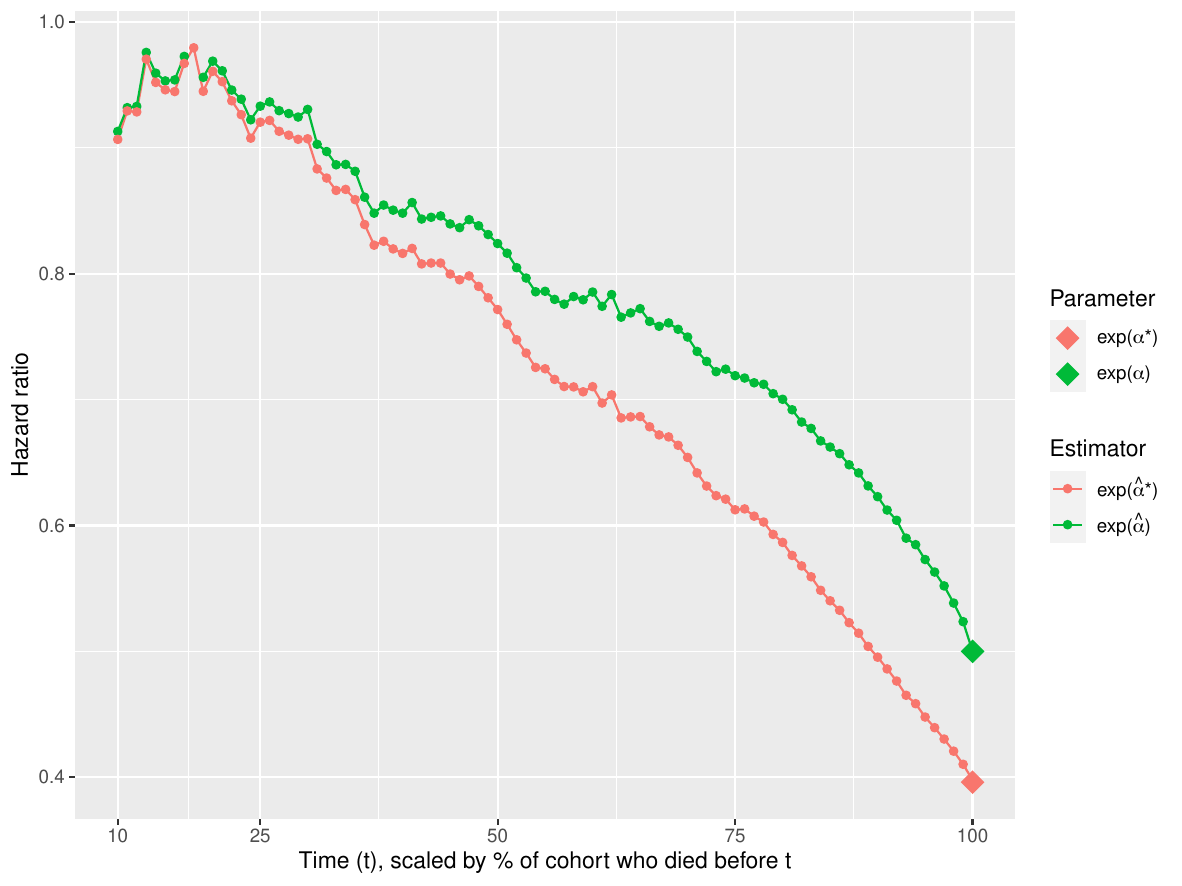}
\caption{MHR = 0.5, no censoring.}
\label{fig:hr_plot_counterfactual_05_uncens}
\end{subfigure}\hfill
\begin{subfigure}{0.49\columnwidth}
\centering
\includegraphics[width=\textwidth]{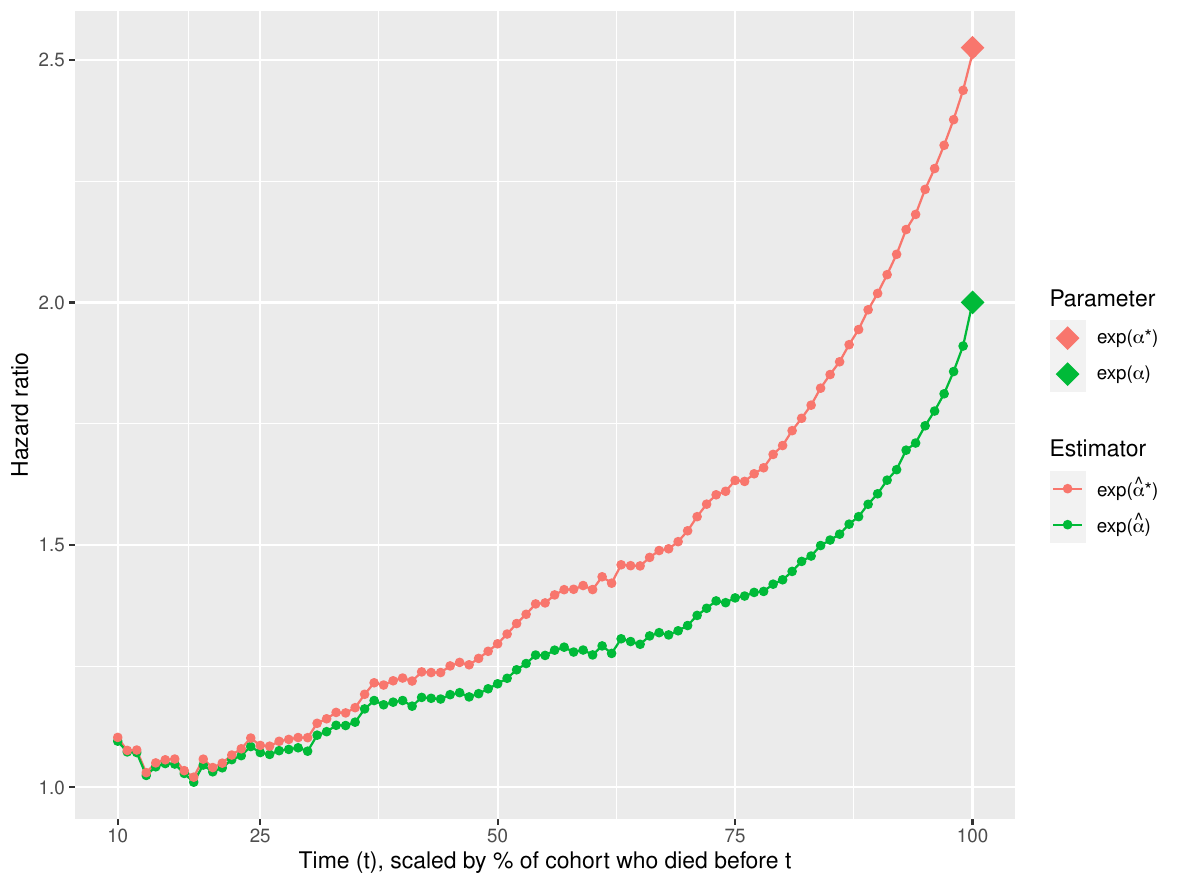}
\caption{MHR = 2.0, no censoring.}
\label{fig:hr_plot_counterfactual_2_uncens}
\end{subfigure}

\medskip
\begin{subfigure}{0.49\columnwidth}
\centering
\includegraphics[width=\textwidth]{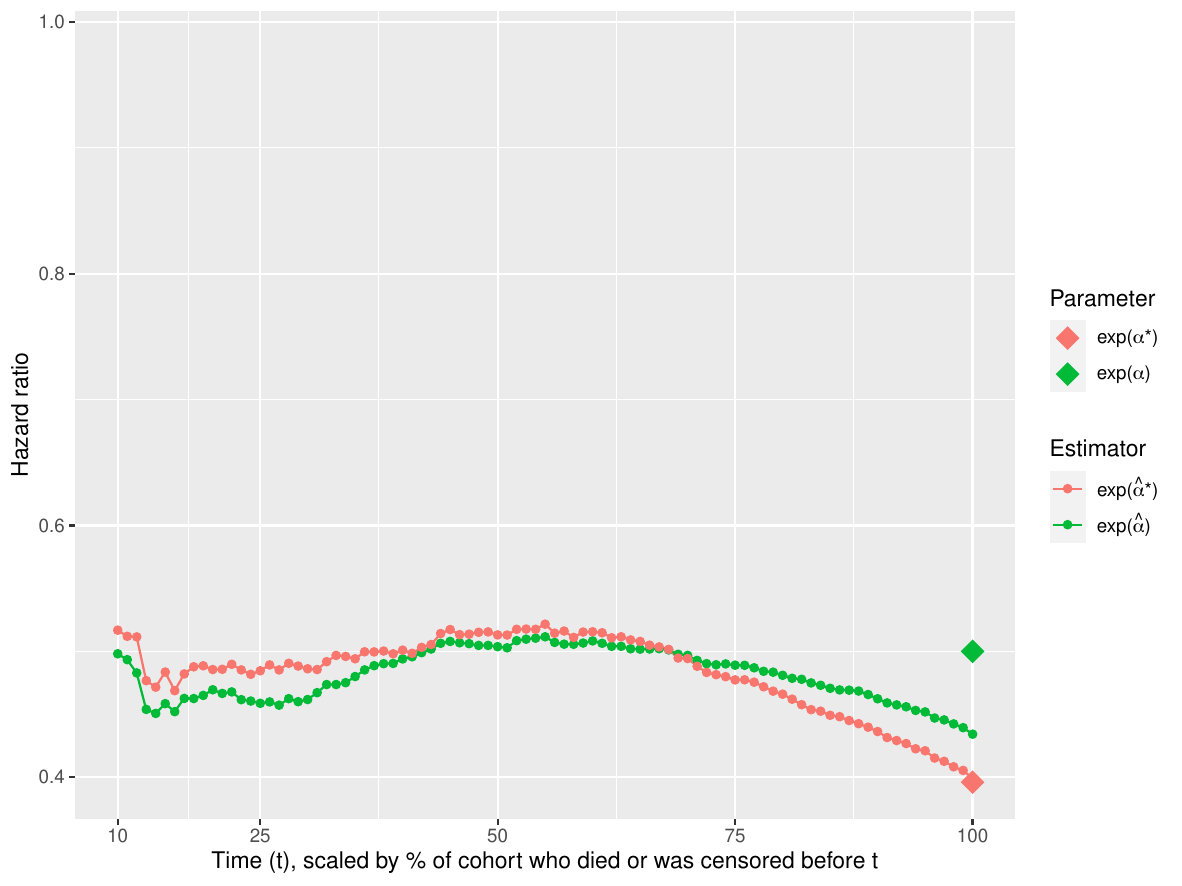}
\caption{MHR = 0.5, 80\% censoring.}
\label{fig:hr_plot_counterfactual_05_cens}
\end{subfigure}
\begin{subfigure}{0.49\columnwidth}
\centering
\includegraphics[width=\textwidth]{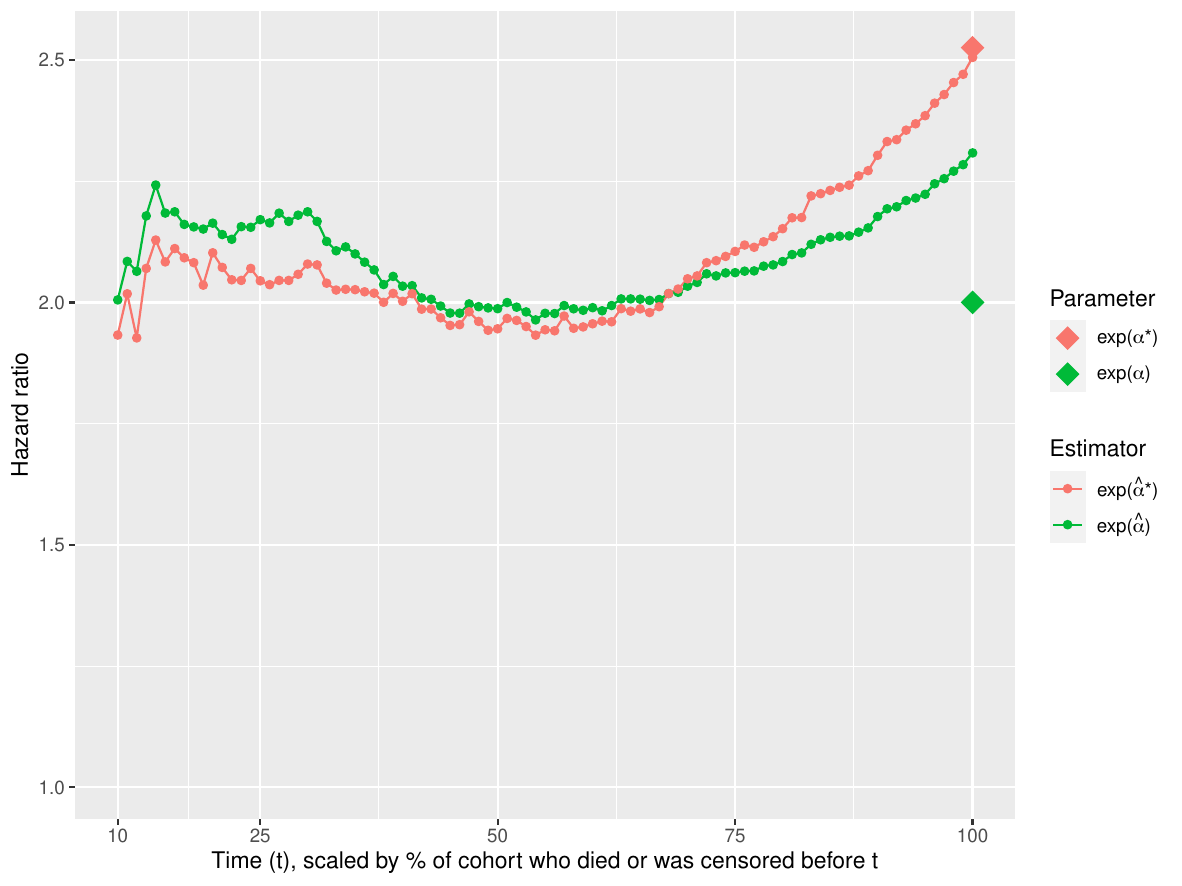}
\caption{MHR = 2.0, 80\% censoring.}
\label{fig:hr_plot_counterfactual_2_cens}
\end{subfigure}

\caption{The data consists of a counterfactual sample of size $n = 50000$, i.e., for each treated individual there is one untreated individual with identical covariate values. Hazard ratios estimated by unweighted Cox with only treatment indicator included (green curve) and unweighted Cox with treatment and all covariates included (red curve). In the first row there is no censoring and green curve at t = 25 shows the unweighted estimate of MHR from t = 0 until the time when 25\% of the full cohort has had an event. In the second row there is 80\% censoring and green curve at t = 25 shows the unweighted estimate of MHR from t = 0 until the time when 25\% of the full cohort has had an event or has been censored. True conditional and marginal hazard ratios at t = 100 are shown as red and green diamond shapes. See Section \ref{section:simulation} for details of the data generating process.}
\label{fig:hr_plot_counterfactual}

\end{figure}

In Figure \ref{fig:hr_plot_counterfactual} we use a simulated counterfactual sample to illustrate the bias arising when fitting model (\ref{equation:cph}) and (\ref{equation:cph_chr}) in scenarios with and without non-informative censoring. As can be seen in Figure \ref{fig:hr_plot_counterfactual_05_uncens} and \ref{fig:hr_plot_counterfactual_2_uncens}, when there is no censoring both $\hat{\alpha}$ and $\hat{\alpha}^*$ unbiasedly estimate their respective target. However, when censoring is introduced (Figure \ref{fig:hr_plot_counterfactual_05_cens} and \ref{fig:hr_plot_counterfactual_2_cens}) $\hat{\alpha}^*$ remains unbiased but this is not true for $\hat{\alpha}$.

\FloatBarrier

\subsection{PS Based Estimation}
In an observational study, we would fit model (\ref{equation:cph}) with individuals weighted according to PS weights or on a PS matched dataset \citep{austin-3}. The PS of an individual $i$ is defined as $\mbox{Pr}(Z_i = 1|\mathbf{X}_i)$, that is, the PS is the probability of that individual receiving treatment ($Z_i = 1$) conditional on baseline covariates. 

IPTW is a method that works in two phases: first, the PS is estimated, e.g., using logistic regression, and then it is used to create weights which are intended to weight individuals in the original data in such a way that the observed differences in the covariate distributions of the two treatment groups are reduced. The weights to be used for each individual depend on the PS of that individual and on the parameter of interest \citep{att-att-ate-weights}.  In this paper, we consider estimating MHR in the total study population, including both treated and untreated, in which case the weights are defined as

\begin{equation}
\label{equation:w_iptw}
w_i^{PS} = \frac{Z_i}{\mbox{Pr}(Z_i = 1|\mathbf{X}_i)} + \frac{1 - Z_i}{1 - \mbox{Pr}(Z_i = 1|\mathbf{X}_i)},
\end{equation}
\noindent
where $w_i^{PS}$ is the weight assigned to individual $i$.



PSM uses the estimated PS to match individuals in one treatment group to similar individuals in the other treatment group and the result of this process is a matched dataset. If the matching is successful there will be no or only small differences in the covariate distributions of the treatment groups in the matched data. A drawback with matching is that it frequently discards a portion of the original sample, due to not finding any sufficiently close matches in the counterfactual treatment group. A PSM method that both allows estimation of MHR in the total population and has the advantage of using a larger portion of the original dataset compared to other matching strategies is full matching \citep{austin-mis-full}. Full matching works by creating a series of strata containing at least one treated individual and one untreated individual. These strata are created in an optimal manner, that is, the number of strata created and the subjects assigned to each are according to the goal of minimizing the mean absolute PS distances within each strata. In the end, full matching also ends up assigning weights, which are in a manner related to IPTW, although in this case the PS is being used to create the strata, not the weights directly. The full matching weight for individual $i$ is,

\begin{equation}
\label{equation:w_psm}
w_i^{PS} = \frac{1_{\{Z_i = 1\}}p_z(n_1^s + n_0^s)}{n_1^s} + \frac{1_{\{Z_i = 0\}}(1-p_z)(n_1^s + n_0^s)}{n_0^s}
\end{equation}

\noindent
where $p_z$ is the marginal probability of treatment in the full sample, $s$ the stratum that contains individual $i$, $n_1^s$ the number of treated in $s$, and $n_0^s$ the number of untreated in $s$ \citep{austin-mis-full}. From now on, when PSM is referred to we mean full PS matching.

\begin{figure}[htp]
\centering

\begin{subfigure}{0.49\columnwidth}
\centering
\includegraphics[width=\textwidth]{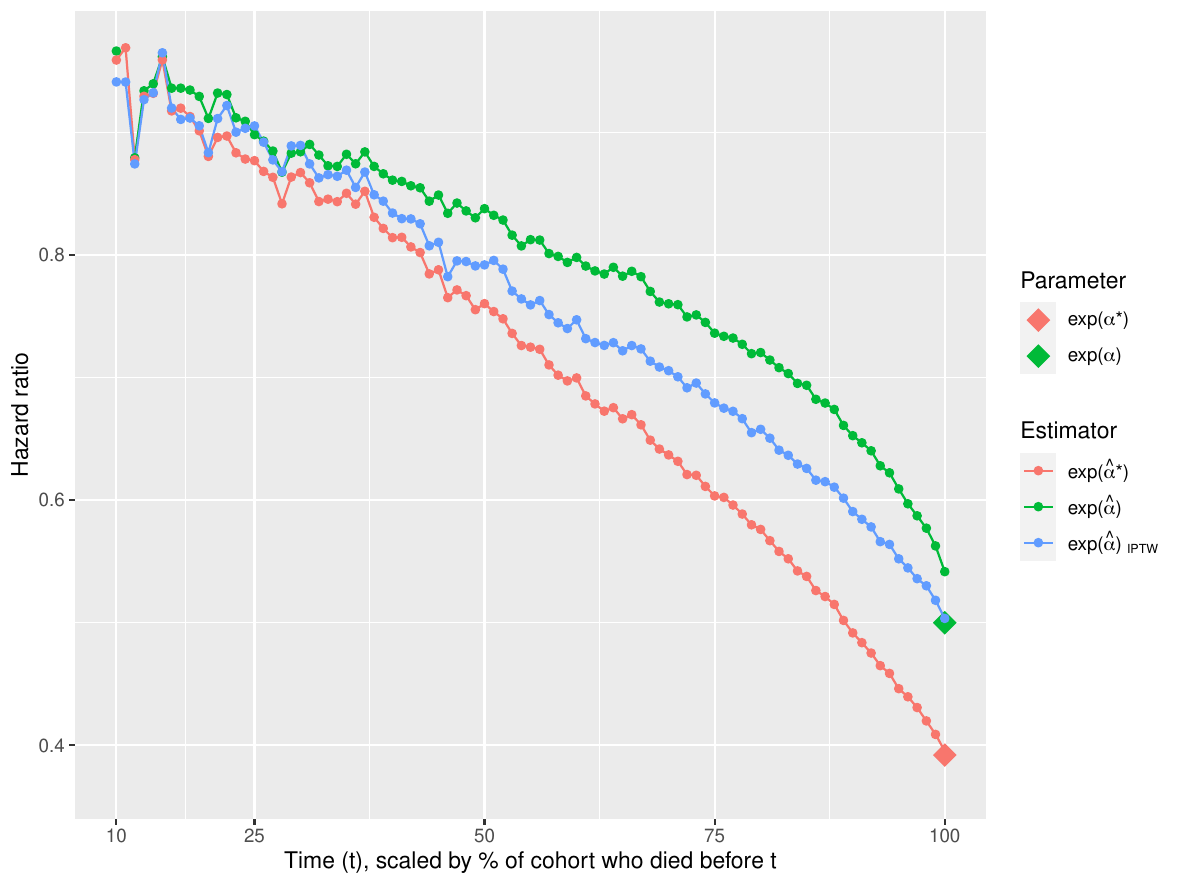}
\caption{MHR = 0.5, no censoring.}
\label{fig:hr_plot_observational_05_uncens}
\end{subfigure}\hfill
\begin{subfigure}{0.49\columnwidth}
\centering
\includegraphics[width=\textwidth]{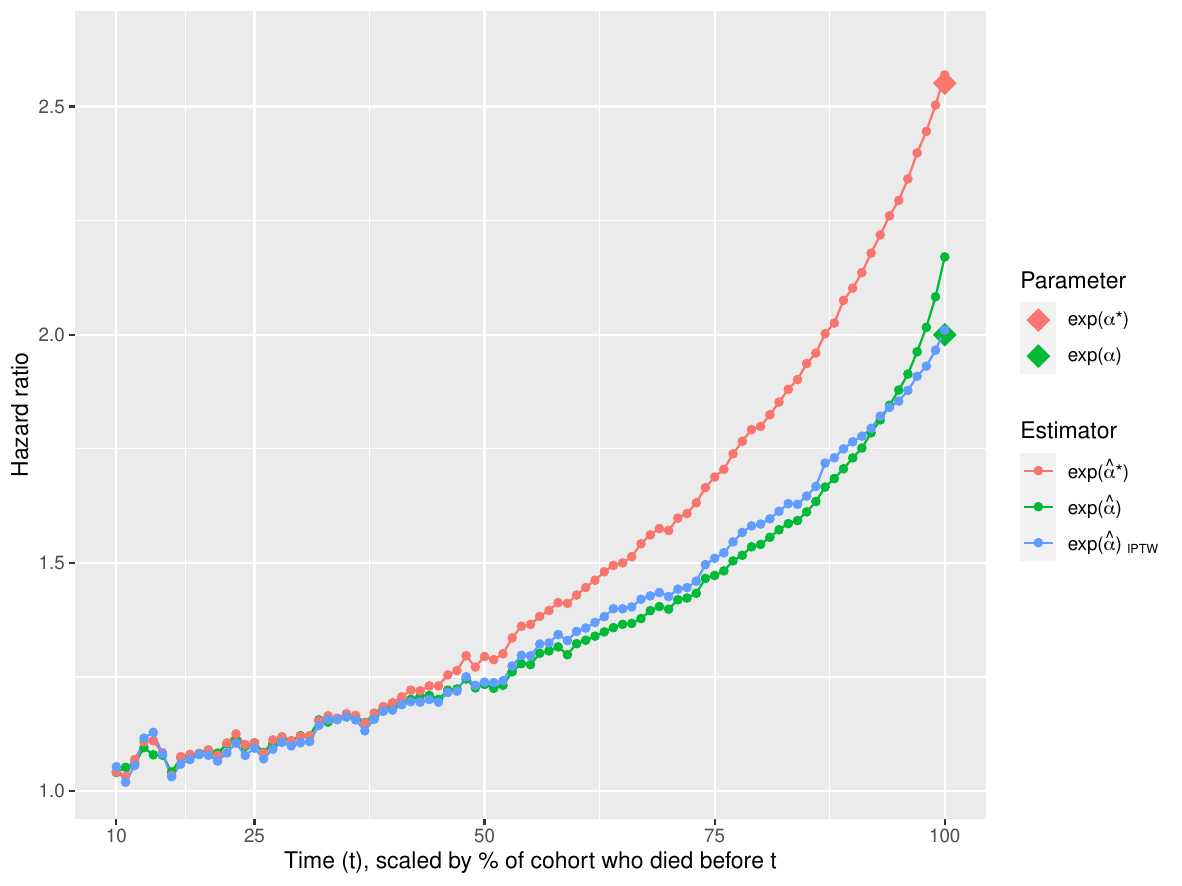}
\caption{MHR = 2.0, no censoring.}
\label{fig:hr_plot_observational_2_uncens}
\end{subfigure}

\medskip
\begin{subfigure}{0.49\columnwidth}
\centering
\includegraphics[width=\textwidth]{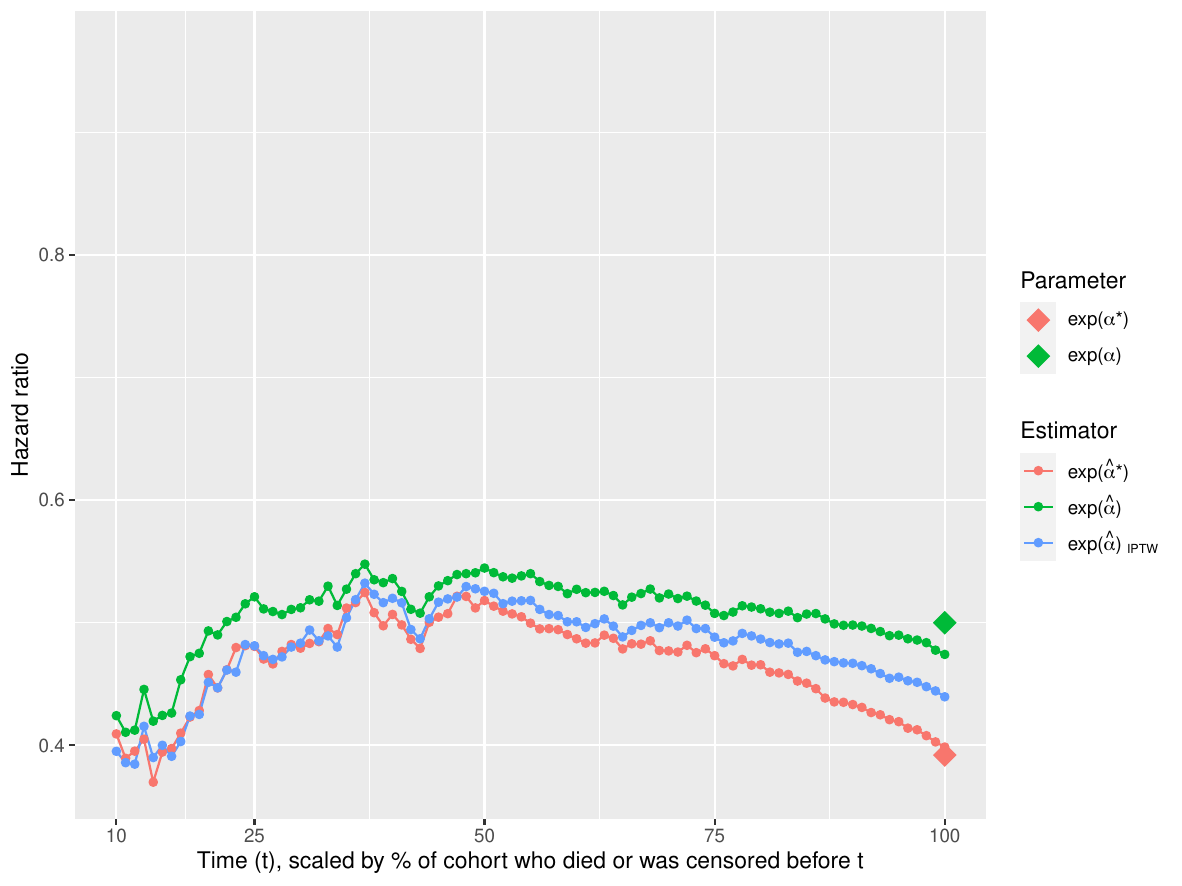}
\caption{MHR = 0.5, 80\% censoring.}
\label{fig:hr_plot_observational_05_cens}
\end{subfigure}
\begin{subfigure}{0.49\columnwidth}
\centering
\includegraphics[width=\textwidth]{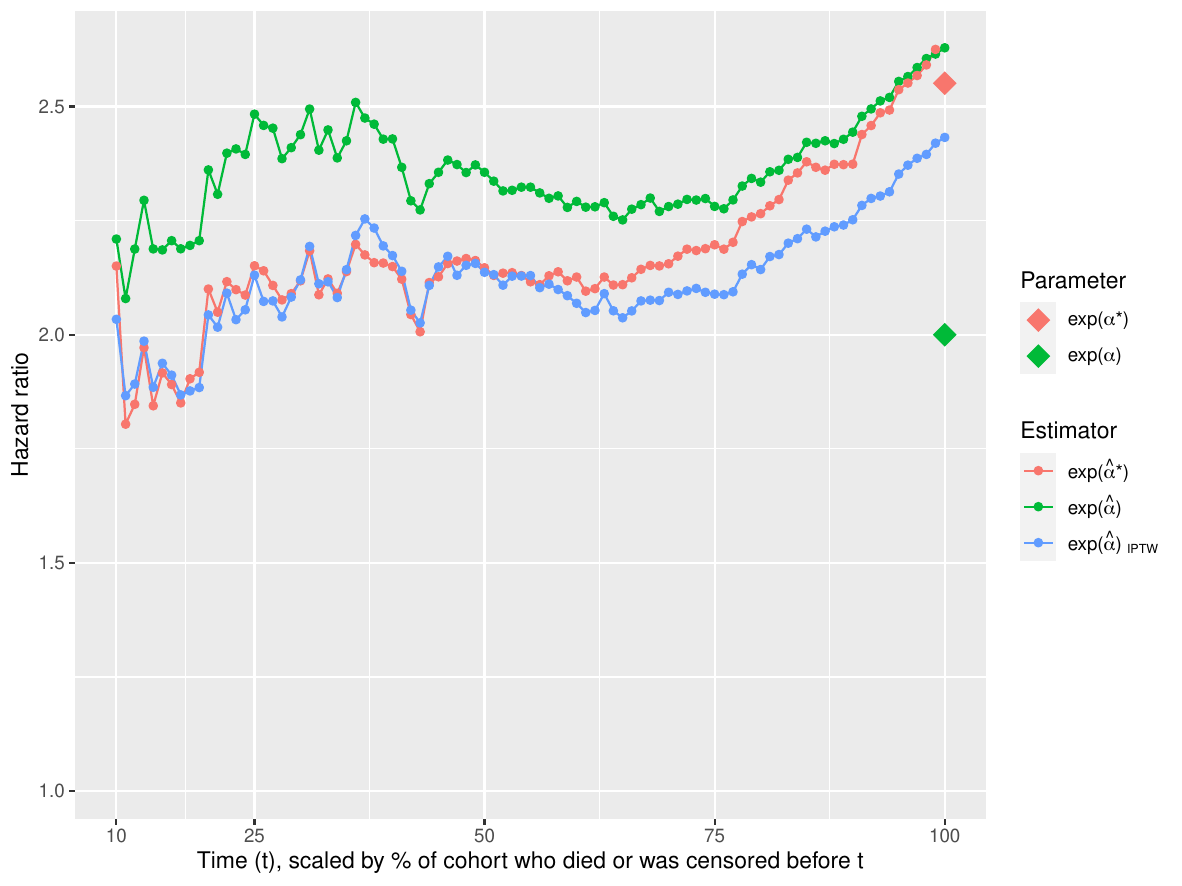}
\caption{MHR = 2.0, 80\% censoring.}
\label{fig:hr_plot_observational_2_cens}
\end{subfigure}

\caption{The data consists of an observational sample of size $n = 50000$. Hazard ratios estimated by unweighted Cox with only treatment indicator included (green curve), unweighted Cox with treatment and all covariates included (red curve), propensity score weighted Cox with only treatment included (blue curve). In the first row there is no censoring and green curve at t = 25 shows the unweighted estimate of MHR from t = 0 until the time when 25\% of the full cohort has had an event. In the second row there is 80\% censoring and green curve at t = 25 shows the unweighted estimate of MHR from t = 0 until the time when 25\% of the full cohort has had an event or has been censored. True conditional and marginal hazard ratios at t = 100 are shown as red and green diamond shapes. See Section \ref{section:simulation} for details of the data generating process.}
\label{fig:hr_plot_observational}

\end{figure}

MHR is estimated by Cox regression, with only the treatment variable as a covariate, and including the weights in \ref{equation:w_iptw} or \ref{equation:w_psm} as case weights. In Figure \ref{fig:hr_plot_observational} we illustrate, with a simulated observational sample, that the PS-weighted Cox model can achieve unbiased estimation of MHR when there is no censoring (Figure \ref{fig:hr_plot_observational_05_uncens} and \ref{fig:hr_plot_observational_2_uncens}). As noted before \citep{hajage_censoring_bias, wyss, fireman} it is not enough to correct for unbalanced baseline covariates to achieve unbiased estimation if there is censoring (Figure \ref{fig:hr_plot_observational_05_cens} and \ref{fig:hr_plot_observational_2_cens}). 

In addition to studying the finite sample properties of the PS based estimators described above we will investigate if it is possible to reduce the finite sample bias by using modified PS weights that take censoring into account, by giving relatively higher weight to individuals with a higher conditional probability of having an event. The modified PS weights are constructed as 

\begin{equation}
\label{equation:w_cens}
w_i^{PS}w_i^{cens} = w_i^{PS}\mbox{Pr}(D_i = 1| Z_i, \mathbf{X}_i)
\end{equation}
where $w_i^{PS}$ is the IPTW or PSM weight.
\noindent

\FloatBarrier

\section{Monte Carlo Simulation}
\label{section:simulation}

Data generation and all computations were performed with the software \texttt{R} \citep{R}. The R package \texttt{survival} \citep{survival} was used for Cox modeling and full matching was implemented using the \texttt{matchit} function from the \texttt{MatchIt} package \citep{HK:06, MatchIt} .

\subsection{Data Generating Process}
\label{subsection:data_gen}



We simulated data for a setting in which there were ten baseline covariates, $X_1$ to $X_{10}$. Of these, six ($X_1$, $X_3$, $X_5$, $X_6$, $X_8$ and $X_9$), were Bernoulli($p=0.5$) distributed and four ($X_2$, $X_4$, $X_7$ and $X_{10}$) were distributed according to a standard normal distribution. The PS was generated according to  $\text{logit}(\mbox{Pr}(Z = 1|\mathbf{X}_i)) = \boldsymbol{\zeta}^T\mathbf{X}_i$ and a linear predictor (LP) was assigned as $LP_i = \alpha^* Z_i + \boldsymbol{\beta}^T\mathbf{X}_i$. Parameter values was set to $\boldsymbol{\zeta} = (0.8, -0.25, 0.6, -0.4, -0.8, -0.5, 0.7, 0, 0, 0)^T$ and $\boldsymbol{\beta} = (0.3, -0.36, -0.73, -0.2, 0, 0, 0, 0.71, -0.19, 0.26)^T$. This simulation design has previously been used in simulation studies involving time-to-event data \citep{austin-mis-full} and is similar to scenario (A) in the framework established by \citet{misspec_orig}.


The time-to-event of each subject was generated as $Y_i = \left( \frac{-\text{log}(u_i)}{\lambda e^{LP_i}} \right)^{1/\eta}$ where $u_i$ was sampled from a standard uniform distribution. The values of $\lambda$ and $\eta$ were, as in \citet{austin-mis-full}, set to 0.00002 and 2, respectively. This procedure results in data generated according to a log conditional hazard ratio of $\alpha^*$, but, since we wanted to generate data according to a specific MHR, an iterative bisection method was used, as in \citet{austin-mis-full}, to select an $\alpha^*$ that resulted in the desired $\alpha$, i.e., log MHR. 

In the end, we had a dataset that was analogous to an observational study, in which $X_1$, $X_2$, $X_3$ and $X_4$ were related to both treatment assignment and time-to-event, while the other covariates only affected either treatment assignment or time-to-event, but not both. We call this the observational setting. To generate a dataset that was analogous to an ideal RCT, we followed the same procedure, but in the final dataset each individual was included twice: under treatment and under control regime, each with its corresponding time-to-event value. We call this  the counterfactual setting.


Censoring times were generated according to either   $C_i \sim \mbox{Uniform}(0, \theta)$ or  $C_i \sim \mbox{Weibull}(\eta, \theta)$, distributions commonly assumed for non-informative censoring times. 
To achieve a pre-specified censoring proportion the value of $\theta$ can be manually tuned, but we used a more precise method introduced by \citet{Wan2016}. This method consists in solving the integral

\begin{equation*}
\gamma(\theta|\pi) = \int_{D'} \mbox{Pr}(\omega = 1 | u, \theta) f_{\tau_i}(u) du - \pi,
\end{equation*}

\noindent
where $\pi$ is the desired censoring proportion, $D'$ is the domain of $\tau_i = \frac{\text{exp}({LP_i/\eta})}{\sqrt {\lambda}}$, $\omega = 1_{\{Y\geq C\}}$ is an indicator variable for censoring and $f_{\tau_i}(\cdot)$ is the density function of $\tau_i$. 

If $ C \sim \mbox{Uniform}(0, \theta)$, the individual censoring probability can be expressed as $\mbox{Pr}(\omega = 1 | \tau_i, \eta ,\theta) = \frac{\tau_i}{\eta \theta} \Gamma \left( \frac{1}{\eta}, (\theta/\tau_i)^\eta \right)$ while if $C \sim \mbox{Weibull}(\eta, \theta)$ it can be expressed as $\mbox{Pr}(\omega = 1 | \tau_i, \eta ,\theta) = \frac{1}{1+(\theta/\tau_i)^{\eta}}$, where $\Gamma(\cdot,\cdot)$ is the lower incomplete gamma function. Since our covariates were a mix of Bernoulli and normally distributed variables, it was not possible to explicitly find a value for $f_{\tau_i}(u)$. Instead, as recommended and described by \citet{Wan2016}, we estimated $f_{\tau_i}(u)$ using kernel methods to find a value for $\theta$ which resulted in a specified censoring proportion of the data. 

We allowed the following factors to vary in our Monte Carlo simulations: the percentage of subjects that were censored (10\%, 20\%, 30\%, 40\%, 50\%, 60\%, 70\%, 80\%, 90\%), the true MHR (0.5, 0.8, 1, 1.25, 2), sample size $N$ (2000, 6000, 10000) and distribution of the censoring mechanism (uniform or Weibull). We thus examined 270 scenarios (9 censoring proportions $\times$ 5 MHRs $\times$ 3 sample sizes $\times$ 2 censoring distributions). In each scenario, we simulated 1000 datasets.

\subsection{Statistical Analysis in the Simulated Datasets}
The PS was estimated by logistic regression including all 10 covariates as main effects. The conditional probability of event was estimated, using unweighted data, by logistic regression including the treatment variable and all 10 covariates as main effects. Weights based on the estimated PS was computed according to \ref{equation:w_iptw} and \ref{equation:w_psm} and weights based on the estimated PS and true or estimated conditional probability of event according to \ref{equation:w_cens}. MHRs was estimated according to \ref{equation:cph} using partial likelihood regression (function \texttt{coxph} in the \texttt{survival} R package) and including the computed weights as case weights. When describing the simulation results we will refer to results based on weights according to \ref{equation:w_iptw} as IPTW, according to \ref{equation:w_psm} as PSM, according to \ref{equation:w_cens} using the true conditional probability of event as IPTW\_PEW1 or PSM\_PEW1 and according to \ref{equation:w_cens} using the estimated conditional probability of event as IPTW\_PEW2 or PSM\_PEW2.

It should be noted that the PS is constant in the counterfactual setting, since every individual has a perfect analogue, and thus the IPTW and PSM estimation of MHR reduces to fitting an unweighted CPH model.

 

 



Let $\mbox{exp}(\alpha)$ denote the true MHR and let $\mbox{exp}(\hat{\alpha}_r)$ denote the estimated MHR, in the $r$th simulated dataset ($r = 1, \ldots, 1000$). Then, with the mean estimated MHR calculated as $\overline{\mbox{MHR}}=\frac{1}{1000}\sum_{r = 1}^{1000} \mbox{exp}(\hat{\alpha}_r)$, bias was estimated as $\mbox{Bias} = \overline{\mbox{MHR}} - \mbox{exp}(\alpha)$; the Monte Carlo standard error as $\mbox{SD} =\frac{1}{1000 - 1}\sum_{r = 1}^{1000} (\mbox{exp}(\hat{\alpha}_r) - \overline{\mbox{MHR}})^2$; $\mbox{RMSE} = \sqrt{\mbox{Bias}^2 + \mbox{SD}^2}$; relative bias as $\mbox{Rel. Bias} = \mbox{Bias}/\mbox{exp}(\alpha)$.
Also, using a robust sandwich type variance estimator \citep{austin-variance} 95\% confidence intervals were constructed for each estimate of MHR, and the proportion of confidence intervals that contained the true MHR was determined (Coverage).



\section{Results}
\label{section:results}

The results for the uniform and Weibull censoring mechanisms were similar and the conclusions equivalent, to save space only results pertaining to uniform censoring will be presented. Likewise, results relating to MHR = 0.8, 1.25 and censoring rates below 0.3 are only presented in figures and not in tables.

\begin{figure}[htp]
\centering

\begin{subfigure}{0.49\columnwidth}
\centering
\includegraphics[width=\textwidth]{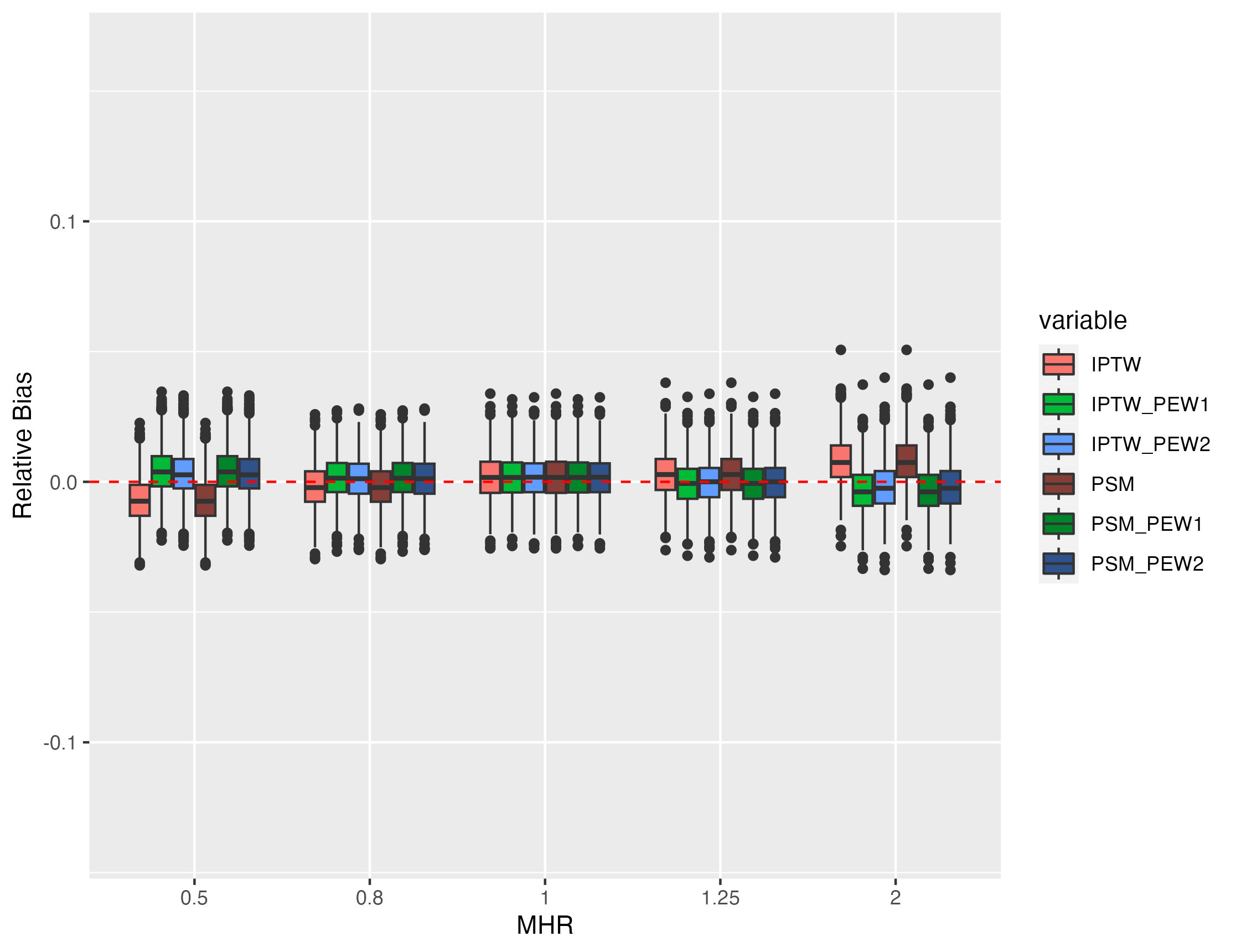}
\caption{Relative bias with censoring rate = 0.1}
\label{fig:n6000_counterfactual_censoring0.1}
\end{subfigure}\hfill
\begin{subfigure}{0.49\columnwidth}
\centering
\includegraphics[width=\textwidth]{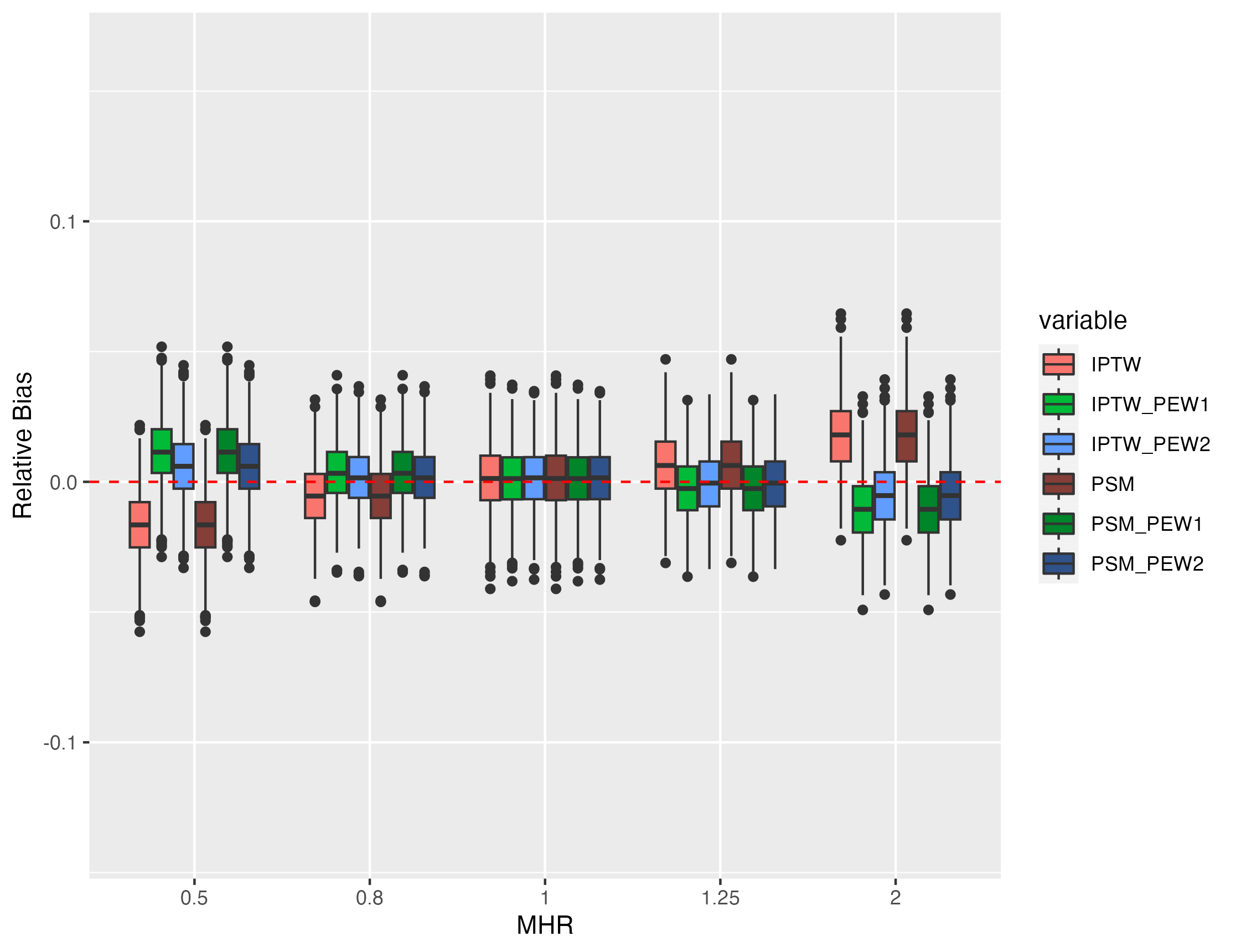}
\caption{Relative bias with censoring rate = 0.2}
\label{fig:n6000_counterfactual_censoring0.2}
\end{subfigure}

\medskip

\begin{subfigure}{0.49\columnwidth}
\centering
\includegraphics[width=\textwidth]{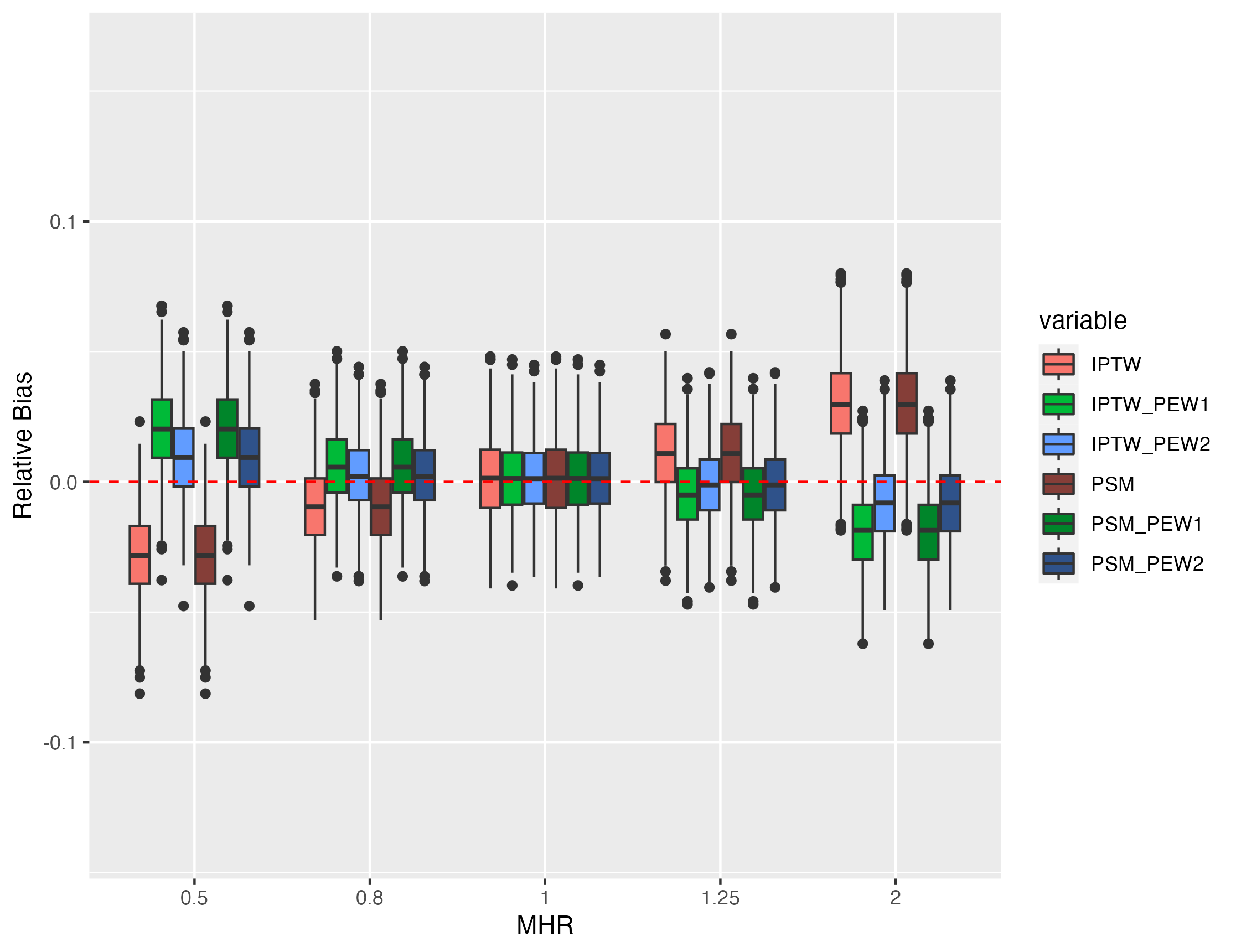}
\caption{Relative bias with censoring rate = 0.3}
\label{fig:n6000_counterfactual_censoring0.3}
\end{subfigure}\hfill
\begin{subfigure}{0.49\columnwidth}
\centering
\includegraphics[width=\textwidth]{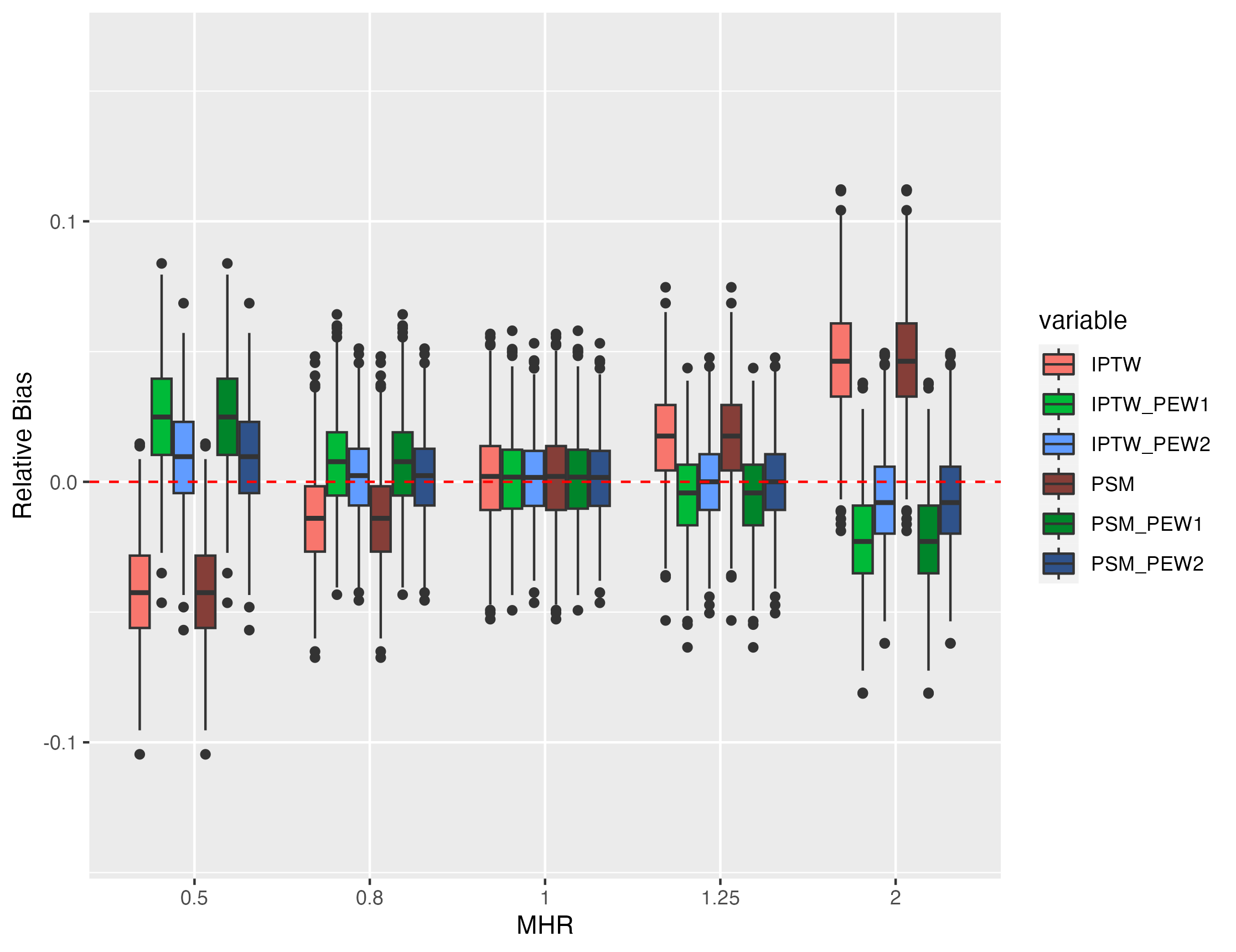}
\caption{Relative bias with censoring rate = 0.4}
\label{fig:n6000_counterfactual_censoring0.4}
\end{subfigure}

\medskip

\begin{subfigure}{0.49\columnwidth}
\centering
\includegraphics[width=\textwidth]{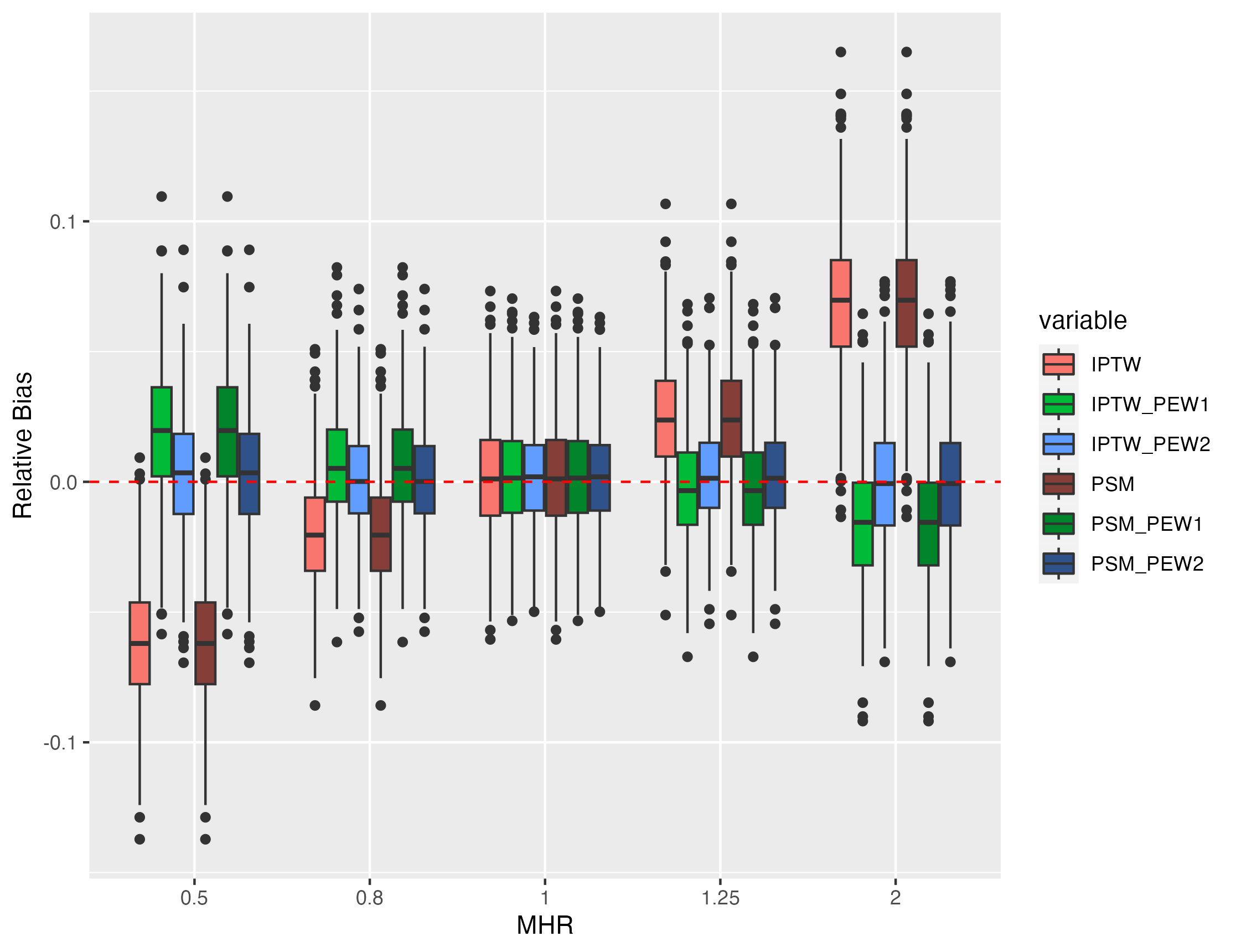}
\caption{Relative bias with censoring rate = 0.5}
\label{fig:n6000_counterfactual_censoring0.5}
\end{subfigure}

\caption{Counterfactual setting, N = 6000, censoring rates = (0.1, 0.2, 0.3, 0.4, 0.5). Relative bias of estimation of MHR under several true values of MHR. Results based on 1000 simulation replicates.}
\label{fig:n6000_counterfactual_censoring_several_01_05}

\end{figure}

\begin{figure}[htp]
\centering

\begin{subfigure}{0.49\columnwidth}
\centering
\includegraphics[width=\textwidth]{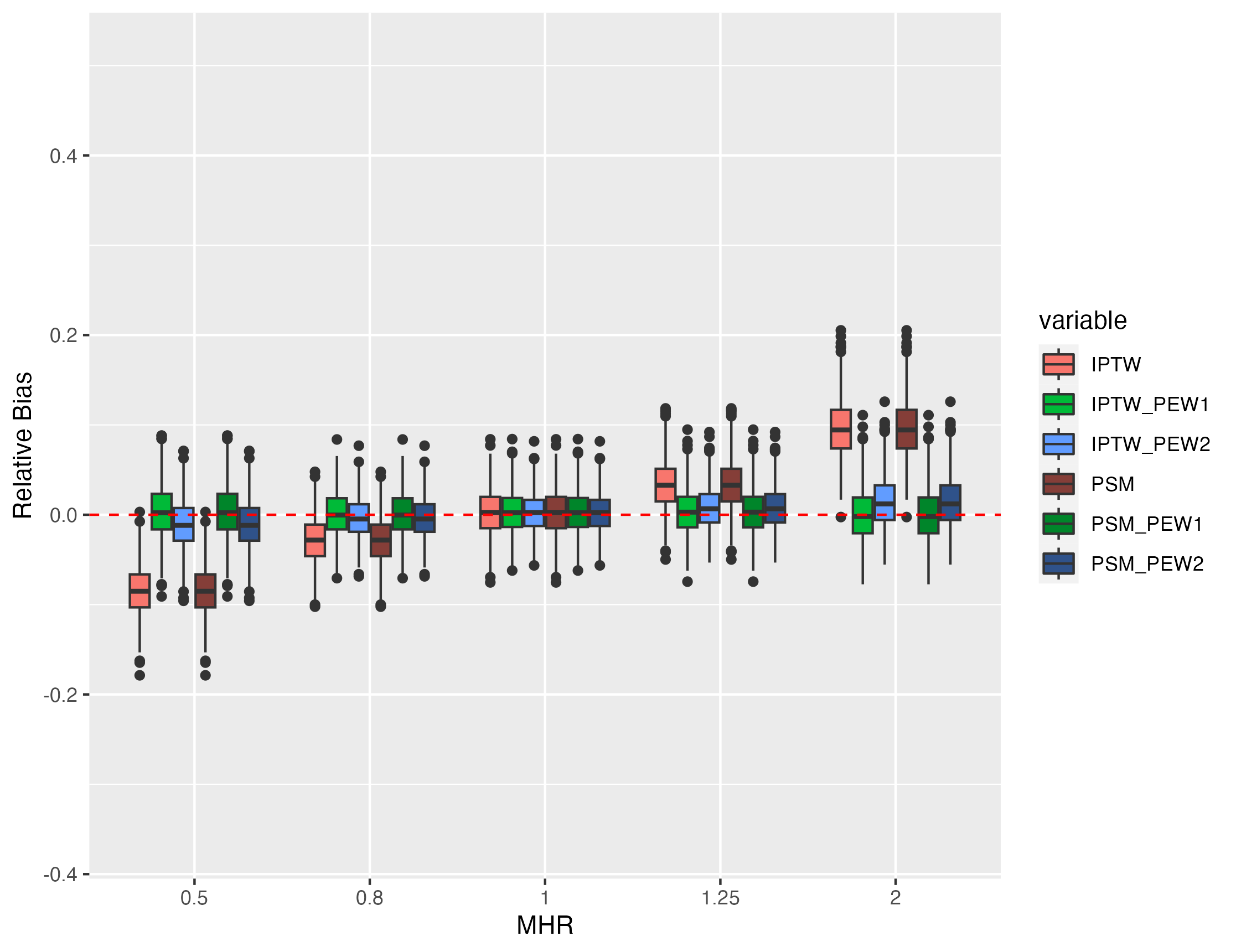}
\caption{Relative bias with censoring rate = 0.6}
\label{fig:n6000_counterfactual_censoring0.6}
\end{subfigure}\hfill
\begin{subfigure}{0.49\columnwidth}
\centering
\includegraphics[width=\textwidth]{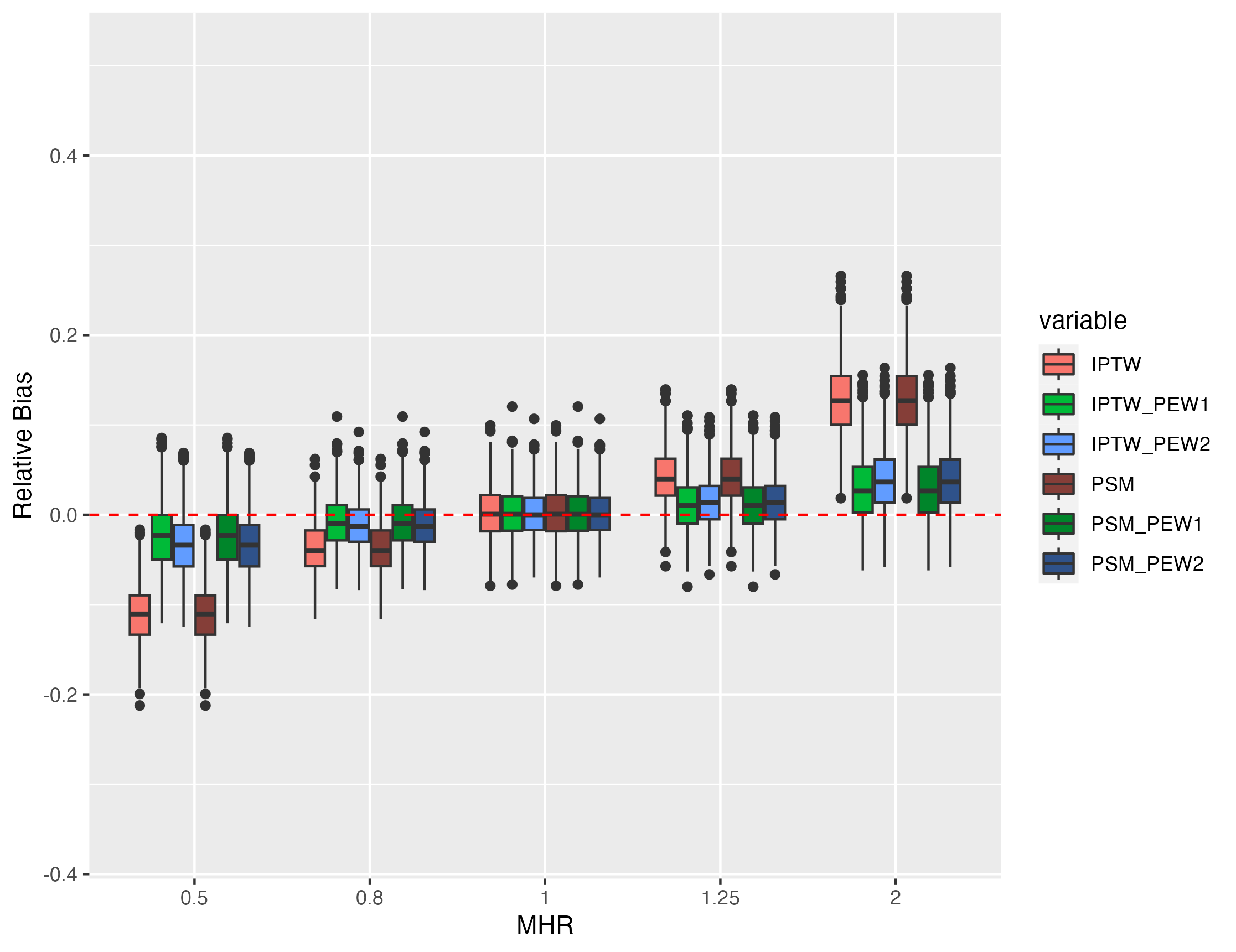}
\caption{Relative bias with censoring rate = 0.7}
\label{fig:n6000_counterfactual_censoring0.7}
\end{subfigure}

\medskip

\begin{subfigure}{0.49\columnwidth}
\centering
\includegraphics[width=\textwidth]{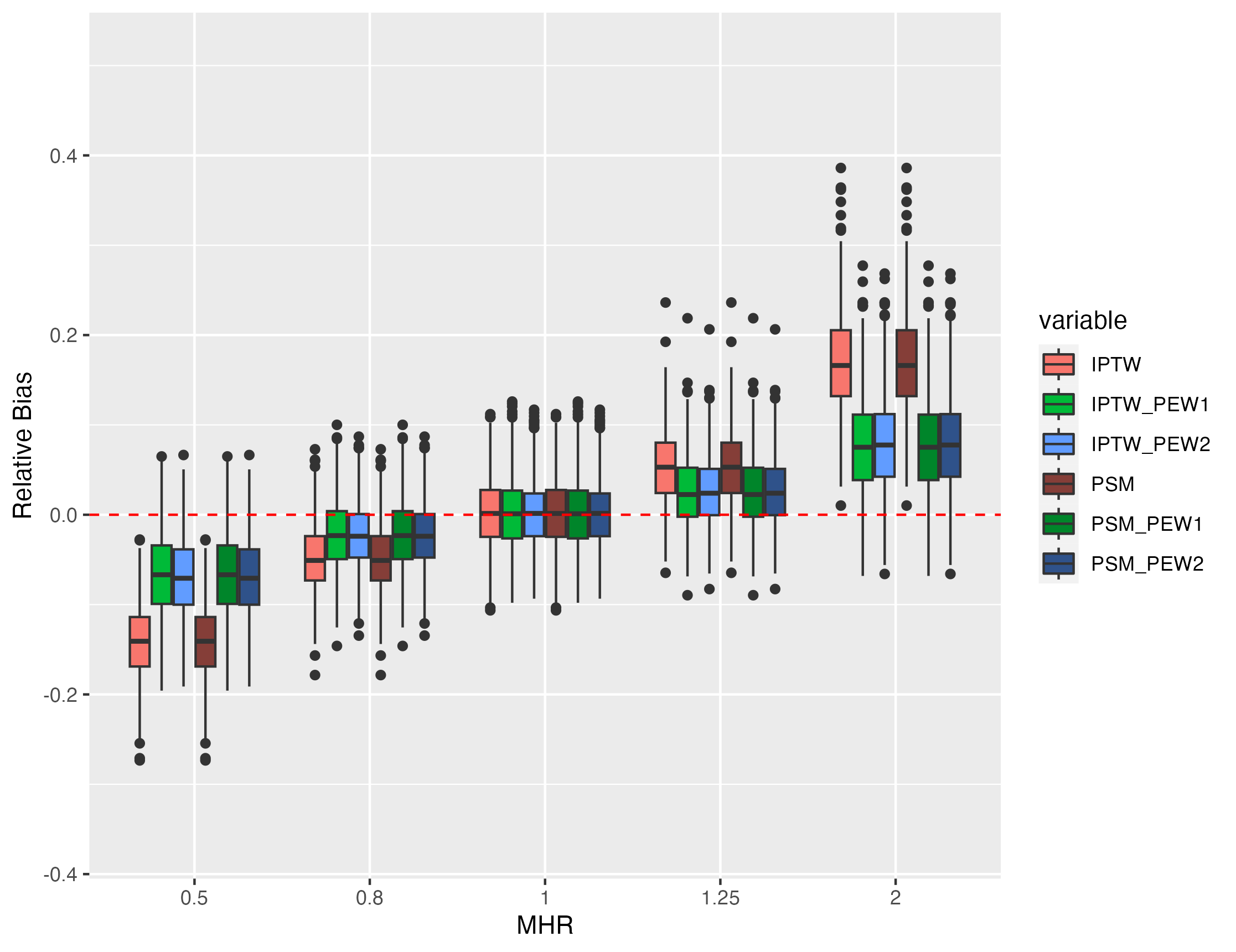}
\caption{Relative bias with censoring rate = 0.8}
\label{fig:n6000_counterfactual_censoring0.8}
\end{subfigure}\hfill
\begin{subfigure}{0.49\columnwidth}
\centering
\includegraphics[width=\textwidth]{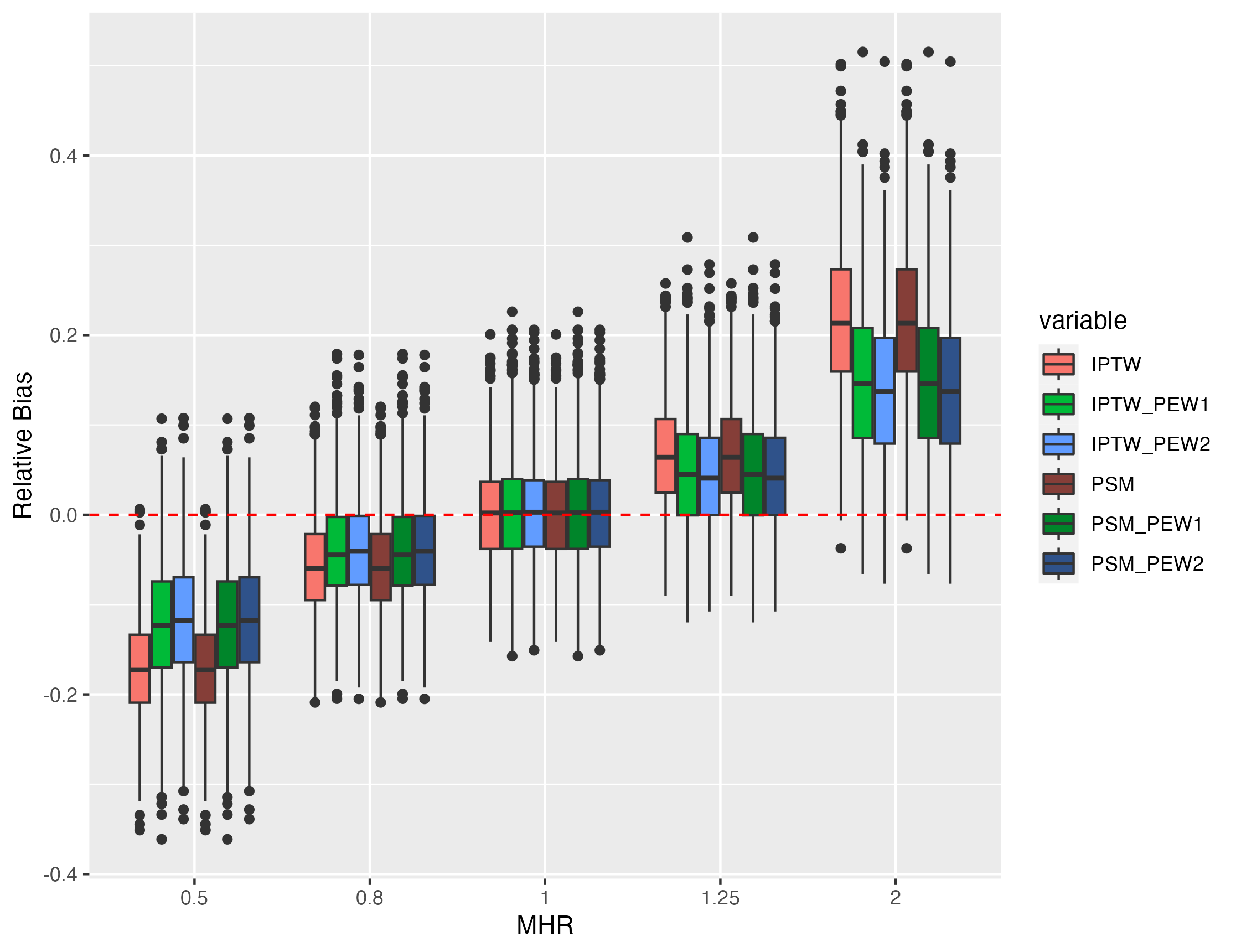}
\caption{Relative bias with censoring rate = 0.9}
\label{fig:n6000_counterfactual_censoring0.9}
\end{subfigure}

\caption{Counterfactual setting, N = 6000, censoring rates = (0.6, 0.7, 0.8, 0.9). Relative bias of estimation of MHR under several true values of MHR. Results based on 1000 simulation replicates.}
\label{fig:n6000_counterfactual_censoring_several_06_09}

\end{figure}

\begin{figure}[htp]
\centering

\begin{subfigure}{0.49\columnwidth}
\centering
\includegraphics[width=\textwidth]{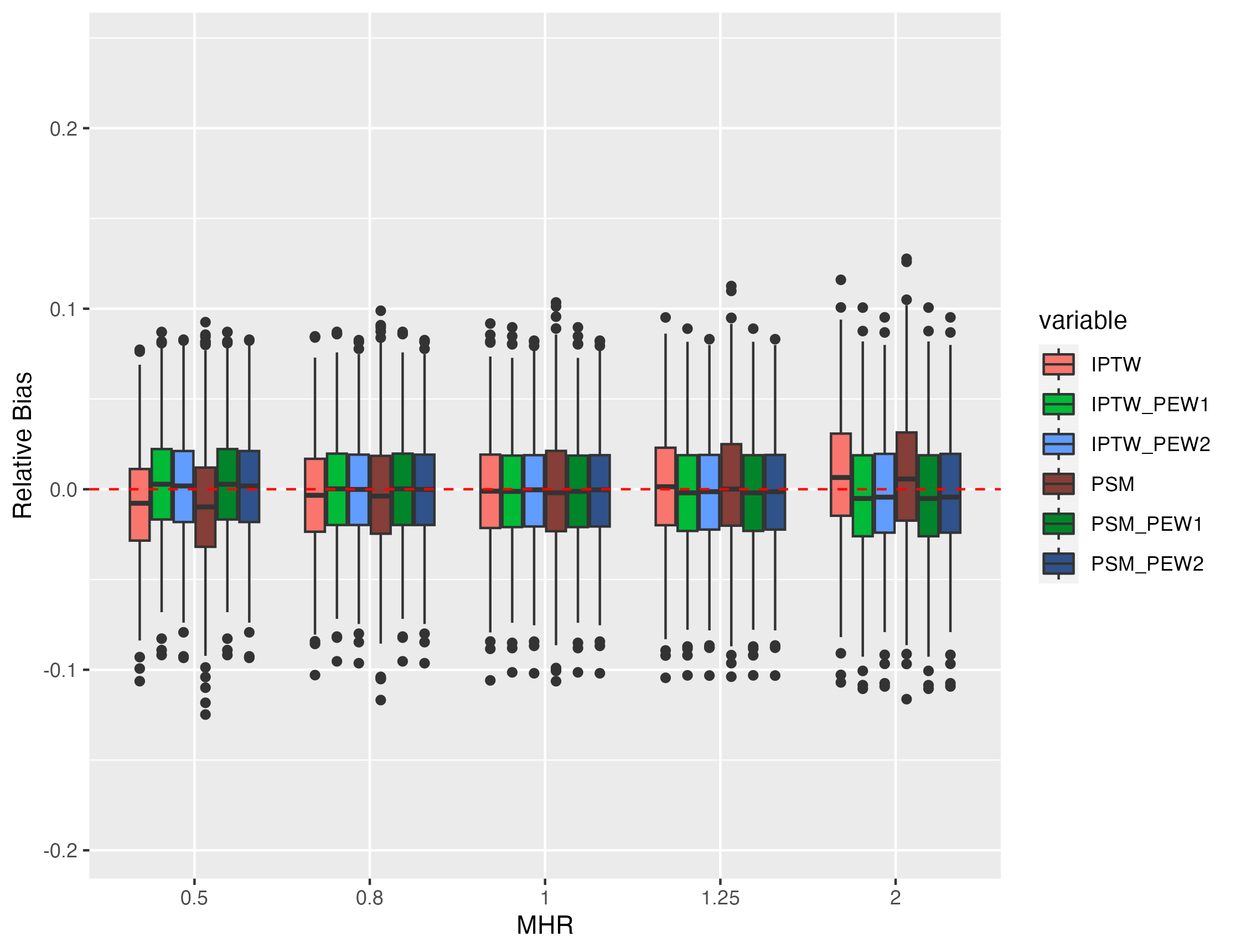}
\caption{Relative bias with censoring rate = 0.1}
\label{fig:n6000_notcounterfactual_censoring0.1}
\end{subfigure}\hfill
\begin{subfigure}{0.49\columnwidth}
\centering
\includegraphics[width=\textwidth]{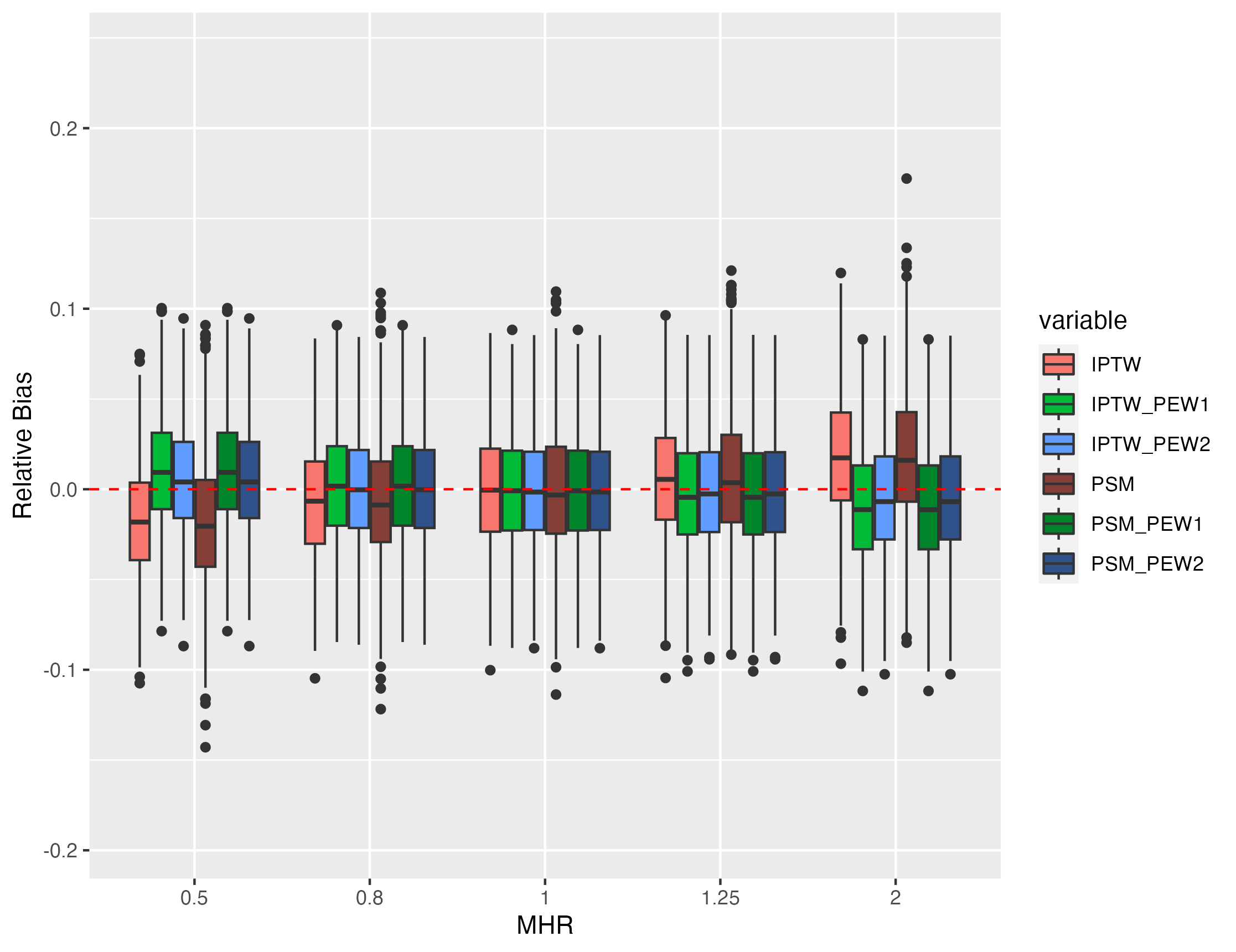}
\caption{Relative bias with censoring rate = 0.2}
\label{fig:n6000_notcounterfactual_censoring0.2}
\end{subfigure}

\medskip

\begin{subfigure}{0.49\columnwidth}
\centering
\includegraphics[width=\textwidth]{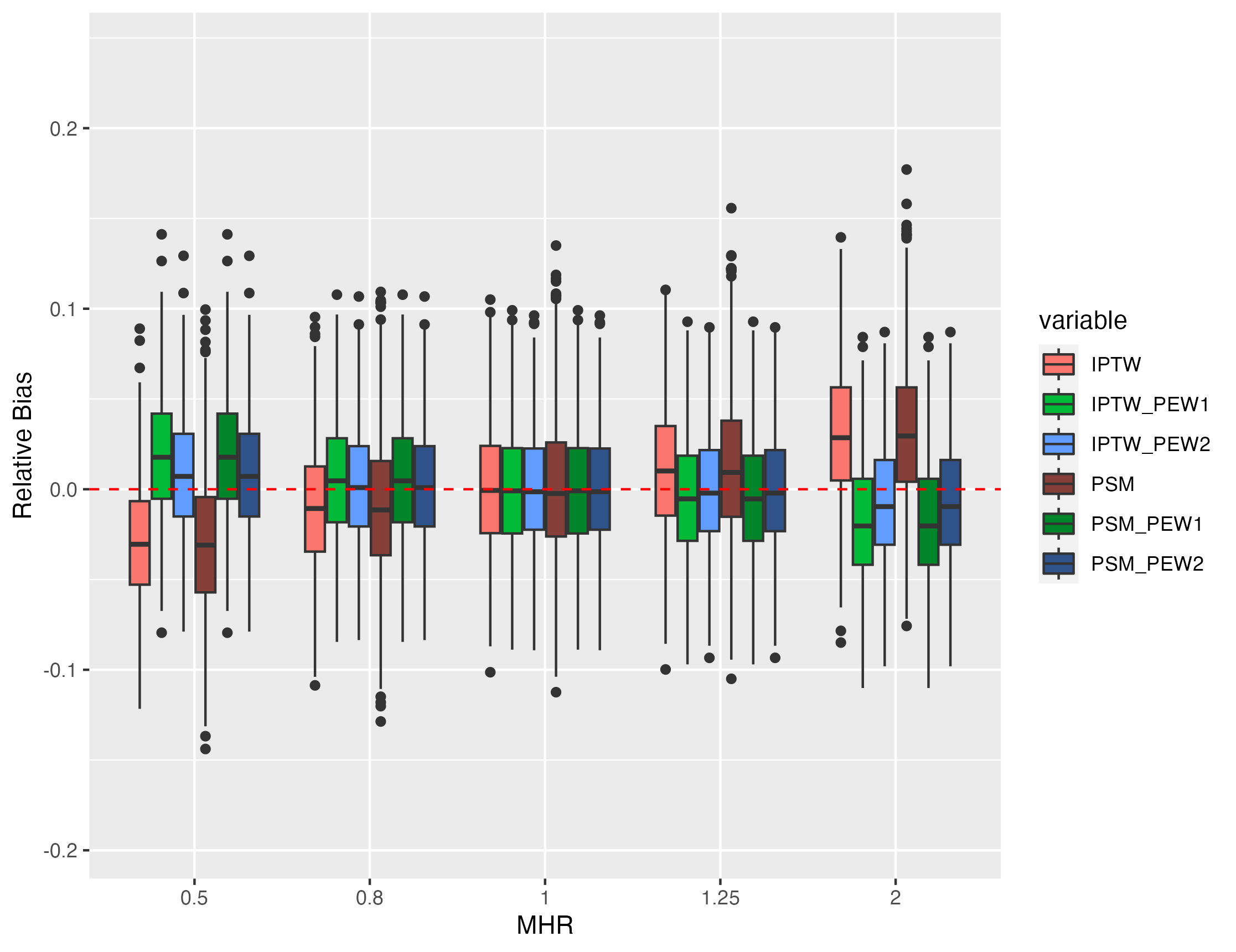}
\caption{Relative bias with censoring rate = 0.3}
\label{fig:n6000_notcounterfactual_censoring0.3}
\end{subfigure}\hfill
\begin{subfigure}{0.49\columnwidth}
\centering
\includegraphics[width=\textwidth]{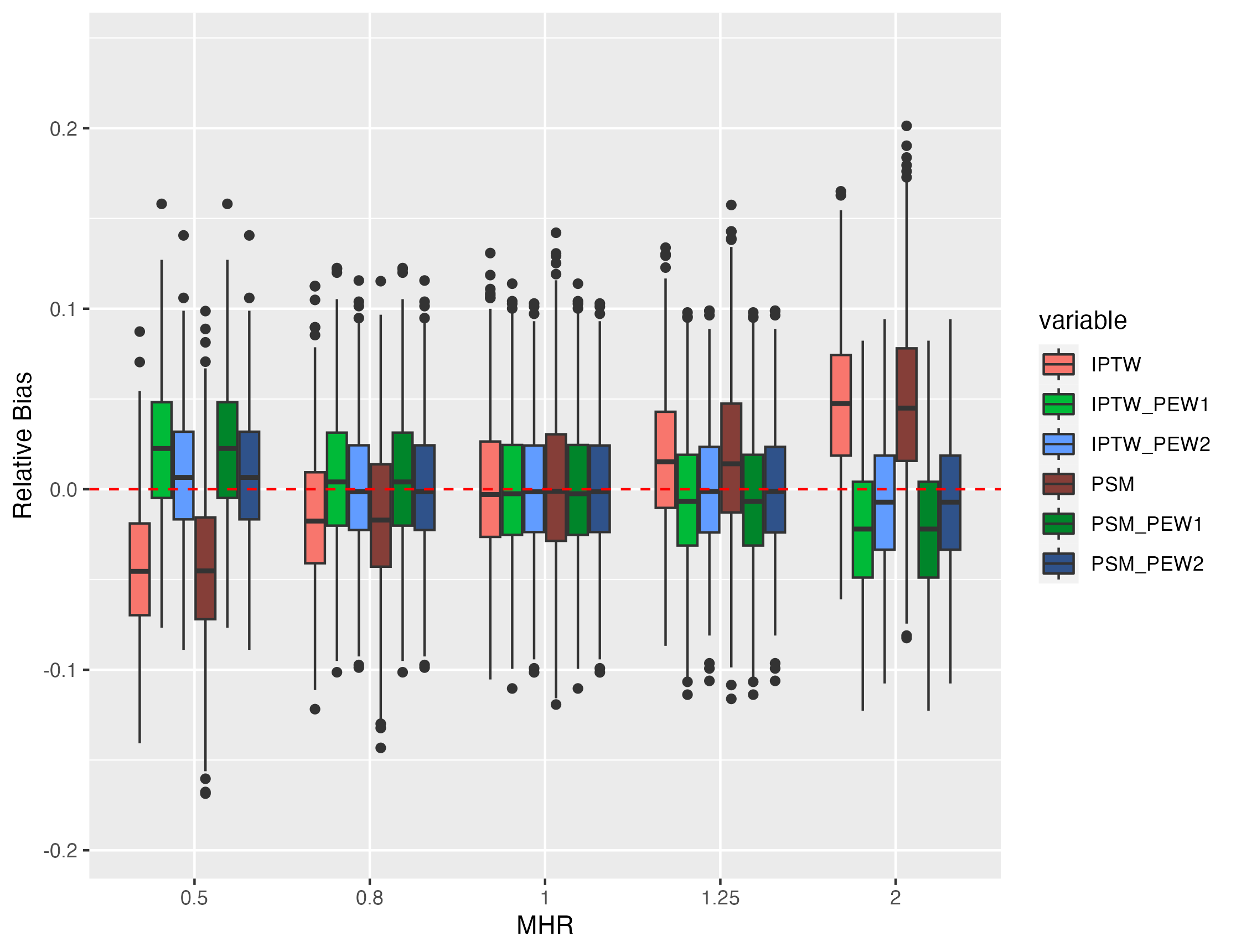}
\caption{Relative bias with censoring rate = 0.4}
\label{fig:n6000_notcounterfactual_censoring0.4}
\end{subfigure}

\medskip

\begin{subfigure}{0.49\columnwidth}
\centering
\includegraphics[width=\textwidth]{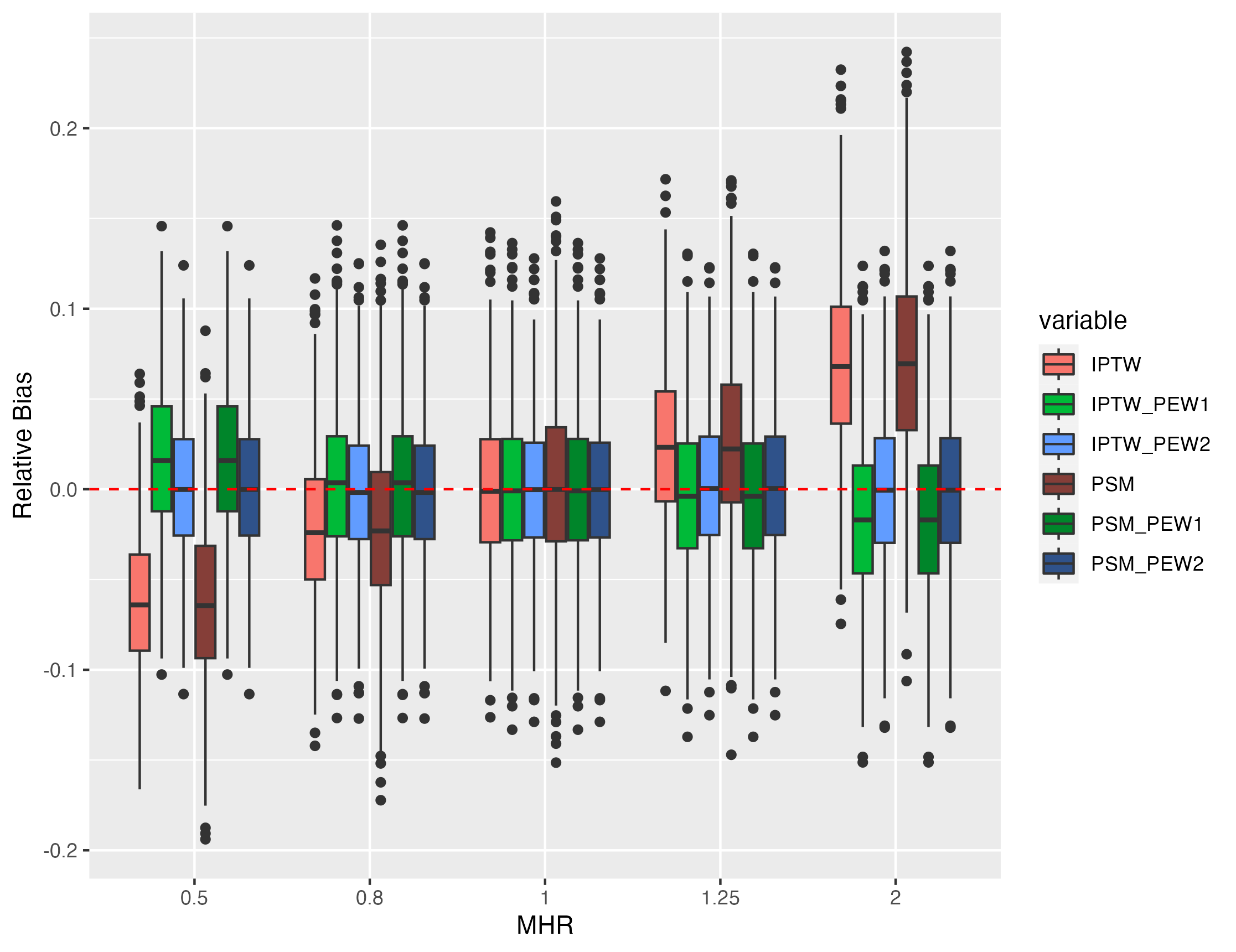}
\caption{Relative bias with censoring rate = 0.5}
\label{fig:n6000_notcounterfactual_censoring0.5}
\end{subfigure}

\caption{Observational setting, N = 6000, censoring rates = (0.1, 0.2, 0.3, 0.4, 0.5). Relative bias of estimation of MHR under several true values of MHR. Results based on 1000 simulation replicates.}
\label{fig:n6000_notcounterfactual_censoring_several_01_05}

\end{figure}

\begin{figure}[htp]
\centering

\begin{subfigure}{0.49\columnwidth}
\centering
\includegraphics[width=\textwidth]{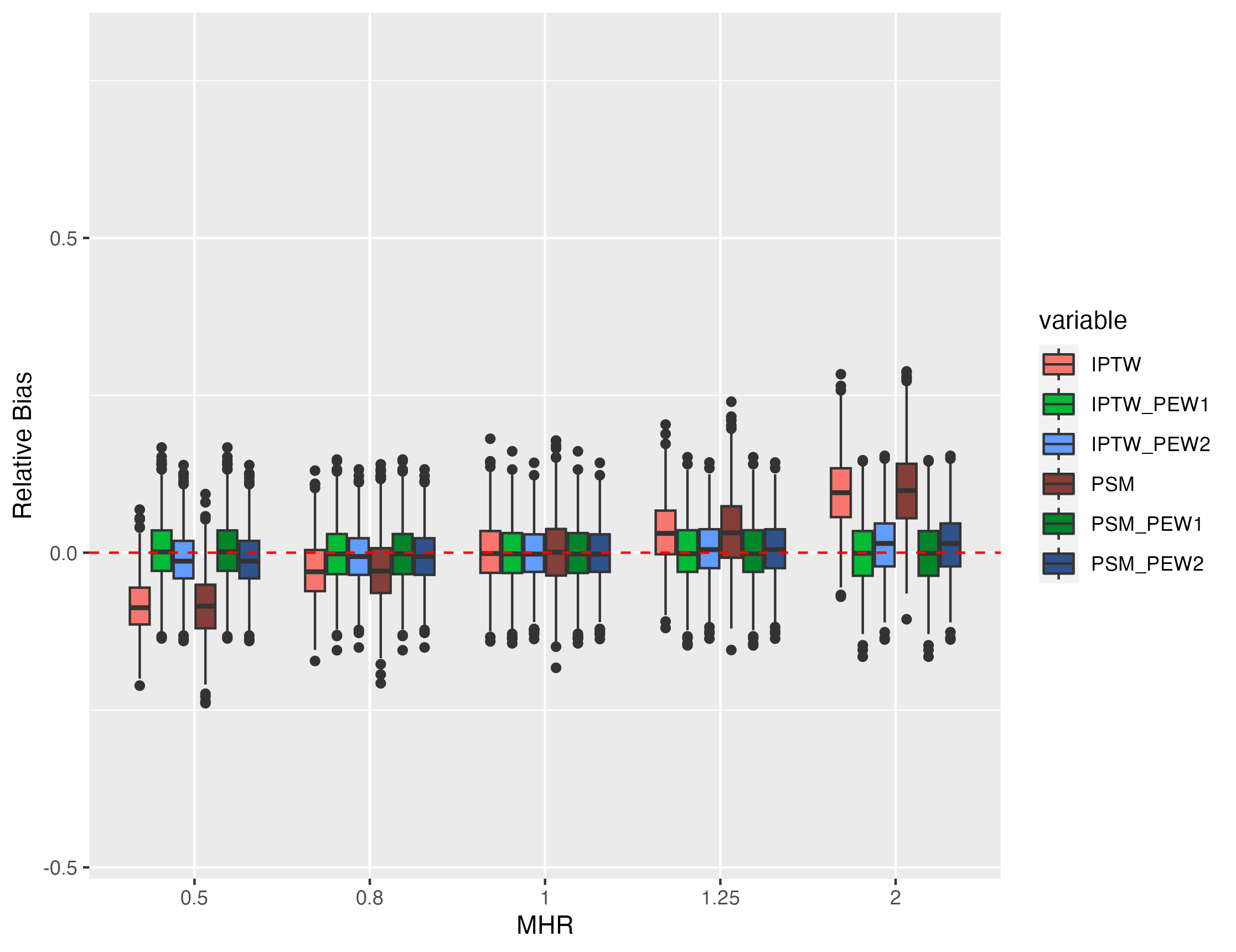}
\caption{Relative bias with censoring rate = 0.6}
\label{fig:n6000_notcounterfactual_censoring0.6}
\end{subfigure}\hfill
\begin{subfigure}{0.49\columnwidth}
\centering
\includegraphics[width=\textwidth]{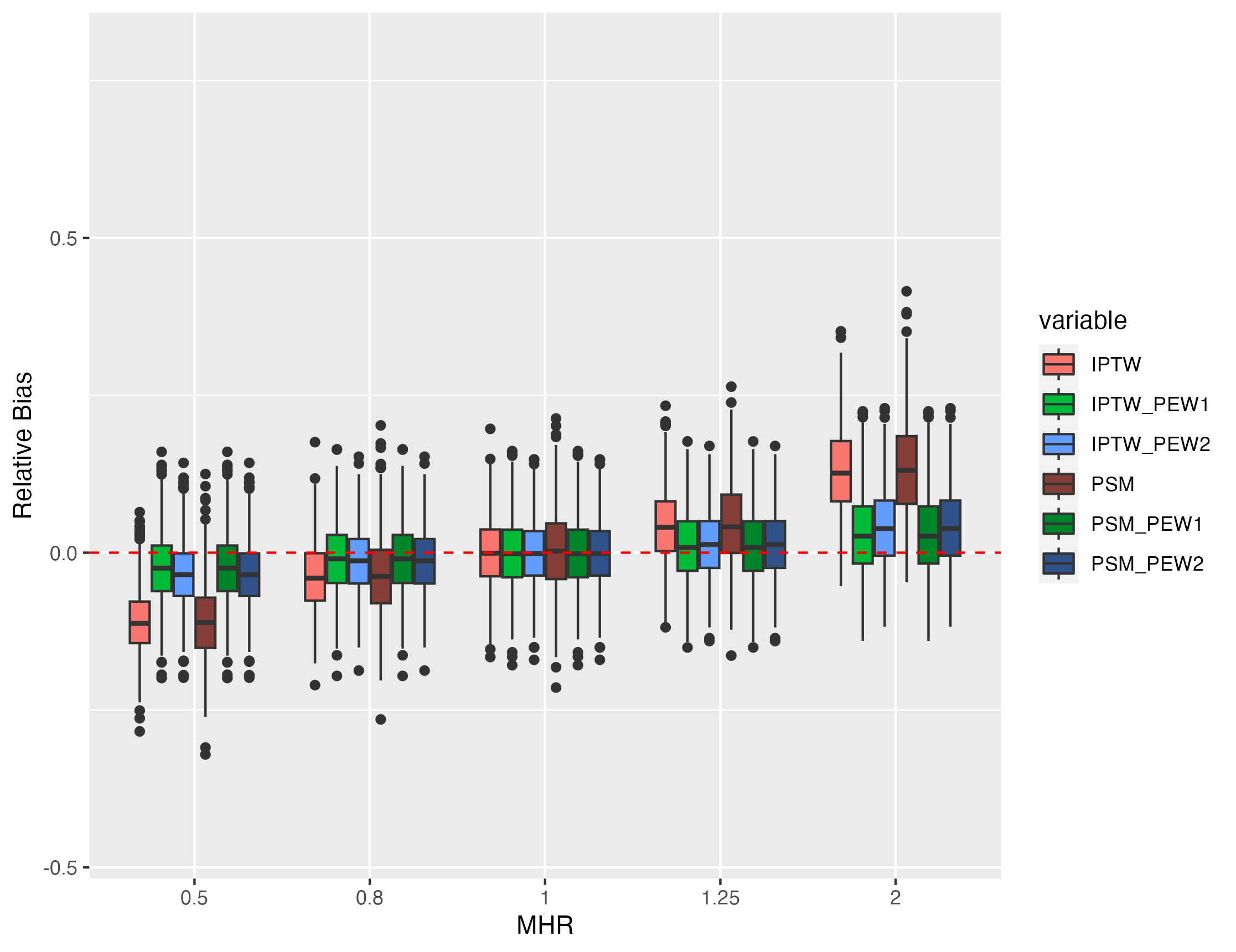}
\caption{Relative bias with censoring rate = 0.7}
\label{fig:n6000_notcounterfactual_censoring0.7}
\end{subfigure}

\medskip

\begin{subfigure}{0.49\columnwidth}
\centering
\includegraphics[width=\textwidth]{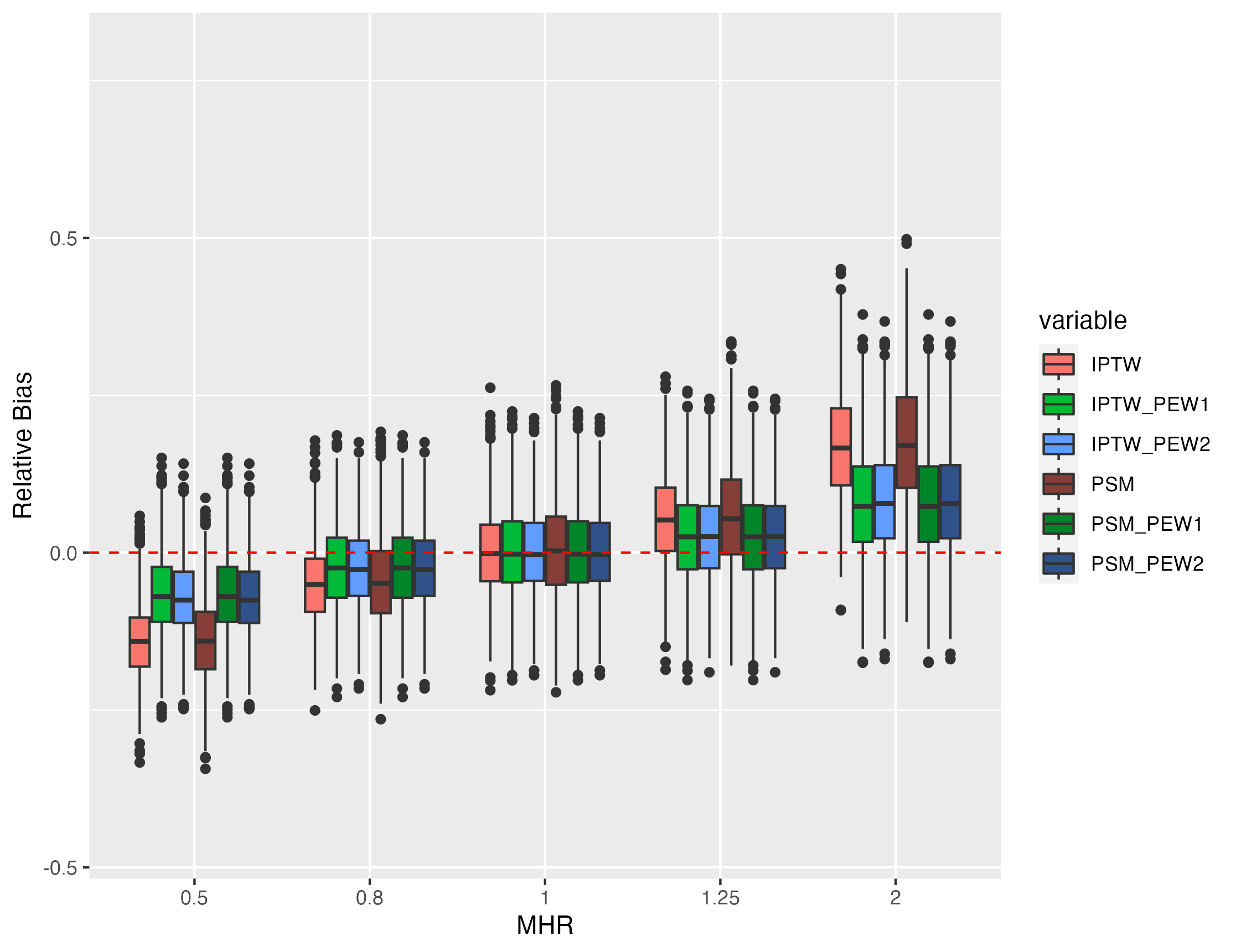}
\caption{Relative bias with censoring rate = 0.8}
\label{fig:n6000_notcounterfactual_censoring0.8}
\end{subfigure}\hfill
\begin{subfigure}{0.49\columnwidth}
\centering
\includegraphics[width=\textwidth]{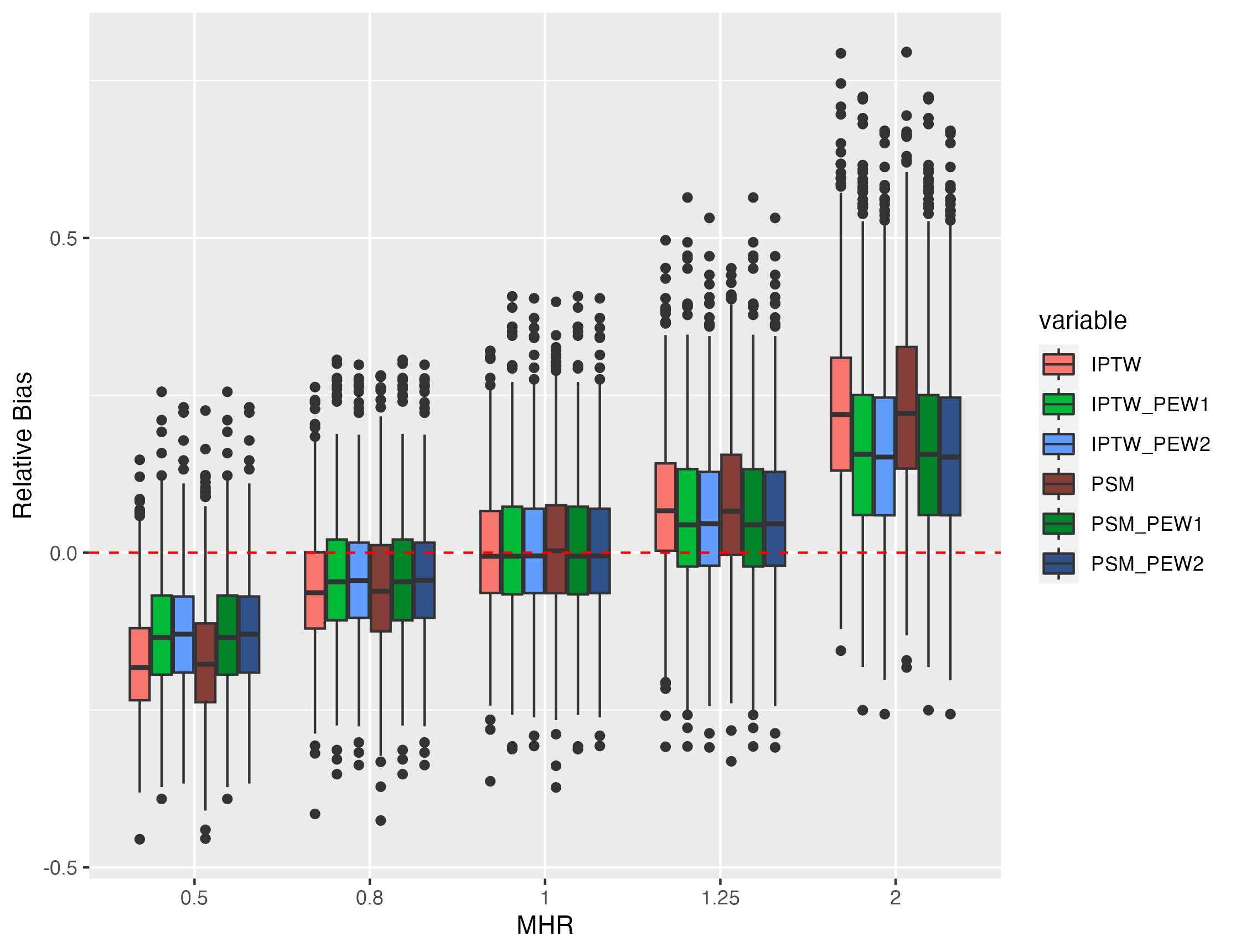}
\caption{Relative bias with censoring rate = 0.9}
\label{fig:n6000_notcounterfactual_censoring0.9}
\end{subfigure}

\caption{Observational setting, N = 6000, censoring rates = (0.6, 0.7, 0.8, 0.9). Relative bias of estimation of MHR under several true values of MHR. Results based on 1000 simulation replicates.}
\label{fig:n6000_notcounterfactual_censoring_several_06_09}

\end{figure}

Results describing estimation of MHR in the counterfactual setting are reported in Figures \ref{fig:n6000_counterfactual_censoring_several_01_05} - \ref{fig:n6000_counterfactual_censoring_several_06_09} ($N = 6000$) and Figures \ref{fig:n2000_counterfactual_censoring_several_01_05} - \ref{fig:n10000_counterfactual_censoring_several_06_09} ($N = 2000, 10000$) and Tables \ref{table:uniform_2000_MHR0.5_counterfactual} - \ref{table:uniform_10000_MHR2_counter} in Appendix \ref{section:AppendixA}. In the counterfactual settings, the two treatment groups were identical at baseline so any bias in the estimation of MHR was related to what happened after baseline. The results from the observational setting can be seen in Figures \ref{fig:n6000_notcounterfactual_censoring_several_01_05} - \ref{fig:n6000_notcounterfactual_censoring_several_06_09} ($N = 6000$) and Figures \ref{fig:n2000_notcounterfactual_censoring_several_01_05} - \ref{fig:n10000_notcounterfactual_censoring_several_06_09} ($N = 2000, 10000$) and Tables \ref{table:uniform_2000_MHR0.5} - \ref{table:uniform_10000_MHR2_notcounterfactual} in Appendix \ref{section:AppendixB}. 

We start by discussing the results from the counterfactual setting. 
First off, as seen elsewhere in the literature \citep{austin-mis-full}, IPTW and PSM had almost identical performance, with similar values for Bias, SD, RMSE and Coverage. 
Secondly, there were two important factors that affected the bias: the censoring rate and the MHR value. The bias was tied directly to the amount of censoring, with very low bias in settings with low censoring proportions and high censoring proportions resulted in considerable bias. When MHR = 1 (no treatment effect) the PS based methods were unbiased regardless of censoring rate and the sample of individuals at risk in the end is equivalent to the baseline sample, since there is no depletion of susceptibles \citep{wyss}. However, when MHR was farther from 1, both for positive and negative effects on the time-to-event, the bias increased, since the final individuals at risk and baseline sample are substantially different due to the treatment effect.

Even with perfect balance at baseline, IPTW and PSM resulted in substantial bias when there was moderate to high censoring (40-50\% or higher). This is an important insight related to reporting results from RCTs. Modifying the PS based methods by incorporating PE weights showed an improvement in performance when estimating MHR. Using the modified weights with either the true conditional probability of event (IPTW\_PEW1 and MATCH\_PEW1) or the estimated, with correct model, probability (IPTW\_PEW2 and MATCH\_PEW2) managed to reduce bias to a similar extent.

It should be noted that the PE modified methods had results comparable to the conventional PS based methods in settings where the latter were unbiased, while the modified methods reduced bias in the scenarios where the PS based methods were biased. However, PE modified methods still resulted in considerable bias in scenarios with high censoring rates (70-80\% or higher). 

IPTW and PSM resulted in empirical coverage above the nominal 0.95 for censoring proportions 40-50\% or lower when $N = 2000$ and for censoring proportions of 30\% or lower when $N = 6000$. It is known that the robust sandwich type variance estimator can result in conservative confidence intervals at lower censoring rates \citep{austin-variance}, but in scenarios with high rates of censoring even such an overestimation of the variance was not enough to achieve coverage of 0.95.  The PE modified methods had empirical coverage above 0.95 for all censoring rates when $N = 2000$ and for censoring proportions 70\% or lower when $N = 6000$ and showed equivalent improvement for both matching and weighting methods.

Simulations with sample size $N = 10000$ (Table \ref{table:uniform_10000_MHR0.5_counter} - \ref{table:uniform_10000_MHR2_counter}) revealed that the bias had stabilized at $N = 6000$. Increasing the sample size resulted in even lower empirical coverages as the confidence intervals got shorter, due to the reduced variance, and were centered around a biased estimate of MHR.

The results from the observational setting generally followed the same pattern as in the counterfactual setting. The bias in the two settings were similar but the Monte Carlo SD and hence RMSE were larger in the observational setting. 
The empirical coverage for IPTW and PSM was at or above 0.95 for censoring proportions 30\% or lower when $N = 2000$ and 20\% or lower when $N = 6000$. For the PE modified methods the empirical coverage was at or above the nominal for censoring proportions 60\% or lower when $N = 2000, 6000, 10000$.

\FloatBarrier
\section{Discussion}
\label{section:discussion}

The impact of non-informative censoring on PS based estimation of MHR  has largely gone unnoticed in the previous literature, with a few exceptions \citep{wyss, fireman}. In this paper an extensive simulation study (considering a range of scenarios with varying sample sizes, censoring rates, and MHR values) was carried out to explore the impact in a more systematic manner. 

The simulation results revealed that PS based estimation of MHR was biased (in the presence of non-informative censoring) and that the bias exacerbated when the censoring rate increased and when MHR was far from 1. 
Bias was present both in simulations mimicking observational data and counterfactual data (a scenario equivalent to an ideal RCT). However, even in situations with high rate of non-informative censoring, incorrectly concluding that a treatment has an effect (type I error) is unlikely, since the PS based estimators are unbiased when MHR = 1.



In an attempt to correct for such bias, modified PS based estimators (including weights related to the conditional probabilty of suffering an event) were suggested and compared with the conventional PS estimators. The modification was reasonably successful in reducing bias in scenarios with low or moderate censoring, but it only managed to partially reduce bias in scenarios with high censoring. 

The problem of biased PS based MHR estimation under high rate of non-informative censoring is yet to be solved. For now, we have shown that it is possible to reduce such bias with modified PS weighting and recommend further research along the same vein for more successful methods. When estimating MHR in practice, researchers should be aware of the impact of non-informative censoring and we urge researchers to also report other effect measures, such as survival curves and CHR, in order to get a clearer picture of the effect of any proposed treatment.




\section*{Acknowledgements}
The authors are grateful to Associate Professor Anita Lindmark for helpful and constructive comments. This work was supported by the Swedish Research Council 377 (Dnr: 2018–01610).


\nocite{*}

\printbibliography[category=cited]

\FloatBarrier

\newpage
\appendix
\counterwithin{figure}{section}
\counterwithin{table}{section}

\section{Appendix A, Counterfactual, Uniform Censoring}
\label{section:AppendixA}

\begin{figure}[htp]
\centering

\begin{subfigure}{0.49\columnwidth}
\centering
\includegraphics[width=\textwidth]{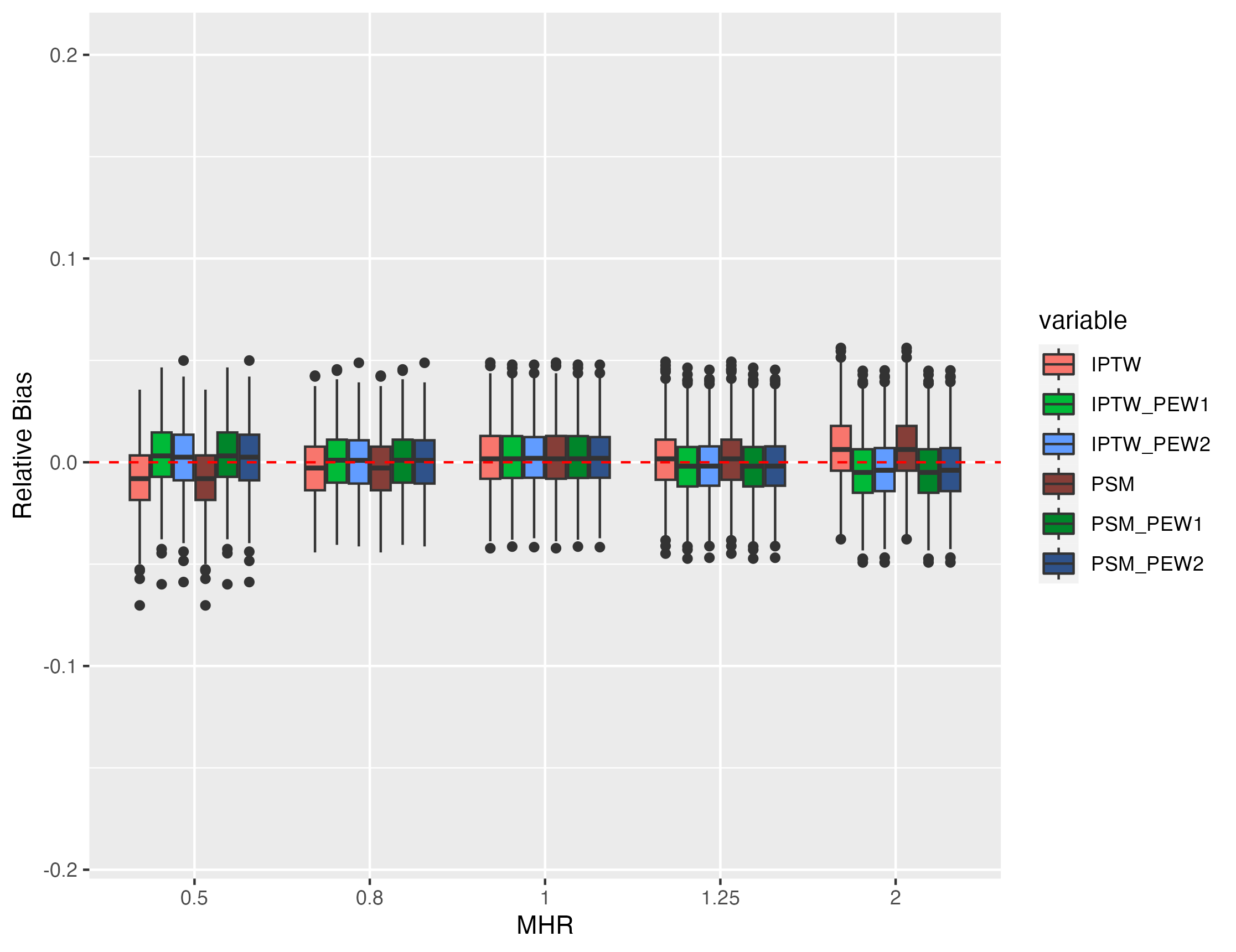}
\caption{Relative bias with censoring rate = 0.1}
\label{fig:n2000_counterfactual_censoring0.1}
\end{subfigure}\hfill
\begin{subfigure}{0.49\columnwidth}
\centering
\includegraphics[width=\textwidth]{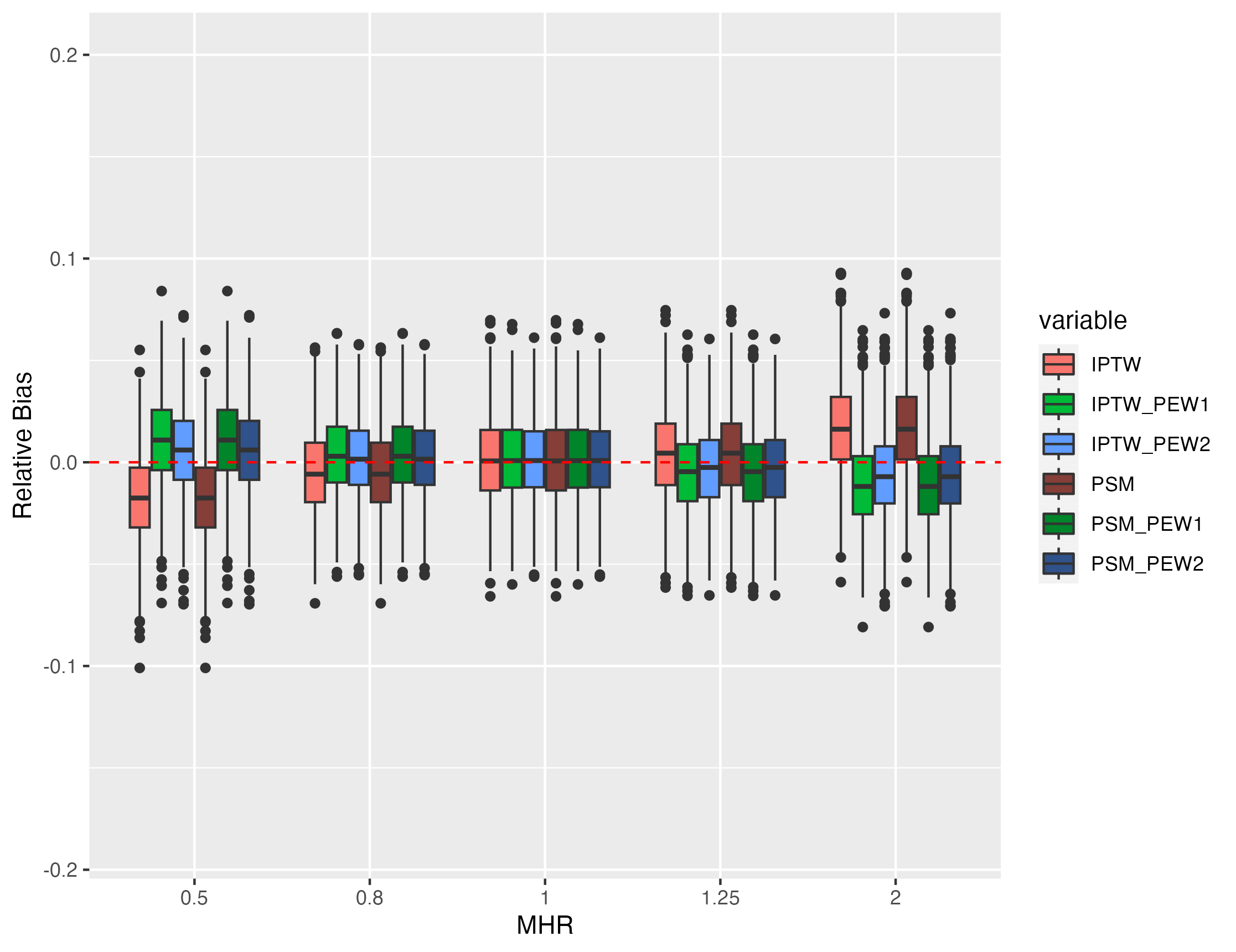}
\caption{Relative bias with censoring rate = 0.2}
\label{fig:n2000_counterfactual_censoring0.2}
\end{subfigure}

\medskip

\begin{subfigure}{0.49\columnwidth}
\centering
\includegraphics[width=\textwidth]{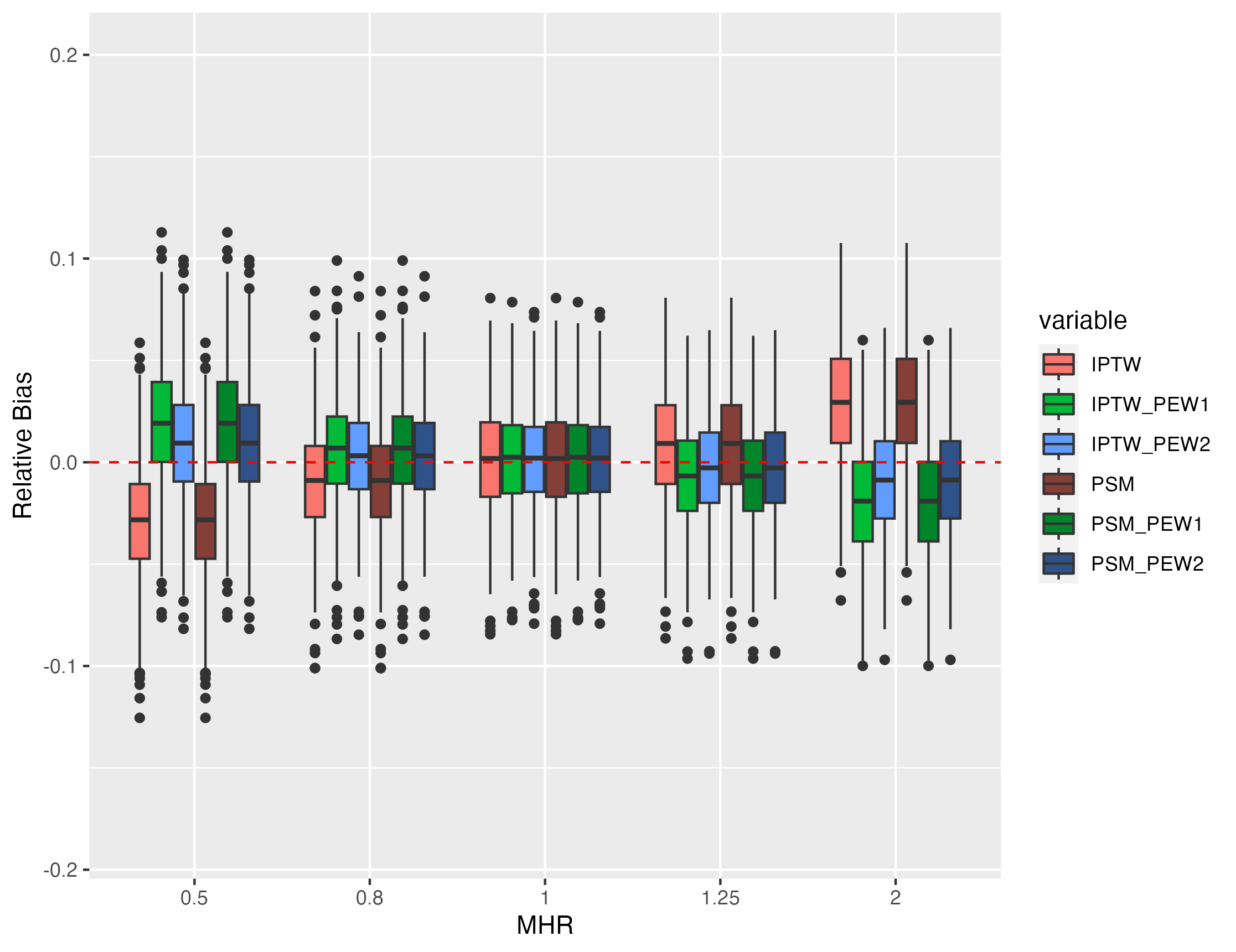}
\caption{Relative bias with censoring rate = 0.3}
\label{fig:n2000_counterfactual_censoring0.3}
\end{subfigure}\hfill
\begin{subfigure}{0.49\columnwidth}
\centering
\includegraphics[width=\textwidth]{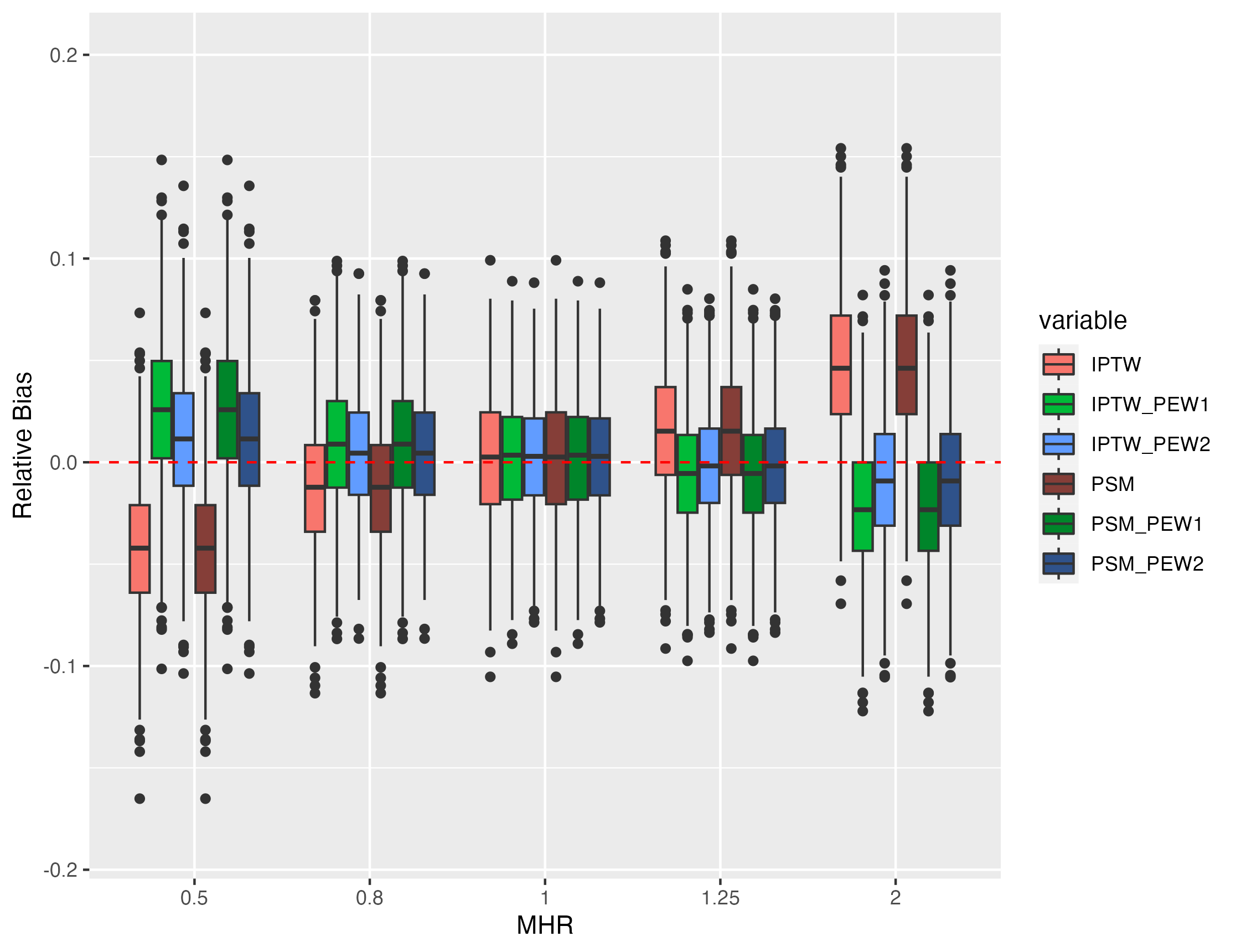}
\caption{Relative bias with censoring rate = 0.4}
\label{fig:n2000_counterfactual_censoring0.4}
\end{subfigure}

\medskip

\begin{subfigure}{0.49\columnwidth}
\centering
\includegraphics[width=\textwidth]{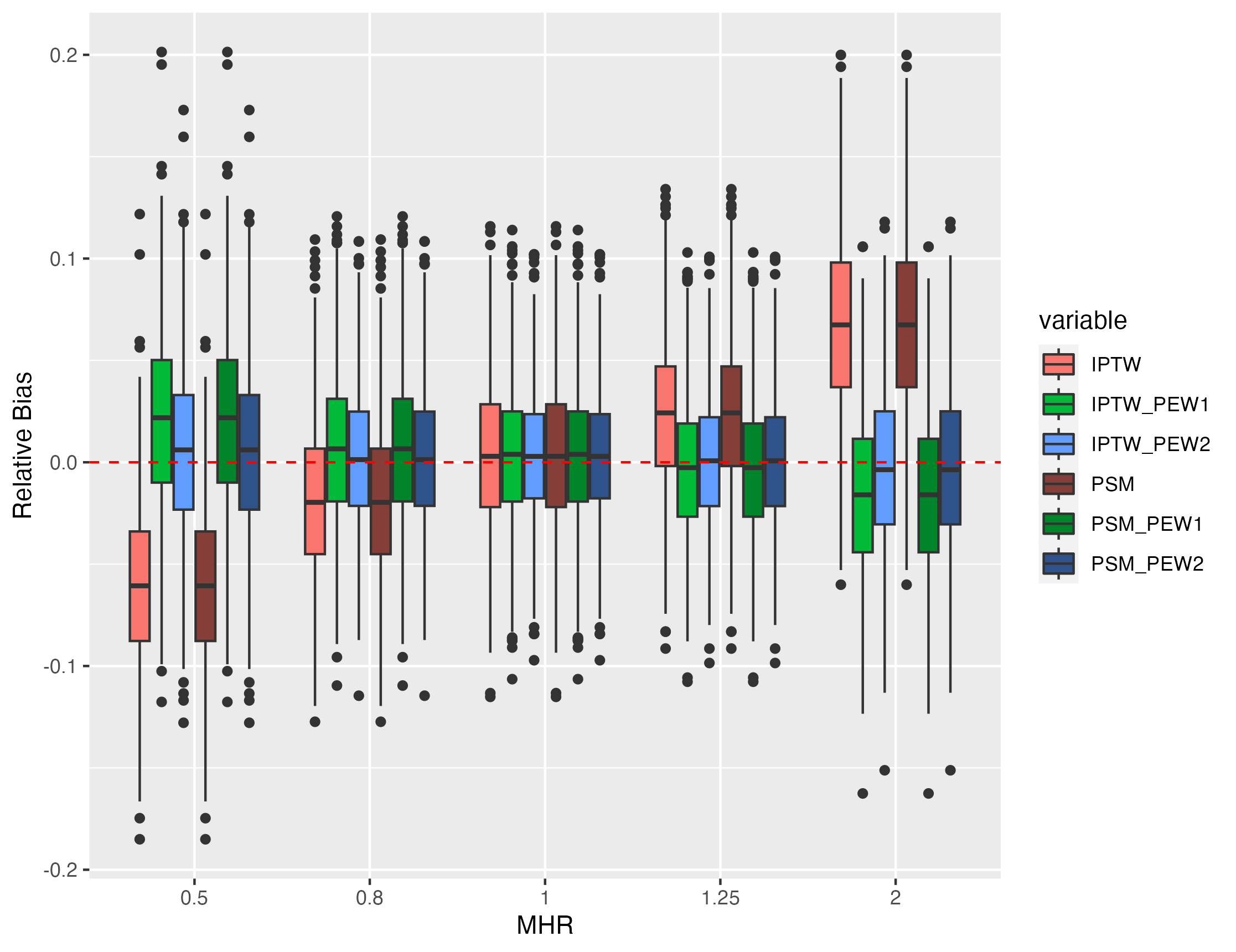}
\caption{Relative bias with censoring rate = 0.5}
\label{fig:n2000_counterfactual_censoring0.5}
\end{subfigure}

\caption{Counterfactual setting, N = 2000, censoring rates = (0.1, 0.2, 0.3, 0.4, 0.5). Relative bias of estimation of MHR under several true values of MHR. Results based on 1000 simulation replicates.}
\label{fig:n2000_counterfactual_censoring_several_01_05}

\end{figure}

\begin{figure}[htp]
\centering

\begin{subfigure}{0.49\columnwidth}
\centering
\includegraphics[width=\textwidth]{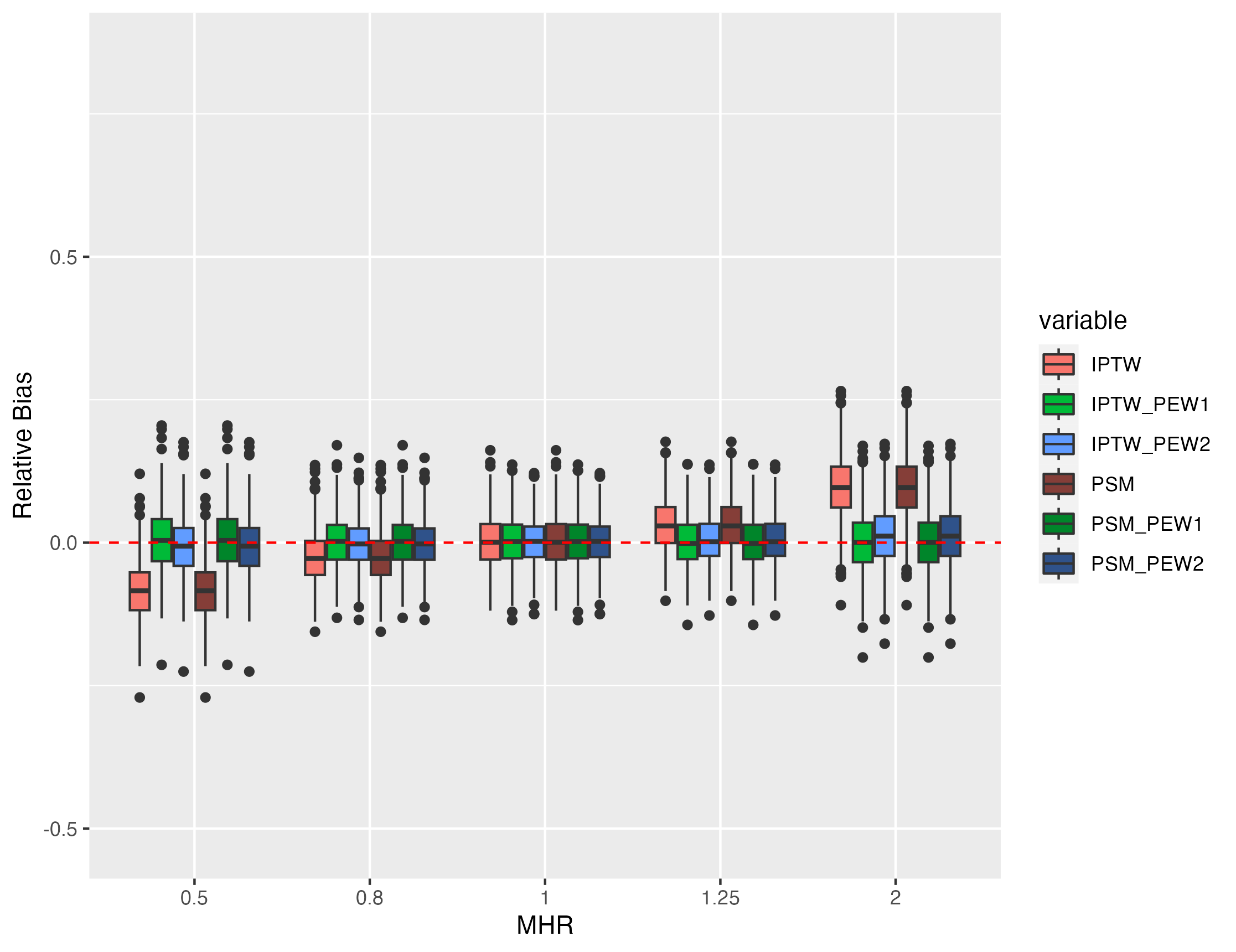}
\caption{Relative bias with censoring rate = 0.6}
\label{fig:n2000_counterfactual_censoring0.6}
\end{subfigure}\hfill
\begin{subfigure}{0.49\columnwidth}
\centering
\includegraphics[width=\textwidth]{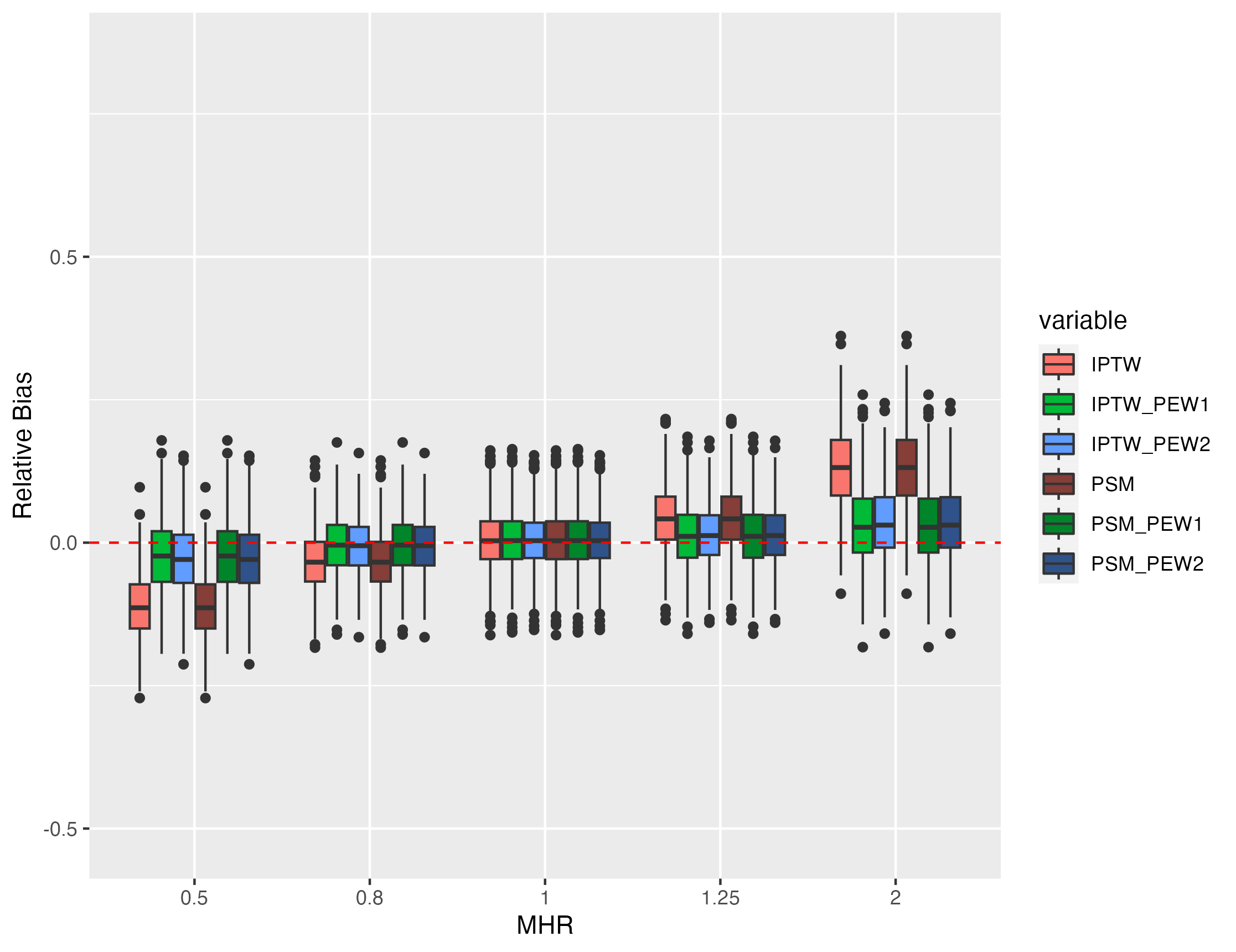}
\caption{Relative bias with censoring rate = 0.7}
\label{fig:n2000_counterfactual_censoring0.7}
\end{subfigure}

\medskip

\begin{subfigure}{0.49\columnwidth}
\centering
\includegraphics[width=\textwidth]{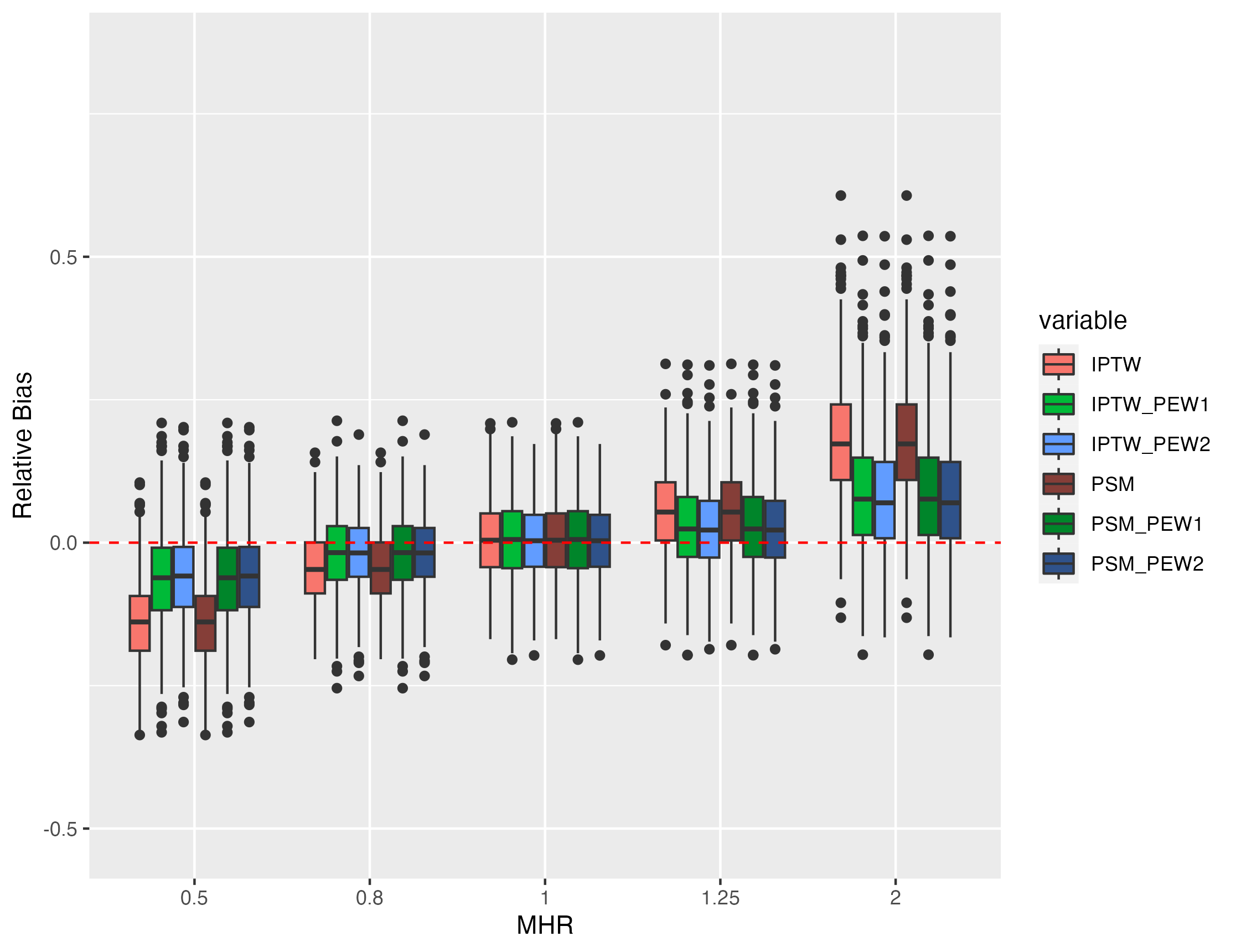}
\caption{Relative bias with censoring rate = 0.8}
\label{fig:n2000_counterfactual_censoring0.8}
\end{subfigure}\hfill
\begin{subfigure}{0.49\columnwidth}
\centering
\includegraphics[width=\textwidth]{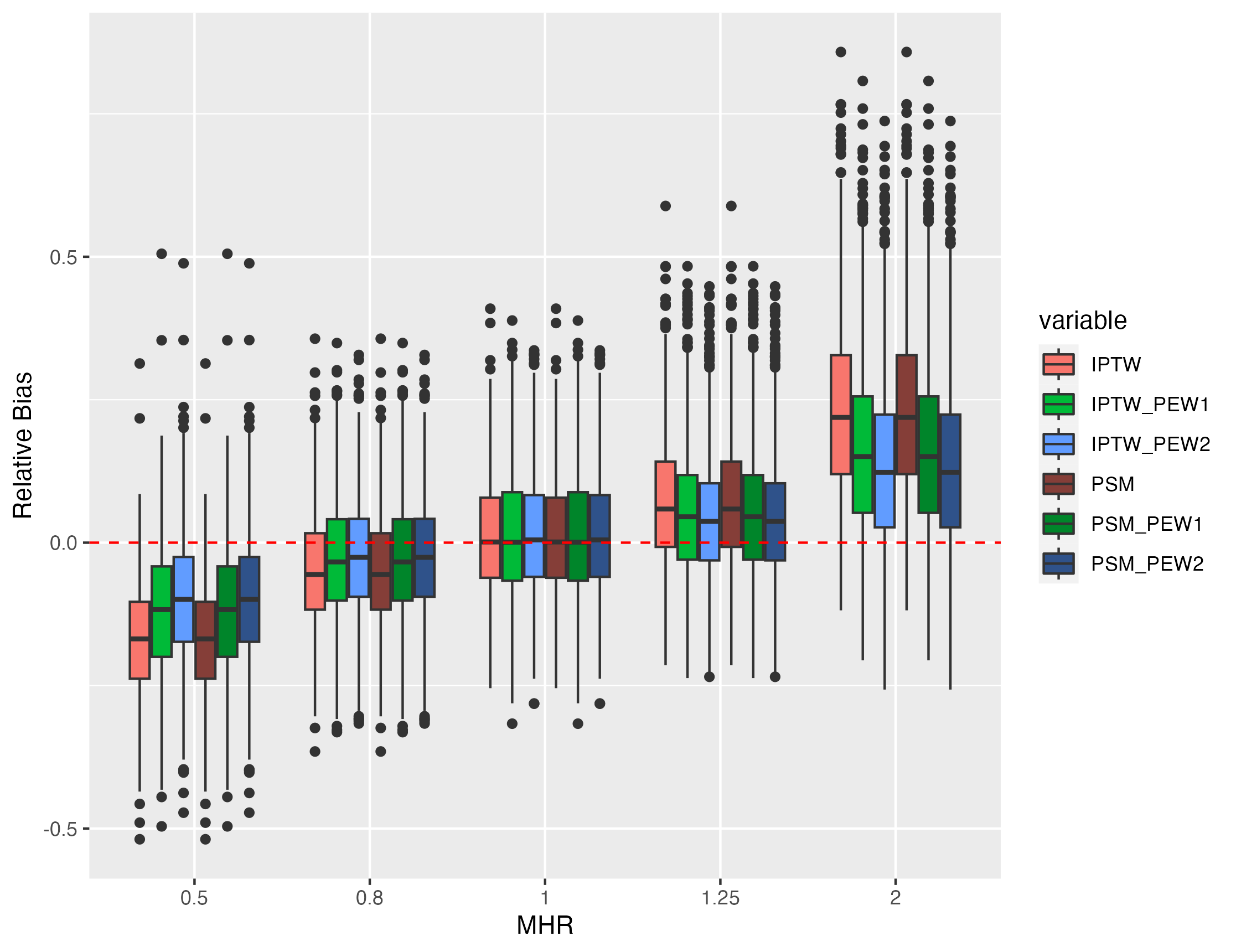}
\caption{Relative bias with censoring rate = 0.9}
\label{fig:n2000_counterfactual_censoring0.9}
\end{subfigure}

\caption{Counterfactual setting, N = 2000, censoring rates = (0.6, 0.7, 0.8, 0.9). Relative bias of estimation of MHR under several true values of MHR. Results based on 1000 simulation replicates.}
\label{fig:n2000_counterfactual_censoring_several_06_09}

\end{figure}

\begin{figure}[htp]
\centering

\begin{subfigure}{0.49\columnwidth}
\centering
\includegraphics[width=\textwidth]{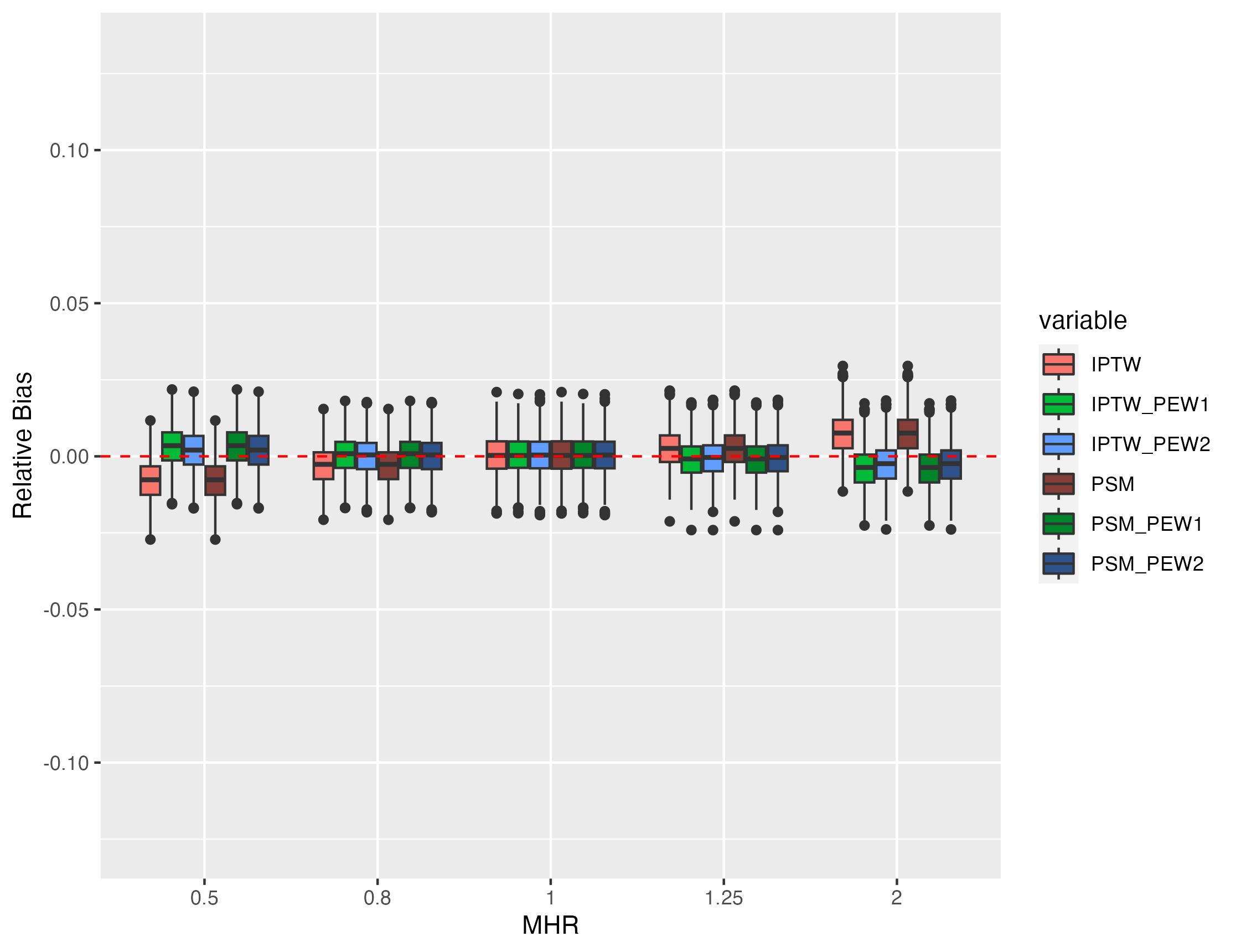}
\caption{Relative bias with censoring rate = 0.1}
\label{fig:n10000_counterfactual_censoring0.1}
\end{subfigure}\hfill
\begin{subfigure}{0.49\columnwidth}
\centering
\includegraphics[width=\textwidth]{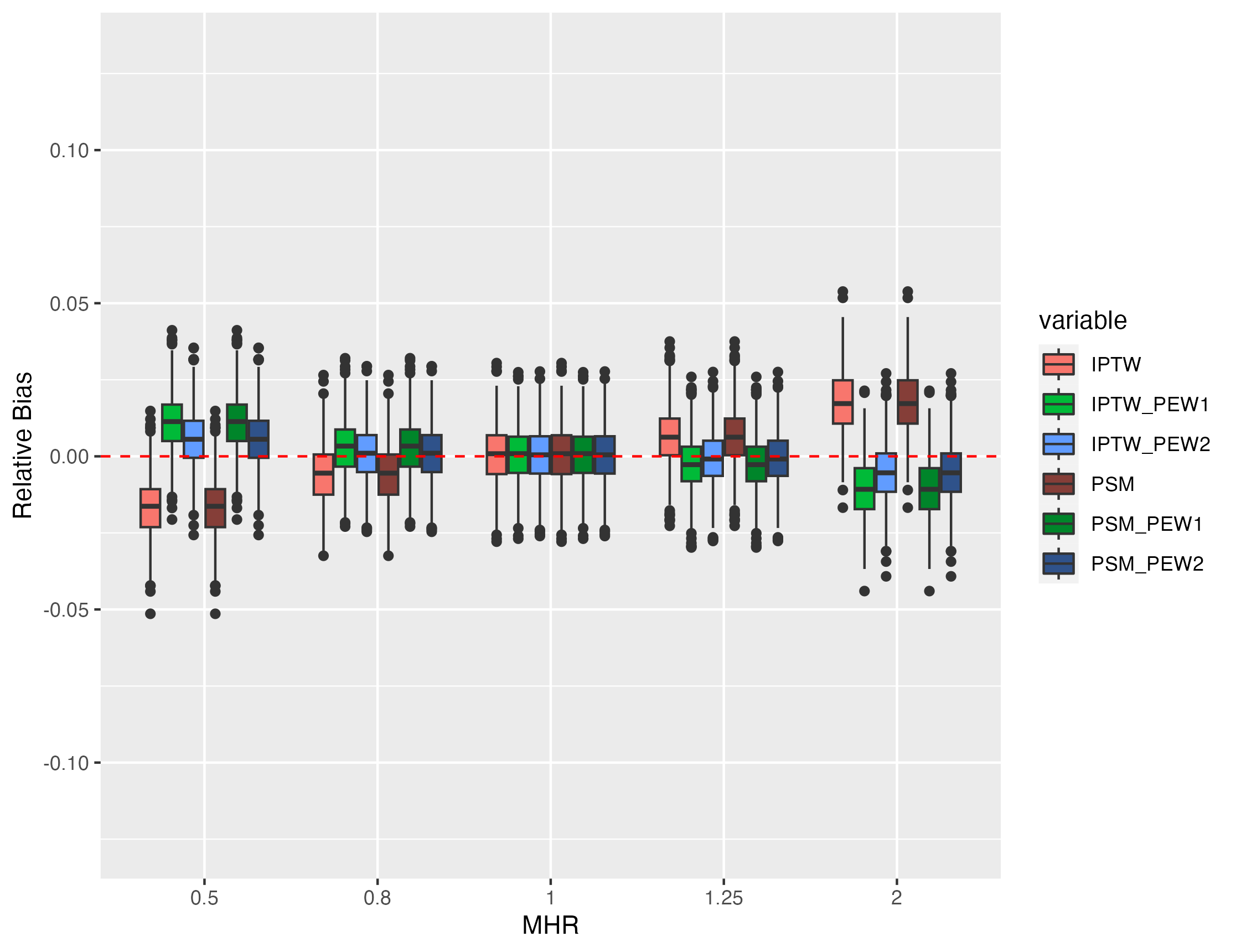}
\caption{Relative bias with censoring rate = 0.2}
\label{fig:n10000_counterfactual_censoring0.2}
\end{subfigure}

\medskip

\begin{subfigure}{0.49\columnwidth}
\centering
\includegraphics[width=\textwidth]{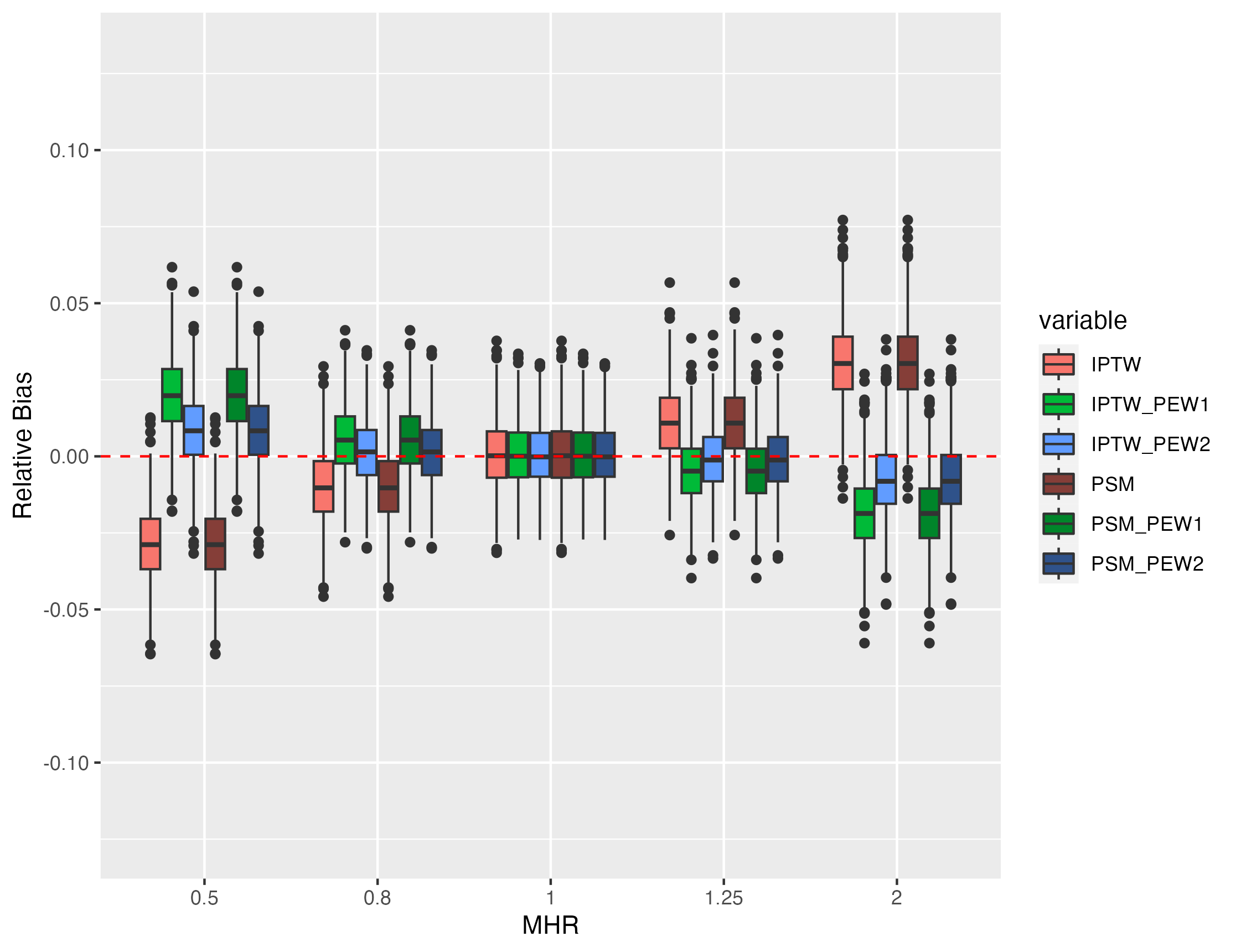}
\caption{Relative bias with censoring rate = 0.3}
\label{fig:n10000_counterfactual_censoring0.3}
\end{subfigure}\hfill
\begin{subfigure}{0.49\columnwidth}
\centering
\includegraphics[width=\textwidth]{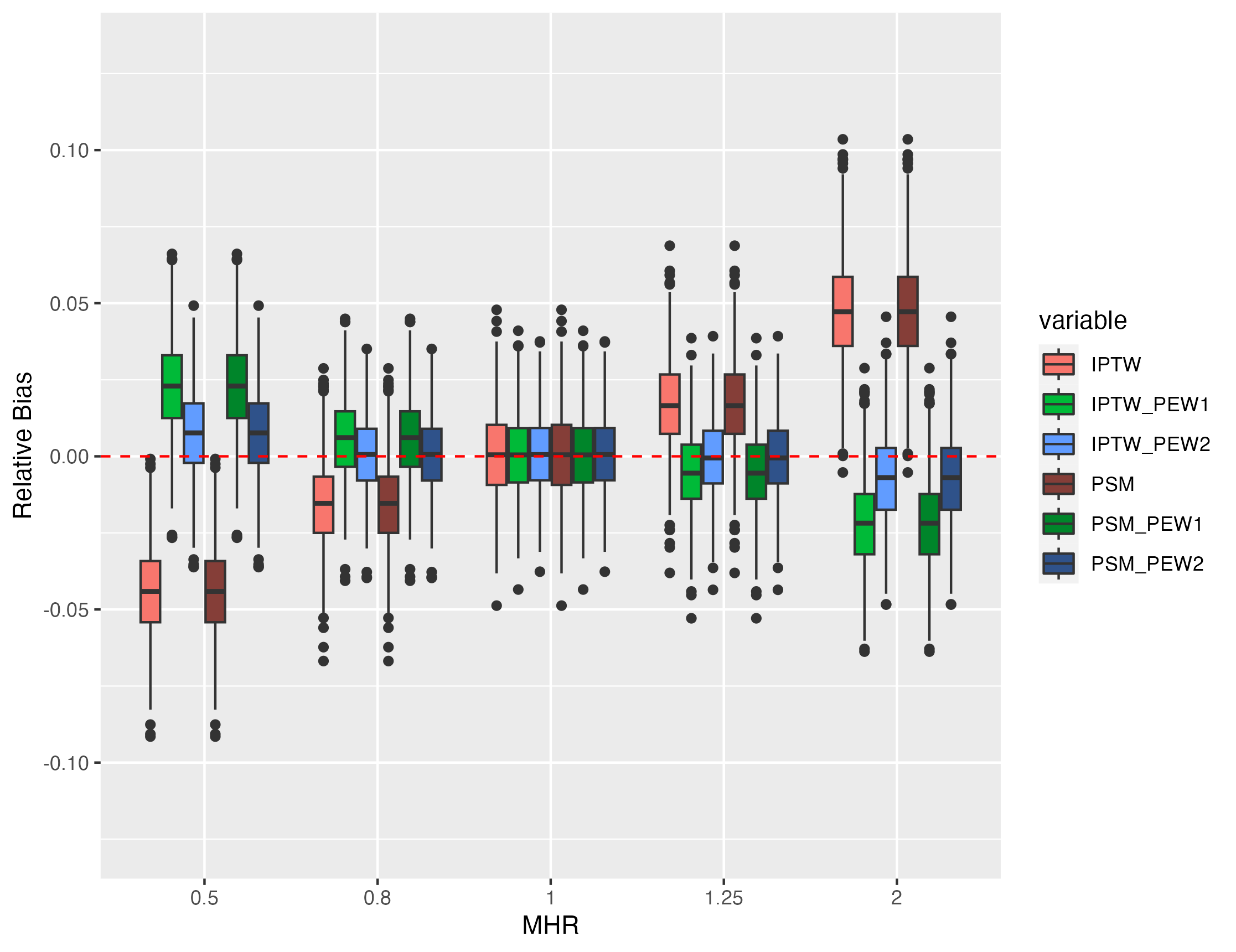}
\caption{Relative bias with censoring rate = 0.4}
\label{fig:n10000_counterfactual_censoring0.4}
\end{subfigure}

\medskip

\begin{subfigure}{0.49\columnwidth}
\centering
\includegraphics[width=\textwidth]{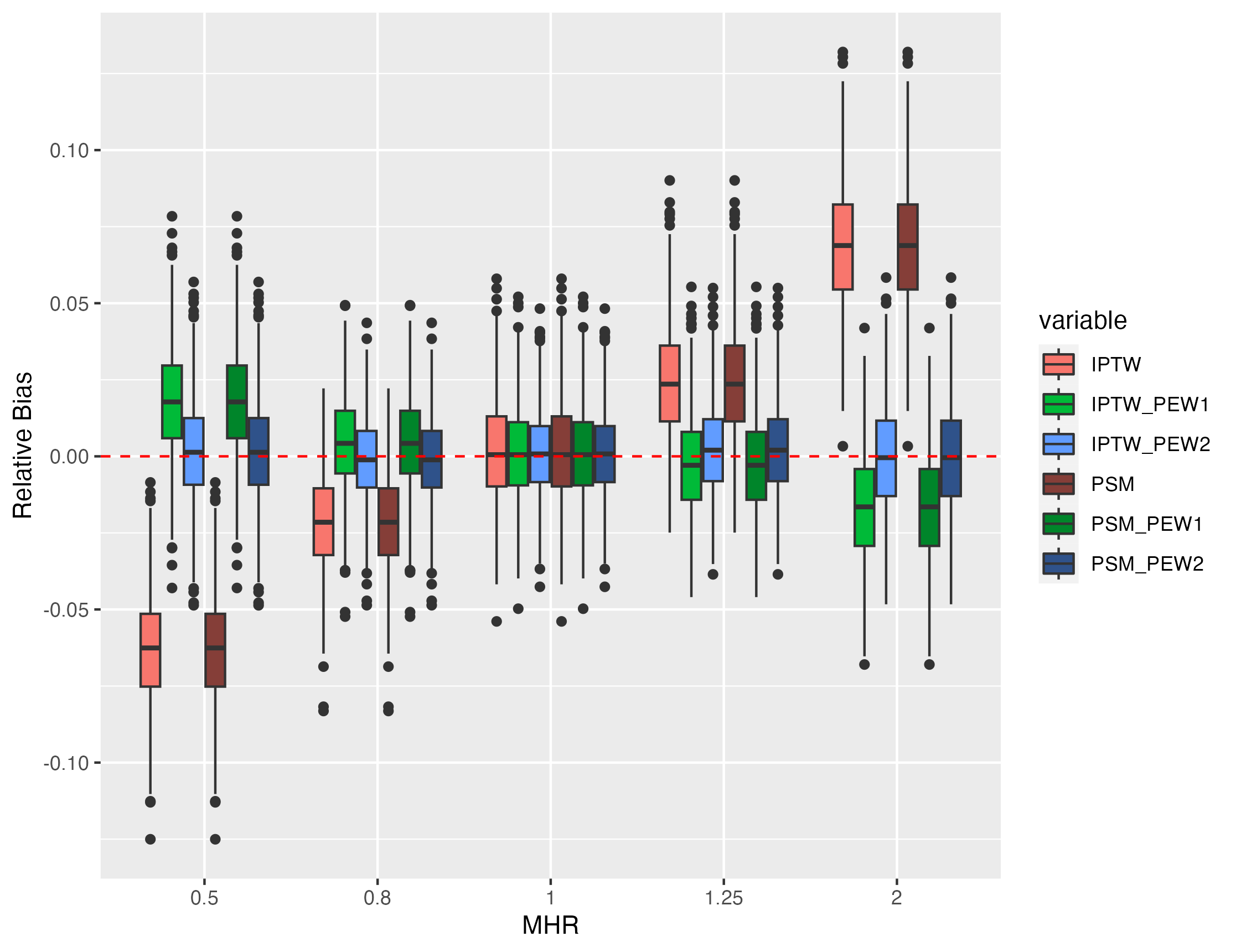}
\caption{Relative bias with censoring rate = 0.5}
\label{fig:n10000_counterfactual_censoring0.5}
\end{subfigure}

\caption{Counterfactual setting, N = 10000, censoring rates = (0.1, 0.2, 0.3, 0.4, 0.5). Relative bias of estimation of MHR under several true values of MHR. Results based on 1000 simulation replicates.}
\label{fig:n10000_counterfactual_censoring_several_01_05}

\end{figure}

\begin{figure}[htp]
\centering

\begin{subfigure}{0.49\columnwidth}
\centering
\includegraphics[width=\textwidth]{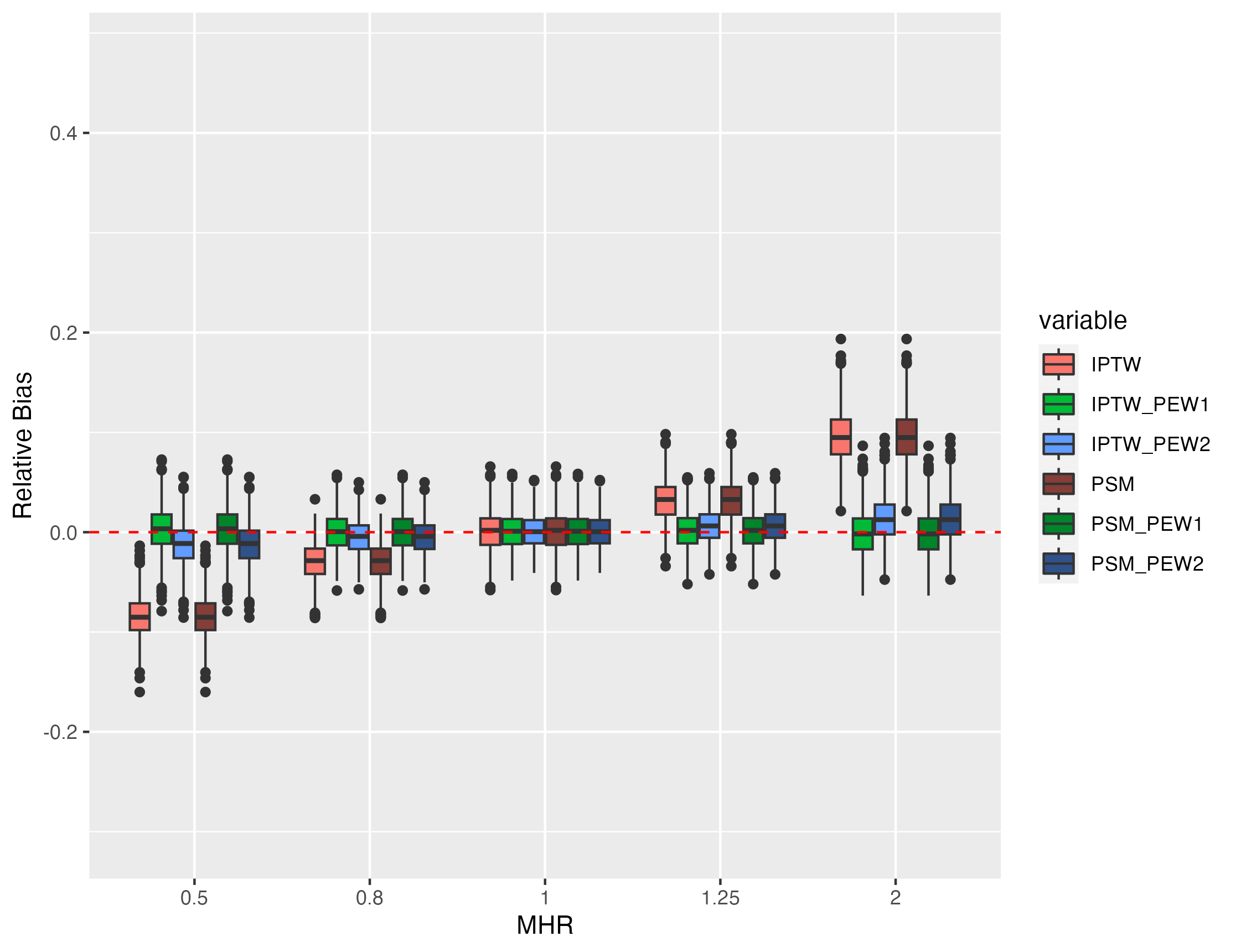}
\caption{Relative bias with censoring rate = 0.6}
\label{fig:n10000_counterfactual_censoring0.6}
\end{subfigure}\hfill
\begin{subfigure}{0.49\columnwidth}
\centering
\includegraphics[width=\textwidth]{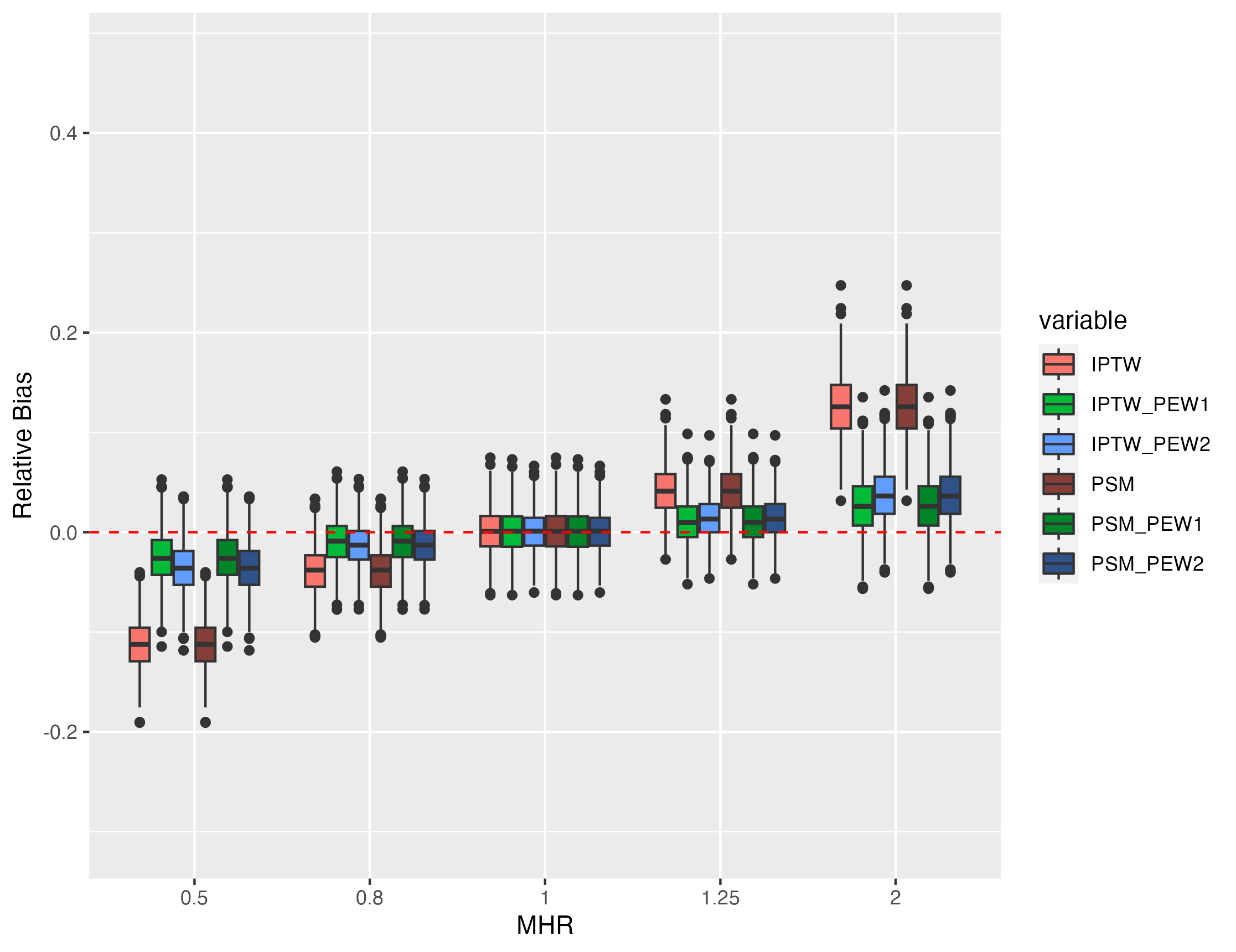}
\caption{Relative bias with censoring rate = 0.7}
\label{fig:n10000_counterfactual_censoring0.7}
\end{subfigure}

\medskip

\begin{subfigure}{0.49\columnwidth}
\centering
\includegraphics[width=\textwidth]{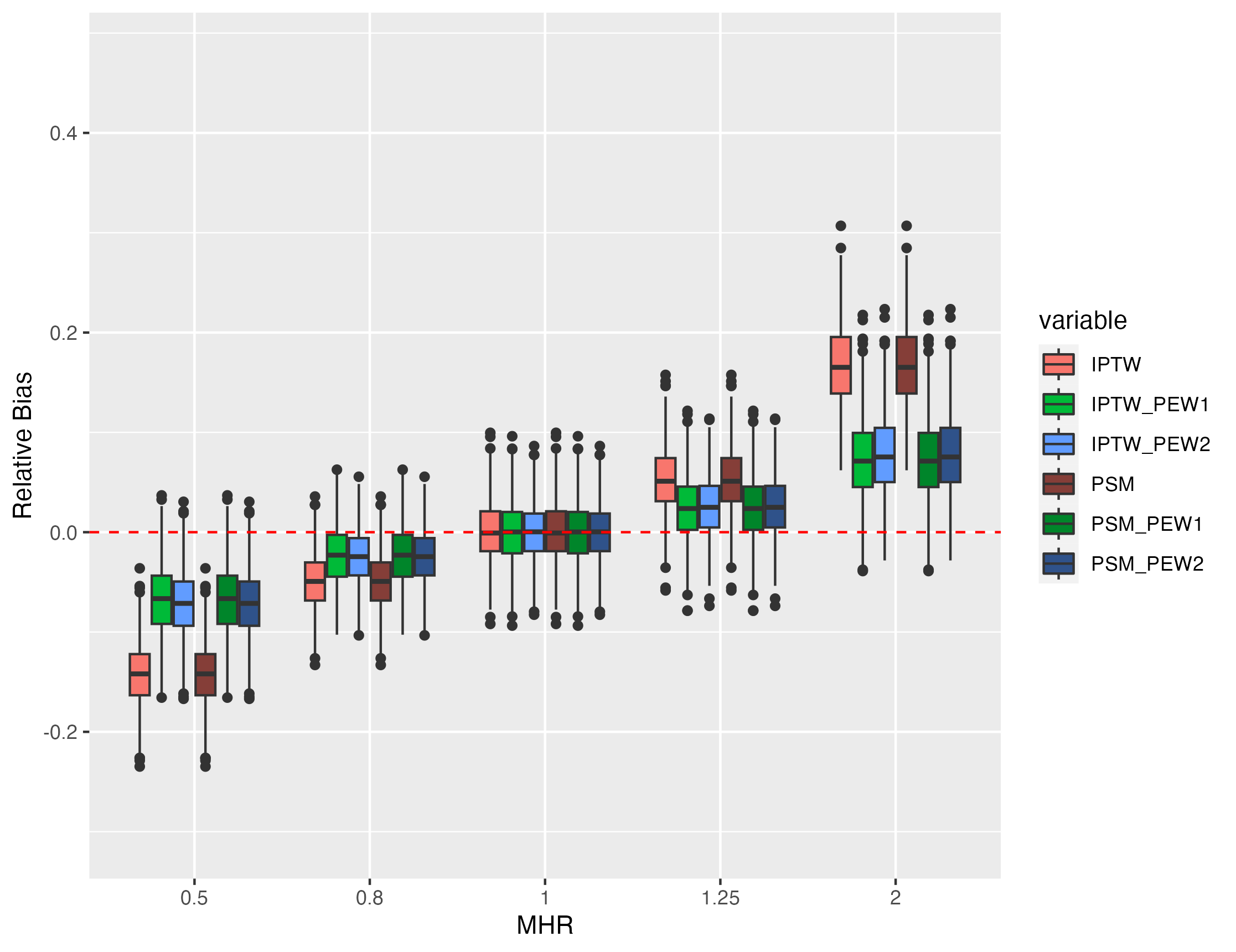}
\caption{Relative bias with censoring rate = 0.8}
\label{fig:n10000_counterfactual_censoring0.8}
\end{subfigure}\hfill
\begin{subfigure}{0.49\columnwidth}
\centering
\includegraphics[width=\textwidth]{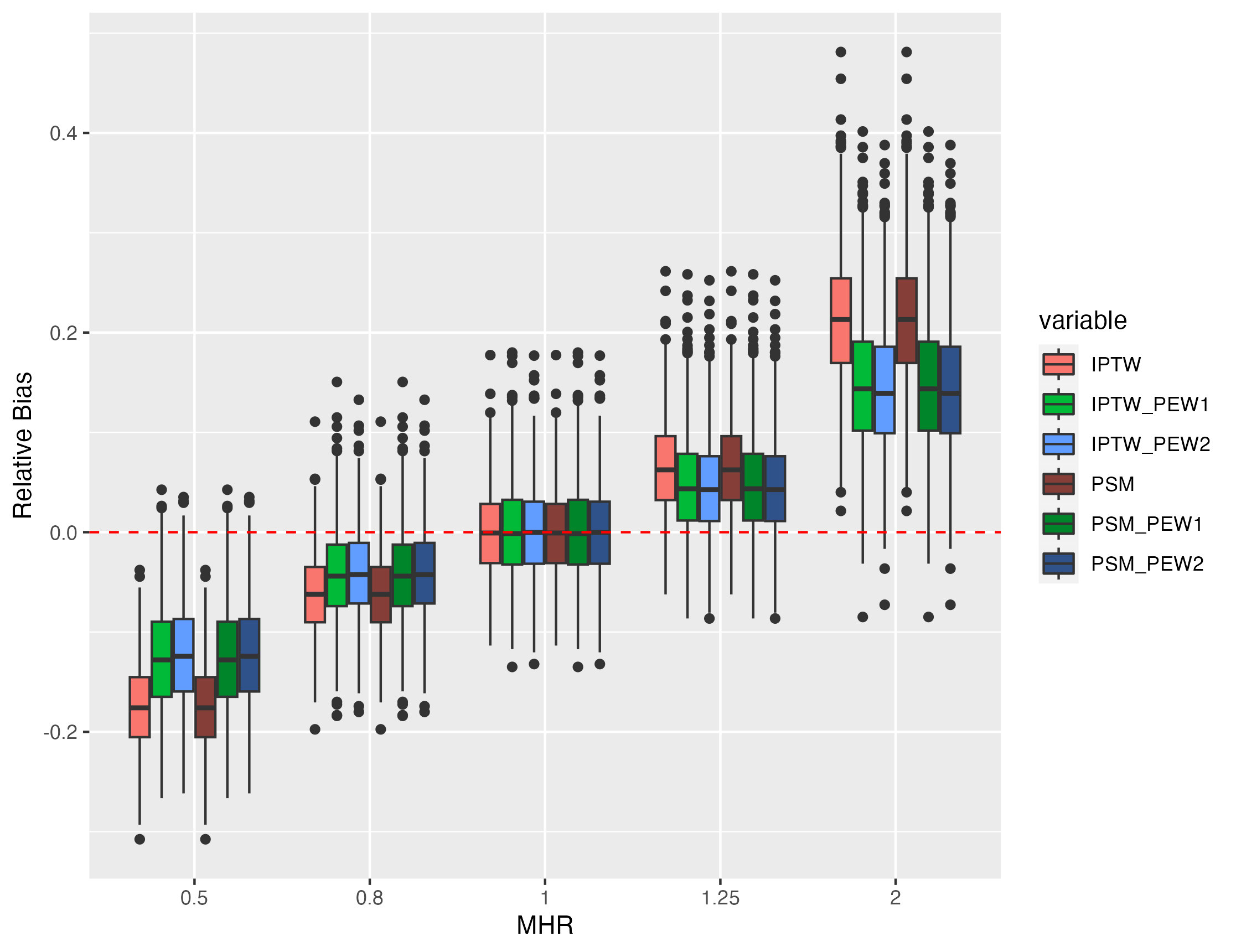}
\caption{Relative bias with censoring rate = 0.9}
\label{fig:n10000_counterfactual_censoring0.9}
\end{subfigure}

\caption{Counterfactual setting, N = 10000, censoring rates = (0.6, 0.7, 0.8, 0.9). Relative bias of estimation of MHR under several true values of MHR. Results based on 1000 simulation replicates.}
\label{fig:n10000_counterfactual_censoring_several_06_09}

\end{figure}

\begin{table}
 \begin{center}
  \caption{Counterfactual setup with sample size $N = 2000$, true MHR = 0.5 and the censoring mechanism is uniform. Results of 1000 simulation replicates for estimators of MHR for IPTW, IPTW\_PEW1, IPTW\_PEW2, PSM, PSM\_PEW1 and PSM\_PEW2. Bias, Monte Carlo standard deviation (SD), root mean-squared error (RMSE), empirical coverage probability of 95\% confidence intervals (Coverage) and relative bias (Rel.Bias).}
 \begin{tabular}{@{}lcccccc@{}}
\hline
  Method & \multicolumn{1}{c}{Censoring} &  \multicolumn{1}{c}{Bias} &  \multicolumn{1}{c}{SD} &  \multicolumn{1}{c}{RMSE} & Rel. Bias & Coverage\\
  \hline
  IPTW & \multirow{6}{*}{0.3} & -0.01 & 0.01 & 0.02 & -0.03 & 0.99 \\ 
  IPTW\_PEW1 & & 0.01 & 0.01 & 0.02 & 0.02 & 1.00 \\ 
  IPTW\_PEW2 & & 0.00 & 0.01 & 0.01 & 0.01 & 1.00 \\ 
  PSM & & -0.01 & 0.01 & 0.02 & -0.03 & 0.98 \\ 
  PSM\_PEW1 & & 0.01 & 0.01 & 0.02 & 0.02 & 0.99 \\ 
  PSM\_PEW2 & & 0.00 & 0.01 & 0.01 & 0.01 & 1.00 \\ 
  \hline
  IPTW & \multirow{6}{*}{0.4} & -0.02 & 0.02 & 0.03 & -0.04 & 0.98 \\ 
  IPTW\_PEW1 & & 0.01 & 0.02 & 0.02 & 0.02 & 0.99 \\ 
  IPTW\_PEW2 & & 0.01 & 0.02 & 0.02 & 0.01 & 1.00 \\ 
  PSM & & -0.02 & 0.02 & 0.03 & -0.04 & 0.95 \\ 
  PSM\_PEW1 & & 0.01 & 0.02 & 0.02 & 0.02 & 0.98 \\ 
  PSM\_PEW2 & & 0.01 & 0.02 & 0.02 & 0.01 & 0.99 \\ 
  \hline
  IPTW & \multirow{6}{*}{0.5} & -0.03 & 0.02 & 0.04 & -0.06 & 0.94 \\ 
  IPTW\_PEW1 & & 0.01 & 0.02 & 0.02 & 0.02 & 0.99 \\ 
  IPTW\_PEW2 & & 0.00 & 0.02 & 0.02 & 0.01 & 1.00 \\ 
  PSM & & -0.03 & 0.02 & 0.04 & -0.06 & 0.90 \\ 
  PSM\_PEW1 & & 0.01 & 0.02 & 0.02 & 0.02 & 0.98 \\ 
  PSM\_PEW2 & & 0.00 & 0.02 & 0.02 & 0.01 & 0.99 \\ 
  \hline
  IPTW & \multirow{6}{*}{0.6} & -0.04 & 0.02 & 0.05 & -0.08 & 0.84 \\ 
  IPTW\_PEW1 & & 0.00 & 0.03 & 0.03 & 0.00 & 0.99 \\ 
  IPTW\_PEW2 & & -0.00 & 0.02 & 0.03 & -0.01 & 0.99 \\ 
  PSM & & -0.04 & 0.02 & 0.05 & -0.08 & 0.79 \\ 
  PSM\_PEW1 & & 0.00 & 0.03 & 0.03 & 0.00 & 0.99 \\ 
  PSM\_PEW2 & & -0.00 & 0.02 & 0.03 & -0.01 & 0.99 \\ 
  \hline
  IPTW & \multirow{6}{*}{0.7} & -0.06 & 0.03 & 0.06 & -0.11 & 0.78 \\ 
  IPTW\_PEW1 & & -0.01 & 0.03 & 0.03 & -0.02 & 0.99 \\ 
  IPTW\_PEW2 & & -0.01 & 0.03 & 0.03 & -0.03 & 0.99 \\ 
  PSM & & -0.06 & 0.03 & 0.06 & -0.11 & 0.72 \\ 
  PSM\_PEW1 & & -0.01 & 0.03 & 0.03 & -0.02 & 0.98 \\ 
  PSM\_PEW2 & & -0.01 & 0.03 & 0.03 & -0.03 & 0.98 \\ 
  \hline
  IPTW & \multirow{6}{*}{0.8} & -0.07 & 0.03 & 0.08 & -0.14 & 0.77 \\ 
  IPTW\_PEW1 & & -0.03 & 0.04 & 0.05 & -0.06 & 0.95 \\ 
  IPTW\_PEW2 & & -0.03 & 0.04 & 0.05 & -0.06 & 0.96 \\ 
  PSM & & -0.07 & 0.03 & 0.08 & -0.14 & 0.71 \\ 
  PSM\_PEW1 & & -0.03 & 0.04 & 0.05 & -0.06 & 0.94 \\ 
  PSM\_PEW2 & & -0.03 & 0.04 & 0.05 & -0.06 & 0.95 \\ 
  \hline
  IPTW & \multirow{6}{*}{0.9} & -0.09 & 0.05 & 0.10 & -0.17 & 0.84 \\ 
  IPTW\_PEW1 & & -0.06 & 0.06 & 0.08 & -0.12 & 0.94 \\ 
  IPTW\_PEW2 & & -0.05 & 0.06 & 0.08 & -0.10 & 0.96 \\ 
  PSM & & -0.09 & 0.05 & 0.10 & -0.17 & 0.80 \\ 
  PSM\_PEW1 & & -0.06 & 0.06 & 0.08 & -0.12 & 0.92 \\ 
  PSM\_PEW2 & & -0.05 & 0.06 & 0.08 & -0.10 & 0.95 \\ 
   \hline
\end{tabular}
\label{table:uniform_2000_MHR0.5_counterfactual}
\end{center}
\end{table}

\begin{table}
 \begin{center}
  \caption{Counterfactual setup with sample size $n = 2000$, true MHR = 2 and the censoring mechanism is uniform. Results of 1000 simulation replicates for estimators of MHR for IPTW, IPTW\_PEW1, IPTW\_PEW2, PSM, PSM\_PEW1 and PSM\_PEW2. Bias, Monte Carlo standard deviation (SD), root mean-squared error (RMSE), empirical coverage probability of 95\% confidence intervals (Coverage) and relative bias (Rel.Bias). }
   \begin{tabular}{@{}lcccccc@{}}
   \hline
  Method & \multicolumn{1}{c}{Censoring} &  \multicolumn{1}{c}{Bias} &  \multicolumn{1}{c}{SD} &  \multicolumn{1}{c}{RMSE} & Rel. Bias & Coverage\\
 \hline
  IPTW & \multirow{6}{*}{0.3} & 0.06 & 0.06 & 0.08 & 0.03 & 1.00 \\ 
  IPTW\_PEW1 & & -0.04 & 0.05 & 0.07 & -0.02 & 1.00 \\ 
  IPTW\_PEW2 & & -0.02 & 0.05 & 0.06 & -0.01 & 1.00 \\ 
  PSM & & 0.06 & 0.06 & 0.08 & 0.03 & 1.00 \\ 
  PSM\_PEW1 & & -0.04 & 0.05 & 0.07 & -0.02 & 0.99 \\ 
  PSM\_PEW2 & & -0.02 & 0.05 & 0.06 & -0.01 & 1.00 \\ 
  \hline
  IPTW & \multirow{6}{*}{0.4} & 0.09 & 0.07 & 0.12 & 0.05 & 0.99 \\ 
  IPTW\_PEW1 & & -0.04 & 0.07 & 0.08 & -0.02 & 0.99 \\ 
  IPTW\_PEW2 & & -0.02 & 0.06 & 0.07 & -0.01 & 1.00 \\ 
  PSM & & 0.09 & 0.07 & 0.12 & 0.05 & 0.96 \\ 
  PSM\_PEW1 & & -0.04 & 0.07 & 0.08 & -0.02 & 0.99 \\ 
  PSM\_PEW2 & & -0.02 & 0.06 & 0.07 & -0.01 & 1.00 \\ 
  \hline
  IPTW & \multirow{6}{*}{0.5} & 0.14 & 0.09 & 0.16 & 0.07 & 0.94 \\ 
  IPTW\_PEW1 & & -0.03 & 0.08 & 0.09 & -0.02 & 0.99 \\ 
  IPTW\_PEW2 & & -0.01 & 0.08 & 0.08 & -0.00 & 1.00 \\ 
  PSM & & 0.14 & 0.09 & 0.16 & 0.07 & 0.90 \\ 
  PSM\_PEW1 & & -0.03 & 0.08 & 0.09 & -0.02 & 0.99 \\ 
  PSM\_PEW2 & & -0.01 & 0.08 & 0.08 & -0.00 & 1.00 \\ 
  \hline
  IPTW & \multirow{6}{*}{0.6} & 0.20 & 0.11 & 0.22 & 0.10 & 0.86 \\ 
  IPTW\_PEW1 & & 0.00 & 0.10 & 0.10 & 0.00 & 0.99 \\ 
  IPTW\_PEW2 & & 0.02 & 0.10 & 0.10 & 0.01 & 0.99 \\ 
  PSM & & 0.20 & 0.11 & 0.22 & 0.10 & 0.81 \\ 
  PSM\_PEW1 & & 0.00 & 0.10 & 0.10 & 0.00 & 0.99 \\ 
  PSM\_PEW2 & & 0.02 & 0.10 & 0.10 & 0.01 & 0.99 \\ 
  \hline
  IPTW & \multirow{6}{*}{0.7} & 0.26 & 0.14 & 0.30 & 0.13 & 0.77 \\ 
  IPTW\_PEW1 & & 0.06 & 0.13 & 0.15 & 0.03 & 0.99 \\ 
  IPTW\_PEW2 & & 0.07 & 0.13 & 0.15 & 0.04 & 0.99 \\ 
  PSM & & 0.26 & 0.14 & 0.30 & 0.13 & 0.71 \\ 
  PSM\_PEW1 & & 0.06 & 0.13 & 0.15 & 0.03 & 0.98 \\ 
  PSM\_PEW2 & & 0.07 & 0.13 & 0.15 & 0.04 & 0.98 \\ 
  \hline
  IPTW & \multirow{6}{*}{0.8} & 0.35 & 0.19 & 0.40 & 0.18 & 0.73 \\ 
  IPTW\_PEW1 & & 0.16 & 0.20 & 0.26 & 0.08 & 0.96 \\ 
  IPTW\_PEW2 & & 0.15 & 0.19 & 0.24 & 0.08 & 0.97 \\ 
  PSM & & 0.35 & 0.19 & 0.40 & 0.18 & 0.67 \\ 
  PSM\_PEW1 & & 0.16 & 0.20 & 0.26 & 0.08 & 0.94 \\ 
  PSM\_PEW2 & & 0.15 & 0.19 & 0.24 & 0.08 & 0.95 \\ 
  \hline
  IPTW & \multirow{6}{*}{0.9} & 0.47 & 0.31 & 0.56 & 0.23 & 0.83 \\ 
  IPTW\_PEW1 & & 0.33 & 0.32 & 0.46 & 0.16 & 0.95 \\ 
  IPTW\_PEW2 & & 0.27 & 0.31 & 0.41 & 0.14 & 0.96 \\ 
  PSM & & 0.47 & 0.31 & 0.56 & 0.23 & 0.78 \\ 
  PSM\_PEW1 & & 0.33 & 0.32 & 0.46 & 0.16 & 0.92 \\ 
  PSM\_PEW2 & & 0.27 & 0.31 & 0.41 & 0.14 & 0.95 \\ 
  \hline
\end{tabular}
\label{table:uniform_2000_MHR2_counterfactual}
\end{center}
\end{table}

\begin{table}
 \begin{center}
  \caption{Counterfactual setup with sample size $N = 6000$, true MHR = 0.5 and the censoring mechanism is uniform. Results of 1000 simulation replicates for estimators of MHR for IPTW, IPTW\_PEW1, IPTW\_PEW2, PSM, PSM\_PEW1 and PSM\_PEW2. Bias, Monte Carlo standard deviation (SD), root mean-squared error (RMSE), empirical coverage probability of 95\% confidence intervals (Coverage) and relative bias (Rel.Bias). }
 \begin{tabular}{@{}lcccccc@{}}
  \hline
  Method & \multicolumn{1}{c}{Censoring} &  \multicolumn{1}{c}{Bias} &  \multicolumn{1}{c}{SD} &  \multicolumn{1}{c}{RMSE} & Rel. Bias & Coverage\\
 \hline

  IPTW & \multirow{6}{*}{0.3} & -0.01 & 0.01 & 0.02 & -0.03 & 0.98 \\ 
  IPTW\_PEW1 & & 0.01 & 0.01 & 0.01 & 0.02 & 1.00 \\ 
  IPTW\_PEW2 & & 0.00 & 0.01 & 0.01 & 0.01 & 1.00 \\ 
  PSM & & -0.01 & 0.01 & 0.02 & -0.03 & 0.98 \\ 
  PSM\_PEW1 & & 0.01 & 0.01 & 0.01 & 0.02 & 1.00 \\ 
  PSM\_PEW2 & & 0.00 & 0.01 & 0.01 & 0.01 & 1.00 \\ 
  \hline
  IPTW & \multirow{6}{*}{0.4} & -0.02 & 0.01 & 0.02 & -0.04 & 0.86 \\ 
  IPTW\_PEW1 & & 0.01 & 0.01 & 0.02 & 0.02 & 0.98 \\ 
  IPTW\_PEW2 & & 0.00 & 0.01 & 0.01 & 0.01 & 1.00 \\ 
  PSM & & -0.02 & 0.01 & 0.02 & -0.04 & 0.85 \\ 
  PSM\_PEW1 & & 0.01 & 0.01 & 0.02 & 0.02 & 0.98 \\ 
  PSM\_PEW2 & & 0.00 & 0.01 & 0.01 & 0.01 & 1.00 \\ 
  \hline
  IPTW & \multirow{6}{*}{0.5} & -0.03 & 0.01 & 0.03 & -0.06 & 0.65 \\ 
  IPTW\_PEW1 & & 0.01 & 0.01 & 0.02 & 0.02 & 0.99 \\ 
  IPTW\_PEW2 & & 0.00 & 0.01 & 0.01 & 0.00 & 1.00 \\ 
  PSM & & -0.03 & 0.01 & 0.03 & -0.06 & 0.66 \\ 
  PSM\_PEW1 & & 0.01 & 0.01 & 0.02 & 0.02 & 0.99 \\ 
  PSM\_PEW2 & & 0.00 & 0.01 & 0.01 & 0.00 & 1.00 \\ 
  \hline
  IPTW & \multirow{6}{*}{0.6} & -0.04 & 0.01 & 0.04 & -0.08 & 0.42 \\ 
  IPTW\_PEW1 & & 0.00 & 0.01 & 0.01 & 0.00 & 0.99 \\ 
  IPTW\_PEW2 & & -0.01 & 0.01 & 0.01 & -0.01 & 0.99 \\ 
  PSM & & -0.04 & 0.01 & 0.04 & -0.08 & 0.42 \\ 
  PSM\_PEW1 & & 0.00 & 0.01 & 0.01 & 0.00 & 0.99 \\ 
  PSM\_PEW2 & & -0.01 & 0.01 & 0.01 & -0.01 & 0.99 \\ 
  \hline
  IPTW & \multirow{6}{*}{0.7} & -0.06 & 0.02 & 0.06 & -0.11 & 0.27 \\ 
  IPTW\_PEW1 & & -0.01 & 0.02 & 0.02 & -0.02 & 0.97 \\ 
  IPTW\_PEW2 & & -0.02 & 0.02 & 0.02 & -0.03 & 0.96 \\ 
  PSM & & -0.06 & 0.02 & 0.06 & -0.11 & 0.27 \\ 
  PSM\_PEW1 & & -0.01 & 0.02 & 0.02 & -0.02 & 0.97 \\ 
  PSM\_PEW2 & & -0.02 & 0.02 & 0.02 & -0.03 & 0.96 \\ 
  \hline
  IPTW & \multirow{6}{*}{0.8} & -0.07 & 0.02 & 0.07 & -0.14 & 0.25 \\ 
  IPTW\_PEW1 & & -0.03 & 0.02 & 0.04 & -0.07 & 0.88 \\ 
  IPTW\_PEW2 & & -0.03 & 0.02 & 0.04 & -0.07 & 0.87 \\ 
  PSM & & -0.07 & 0.02 & 0.07 & -0.14 & 0.25 \\ 
  PSM\_PEW1 & & -0.03 & 0.02 & 0.04 & -0.07 & 0.88 \\ 
  PSM\_PEW2 & & -0.03 & 0.02 & 0.04 & -0.07 & 0.87 \\ 
  \hline
  IPTW & \multirow{6}{*}{0.9} & -0.09 & 0.03 & 0.09 & -0.17 & 0.43 \\ 
  IPTW\_PEW1 & & -0.06 & 0.04 & 0.07 & -0.12 & 0.79 \\ 
  IPTW\_PEW2 & & -0.06 & 0.04 & 0.07 & -0.12 & 0.82 \\ 
  PSM & & -0.09 & 0.03 & 0.09 & -0.17 & 0.43 \\ 
  PSM\_PEW1 & & -0.06 & 0.04 & 0.07 & -0.12 & 0.79 \\ 
  PSM\_PEW2 & & -0.06 & 0.04 & 0.07 & -0.12 & 0.82 \\ 
 \hline
  \end{tabular}
\label{table:uniform_6000_MHR0.5_counter}
\end{center}
\end{table}

\begin{table}
 \begin{center}
  \caption{Counterfactual setup with sample size $N = 6000$, true MHR = 2 and the censoring mechanism is uniform. Results of 1000 simulation replicates for estimators of MHR for IPTW, IPTW\_PEW1, IPTW\_PEW2, PSM, PSM\_PEW1 and PSM\_PEW2. Bias, Monte Carlo standard deviation (SD), root mean-squared error (RMSE), empirical coverage probability of 95\% confidence intervals (Coverage) and relative bias (Rel.Bias). }
 \begin{tabular}{@{}lcccccc@{}}
  \hline
  Method & \multicolumn{1}{c}{Censoring} &  \multicolumn{1}{c}{Bias} &  \multicolumn{1}{c}{SD} &  \multicolumn{1}{c}{RMSE} & Rel. Bias & Coverage\\
 \hline
  IPTW & \multirow{6}{*}{0.3} & 0.06 & 0.03 & 0.07 & 0.03 & 0.98 \\ 
  IPTW\_PEW1 & & -0.04 & 0.03 & 0.05 & -0.02 & 0.99 \\ 
  IPTW\_PEW2 & & -0.02 & 0.03 & 0.03 & -0.01 & 1.00 \\ 
  PSM & & 0.06 & 0.03 & 0.07 & 0.03 & 0.98 \\ 
  PSM\_PEW1 & & -0.04 & 0.03 & 0.05 & -0.02 & 0.99 \\ 
  PSM\_PEW2 & & -0.02 & 0.03 & 0.03 & -0.01 & 1.00 \\ 
   \hline
  IPTW & \multirow{6}{*}{0.4} & 0.09 & 0.04 & 0.10 & 0.05 & 0.86 \\ 
  IPTW\_PEW1 & & -0.04 & 0.04 & 0.06 & -0.02 & 0.99 \\ 
  IPTW\_PEW2 & & -0.01 & 0.04 & 0.04 & -0.01 & 1.00 \\ 
  PSM & & 0.09 & 0.04 & 0.10 & 0.05 & 0.86 \\ 
  PSM\_PEW1 & & -0.04 & 0.04 & 0.06 & -0.02 & 0.99 \\ 
  PSM\_PEW2 & & -0.01 & 0.04 & 0.04 & -0.01 & 1.00 \\ 
   \hline
  IPTW & \multirow{6}{*}{0.5} & 0.14 & 0.05 & 0.15 & 0.07 & 0.60 \\ 
  IPTW\_PEW1 & & -0.03 & 0.05 & 0.06 & -0.02 & 0.99 \\ 
  IPTW\_PEW2 & & -0.00 & 0.05 & 0.05 & -0.00 & 1.00 \\ 
  PSM & & 0.14 & 0.05 & 0.15 & 0.07 & 0.60 \\ 
  PSM\_PEW1 & & -0.03 & 0.05 & 0.06 & -0.02 & 0.99 \\ 
  PSM\_PEW2 & & -0.00 & 0.05 & 0.05 & -0.00 & 1.00 \\ 
   \hline
  IPTW & \multirow{6}{*}{0.6} & 0.19 & 0.06 & 0.20 & 0.10 & 0.39 \\ 
  IPTW\_PEW1 & & 0.00 & 0.06 & 0.06 & 0.00 & 0.99 \\ 
  IPTW\_PEW2 & & 0.03 & 0.06 & 0.06 & 0.01 & 0.99 \\ 
  PSM & & 0.19 & 0.06 & 0.20 & 0.10 & 0.39 \\ 
  PSM\_PEW1 & & 0.00 & 0.06 & 0.06 & 0.00 & 0.99 \\ 
  PSM\_PEW2 & & 0.03 & 0.06 & 0.06 & 0.01 & 0.99 \\ 
   \hline
  IPTW & \multirow{6}{*}{0.7} & 0.26 & 0.08 & 0.27 & 0.13 & 0.26 \\ 
  IPTW\_PEW1 & & 0.06 & 0.08 & 0.09 & 0.03 & 0.96 \\ 
  IPTW\_PEW2 & & 0.08 & 0.07 & 0.11 & 0.04 & 0.95 \\ 
  PSM & & 0.26 & 0.08 & 0.27 & 0.13 & 0.26 \\ 
  PSM\_PEW1 & & 0.06 & 0.08 & 0.09 & 0.03 & 0.96 \\ 
  PSM\_PEW2 & & 0.08 & 0.07 & 0.11 & 0.04 & 0.95 \\ 
   \hline
  IPTW & \multirow{6}{*}{0.8} & 0.34 & 0.11 & 0.36 & 0.17 & 0.23 \\ 
  IPTW\_PEW1 & & 0.15 & 0.10 & 0.18 & 0.08 & 0.89 \\ 
  IPTW\_PEW2 & & 0.16 & 0.10 & 0.19 & 0.08 & 0.88 \\ 
  PSM & & 0.34 & 0.11 & 0.36 & 0.17 & 0.23 \\ 
  PSM\_PEW1 & & 0.15 & 0.10 & 0.18 & 0.08 & 0.89 \\ 
  PSM\_PEW2 & & 0.16 & 0.10 & 0.19 & 0.08 & 0.88 \\ 
   \hline
  IPTW & \multirow{6}{*}{0.9} & 0.43 & 0.17 & 0.47 & 0.22 & 0.39 \\ 
  IPTW\_PEW1 & & 0.30 & 0.18 & 0.35 & 0.15 & 0.79 \\ 
  IPTW\_PEW2 & & 0.28 & 0.17 & 0.33 & 0.14 & 0.82 \\ 
  PSM & & 0.43 & 0.17 & 0.47 & 0.22 & 0.39 \\ 
  PSM\_PEW1 & & 0.30 & 0.18 & 0.35 & 0.15 & 0.79 \\ 
  PSM\_PEW2 & & 0.28 & 0.17 & 0.33 & 0.14 & 0.82 \\ 
 \hline
  \end{tabular}
\label{table:uniform_6000_MHR2_counter}
\end{center}
\end{table}

\begin{table}
 \begin{center}
  \caption{Counterfactual setup with sample size $N = 10000$, true MHR = 0.5 and the censoring mechanism is uniform. Results of 1000 simulation replicates for estimators of MHR for IPTW, IPTW\_PEW1, IPTW\_PEW2, PSM, PSM\_PEW1 and PSM\_PEW2. Bias, Monte Carlo standard deviation (SD), root mean-squared error (RMSE), empirical coverage probability of 95\% confidence intervals (Coverage) and relative bias (Rel.Bias). }
 \begin{tabular}{@{}lcccccc@{}}
  \hline
  Method & \multicolumn{1}{c}{Censoring} &  \multicolumn{1}{c}{Bias} &  \multicolumn{1}{c}{SD} &  \multicolumn{1}{c}{RMSE} & Rel. Bias & Coverage\\
 \hline
  IPTW & \multirow{6}{*}{0.3} & -0.01 & 0.01 & 0.02 & -0.03 & 0.94 \\ 
  IPTW\_PEW1 & & 0.01 & 0.01 & 0.01 & 0.02 & 0.99 \\ 
  IPTW\_PEW2 & & 0.00 & 0.01 & 0.01 & 0.01 & 1.00 \\ 
  PSM & & -0.01 & 0.01 & 0.02 & -0.03 & 0.94 \\ 
  PSM\_PEW1 & & 0.01 & 0.01 & 0.01 & 0.02 & 0.99 \\ 
  PSM\_PEW2 & & 0.00 & 0.01 & 0.01 & 0.01 & 1.00 \\ 
  \hline
  IPTW & \multirow{6}{*}{0.4} & -0.02 & 0.01 & 0.02 & -0.04 & 0.67 \\ 
  IPTW\_PEW1 & & 0.01 & 0.01 & 0.01 & 0.02 & 0.96 \\ 
  IPTW\_PEW2 & & 0.00 & 0.01 & 0.01 & 0.01 & 1.00 \\ 
  PSM & & -0.02 & 0.01 & 0.02 & -0.04 & 0.66 \\ 
  PSM\_PEW1 & & 0.01 & 0.01 & 0.01 & 0.02 & 0.96 \\ 
  PSM\_PEW2 & & 0.00 & 0.01 & 0.01 & 0.01 & 1.00 \\ 
  \hline
  IPTW & \multirow{6}{*}{0.5} & -0.03 & 0.01 & 0.03 & -0.06 & 0.34 \\ 
  IPTW\_PEW1 & & 0.01 & 0.01 & 0.01 & 0.02 & 0.98 \\ 
  IPTW\_PEW2 & & 0.00 & 0.01 & 0.01 & 0.00 & 1.00 \\ 
  PSM & & -0.03 & 0.01 & 0.03 & -0.06 & 0.34 \\ 
  PSM\_PEW1 & & 0.01 & 0.01 & 0.01 & 0.02 & 0.98 \\ 
  PSM\_PEW2 & & 0.00 & 0.01 & 0.01 & 0.00 & 1.00 \\ 
  \hline
  IPTW & \multirow{6}{*}{0.6} & -0.04 & 0.01 & 0.04 & -0.08 & 0.12 \\ 
  IPTW\_PEW1 & & 0.00 & 0.01 & 0.01 & 0.00 & 0.99 \\ 
  IPTW\_PEW2 & & -0.01 & 0.01 & 0.01 & -0.01 & 0.99 \\ 
  PSM & & -0.04 & 0.01 & 0.04 & -0.08 & 0.12 \\ 
  PSM\_PEW1 & & 0.00 & 0.01 & 0.01 & 0.00 & 0.99 \\ 
  PSM\_PEW2 & & -0.01 & 0.01 & 0.01 & -0.01 & 0.99 \\ 
  \hline
  IPTW & \multirow{6}{*}{0.7} & -0.06 & 0.01 & 0.06 & -0.11 & 0.06 \\ 
  IPTW\_PEW1 & & -0.01 & 0.01 & 0.02 & -0.03 & 0.97 \\ 
  IPTW\_PEW2 & & -0.02 & 0.01 & 0.02 & -0.04 & 0.93 \\ 
  PSM & & -0.06 & 0.01 & 0.06 & -0.11 & 0.06 \\ 
  PSM\_PEW1 & & -0.01 & 0.01 & 0.02 & -0.03 & 0.96 \\ 
  PSM\_PEW2 & & -0.02 & 0.01 & 0.02 & -0.04 & 0.93 \\ 
  \hline
  IPTW & \multirow{6}{*}{0.8} & -0.07 & 0.02 & 0.07 & -0.14 & 0.04 \\ 
  IPTW\_PEW1 & & -0.03 & 0.02 & 0.04 & -0.07 & 0.77 \\ 
  IPTW\_PEW2 & & -0.04 & 0.02 & 0.04 & -0.07 & 0.75 \\ 
  PSM & & -0.07 & 0.02 & 0.07 & -0.14 & 0.04 \\ 
  PSM\_PEW1 & & -0.03 & 0.02 & 0.04 & -0.07 & 0.77 \\ 
  PSM\_PEW2 & & -0.04 & 0.02 & 0.04 & -0.07 & 0.75 \\ 
  \hline
  IPTW & \multirow{6}{*}{0.9} & -0.09 & 0.02 & 0.09 & -0.18 & 0.13 \\ 
  IPTW\_PEW1 & & -0.06 & 0.03 & 0.07 & -0.13 & 0.59 \\ 
  IPTW\_PEW2 & & -0.06 & 0.03 & 0.07 & -0.12 & 0.62 \\ 
  PSM & & -0.09 & 0.02 & 0.09 & -0.18 & 0.13 \\ 
  PSM\_PEW1 & & -0.06 & 0.03 & 0.07 & -0.13 & 0.59 \\ 
  PSM\_PEW2 & & -0.06 & 0.03 & 0.07 & -0.12 & 0.62 \\  
 \hline
  \end{tabular}
\label{table:uniform_10000_MHR0.5_counter}
\end{center}
\end{table}

\begin{table}
 \begin{center}
  \caption{Counterfactual setup with sample size is $N = 10000$, true MHR = 2 and the censoring mechanism is uniform. Results of 1000 simulation replicates for estimators of MHR for IPTW, IPTW\_PEW1, IPTW\_PEW2, PSM, PSM\_PEW1 and PSM\_PEW2. Bias, Monte Carlo standard deviation (SD), root mean-squared error (RMSE), empirical coverage probability of 95\% confidence intervals (Coverage) and relative bias (Rel.Bias).}
 \begin{tabular}{@{}lcccccc@{}}
  \hline
  Method & \multicolumn{1}{c}{Censoring} &  \multicolumn{1}{c}{Bias} &  \multicolumn{1}{c}{SD} &  \multicolumn{1}{c}{RMSE} & Rel. Bias & Coverage\\
 \hline
  IPTW & \multirow{6}{*}{0.3} & 0.06 & 0.03 & 0.07 & 0.03 & 0.93 \\ 
  IPTW\_PEW1 & & -0.04 & 0.02 & 0.04 & -0.02 & 0.99 \\ 
  IPTW\_PEW2 & & -0.01 & 0.02 & 0.03 & -0.01 & 1.00 \\ 
  PSM & & 0.06 & 0.03 & 0.07 & 0.03 & 0.93 \\ 
  PSM\_PEW1 & & -0.04 & 0.02 & 0.04 & -0.02 & 0.99 \\ 
  PSM\_PEW2 & & -0.01 & 0.02 & 0.03 & -0.01 & 1.00 \\ 
  \hline
  IPTW & \multirow{6}{*}{0.4} & 0.09 & 0.03 & 0.10 & 0.05 & 0.65 \\ 
  IPTW\_PEW1 & & -0.04 & 0.03 & 0.05 & -0.02 & 0.96 \\ 
  IPTW\_PEW2 & & -0.01 & 0.03 & 0.03 & -0.01 & 1.00 \\ 
  PSM & & 0.09 & 0.03 & 0.10 & 0.05 & 0.65 \\ 
  PSM\_PEW1 & & -0.04 & 0.03 & 0.05 & -0.02 & 0.96 \\ 
  PSM\_PEW2 & & -0.01 & 0.03 & 0.03 & -0.01 & 1.00 \\ 
  \hline
  IPTW & \multirow{6}{*}{0.5} & 0.14 & 0.04 & 0.14 & 0.07 & 0.31 \\ 
  IPTW\_PEW1 & & -0.03 & 0.04 & 0.05 & -0.02 & 0.98 \\ 
  IPTW\_PEW2 & & -0.00 & 0.04 & 0.04 & -0.00 & 1.00 \\ 
  PSM & & 0.14 & 0.04 & 0.14 & 0.07 & 0.31 \\ 
  PSM\_PEW1 & & -0.03 & 0.04 & 0.05 & -0.02 & 0.98 \\ 
  PSM\_PEW2 & & -0.00 & 0.04 & 0.04 & -0.00 & 1.00 \\ 
  \hline
  IPTW & \multirow{6}{*}{0.6} & 0.19 & 0.05 & 0.20 & 0.10 & 0.12 \\ 
  IPTW\_PEW1 & & -0.00 & 0.05 & 0.05 & -0.00 & 1.00 \\ 
  IPTW\_PEW2 & & 0.03 & 0.04 & 0.05 & 0.01 & 0.99 \\ 
  PSM & & 0.19 & 0.05 & 0.20 & 0.10 & 0.11 \\ 
  PSM\_PEW1 & & -0.00 & 0.05 & 0.05 & -0.00 & 1.00 \\ 
  PSM\_PEW2 & & 0.03 & 0.04 & 0.05 & 0.01 & 0.99 \\ 
  \hline
  IPTW & \multirow{6}{*}{0.7} & 0.25 & 0.06 & 0.26 & 0.13 & 0.05 \\ 
  IPTW\_PEW1 & & 0.05 & 0.06 & 0.08 & 0.03 & 0.96 \\ 
  IPTW\_PEW2 & & 0.07 & 0.06 & 0.09 & 0.04 & 0.93 \\ 
  PSM & & 0.25 & 0.06 & 0.26 & 0.13 & 0.05 \\ 
  PSM\_PEW1 & & 0.05 & 0.06 & 0.08 & 0.03 & 0.96 \\ 
  PSM\_PEW2 & & 0.07 & 0.06 & 0.09 & 0.04 & 0.93 \\ 
  \hline
  IPTW & \multirow{6}{*}{0.8} & 0.33 & 0.08 & 0.34 & 0.17 & 0.05 \\ 
  IPTW\_PEW1 & & 0.15 & 0.08 & 0.17 & 0.07 & 0.78 \\ 
  IPTW\_PEW2 & & 0.16 & 0.08 & 0.17 & 0.08 & 0.75 \\ 
  PSM & & 0.33 & 0.08 & 0.34 & 0.17 & 0.05 \\ 
  PSM\_PEW1 & & 0.15 & 0.08 & 0.17 & 0.07 & 0.78 \\ 
  PSM\_PEW2 & & 0.16 & 0.08 & 0.17 & 0.08 & 0.75 \\ 
  \hline
  IPTW & \multirow{6}{*}{0.9} & 0.43 & 0.13 & 0.45 & 0.21 & 0.14 \\ 
  IPTW\_PEW1 & & 0.30 & 0.14 & 0.33 & 0.15 & 0.65 \\ 
  IPTW\_PEW2 & & 0.29 & 0.13 & 0.32 & 0.14 & 0.66 \\ 
  PSM & & 0.43 & 0.13 & 0.45 & 0.21 & 0.14 \\ 
  PSM\_PEW1 & & 0.30 & 0.14 & 0.33 & 0.15 & 0.65 \\ 
  PSM\_PEW2 & & 0.29 & 0.13 & 0.32 & 0.14 & 0.66 \\ 
 \hline
  \end{tabular}
\label{table:uniform_10000_MHR2_counter}
\end{center}
\end{table}

\FloatBarrier

\section{Appendix B, Observational, Uniform Censoring}
\label{section:AppendixB}

\begin{figure}[htp]
\centering

\begin{subfigure}{0.49\columnwidth}
\centering
\includegraphics[width=\textwidth]{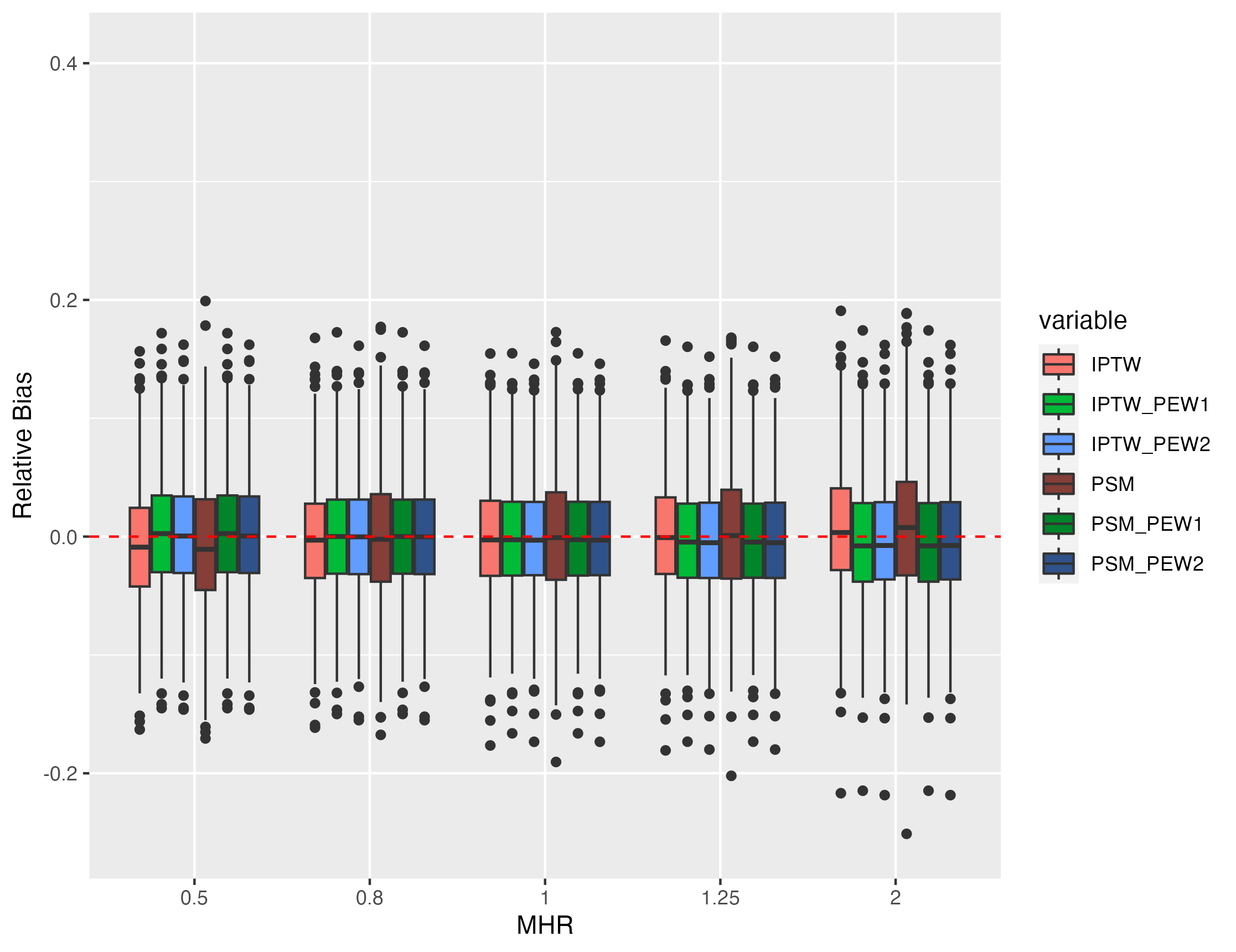}
\caption{Relative bias with censoring rate = 0.1}
\label{fig:n2000_notcounterfactual_censoring0.1}
\end{subfigure}\hfill
\begin{subfigure}{0.49\columnwidth}
\centering
\includegraphics[width=\textwidth]{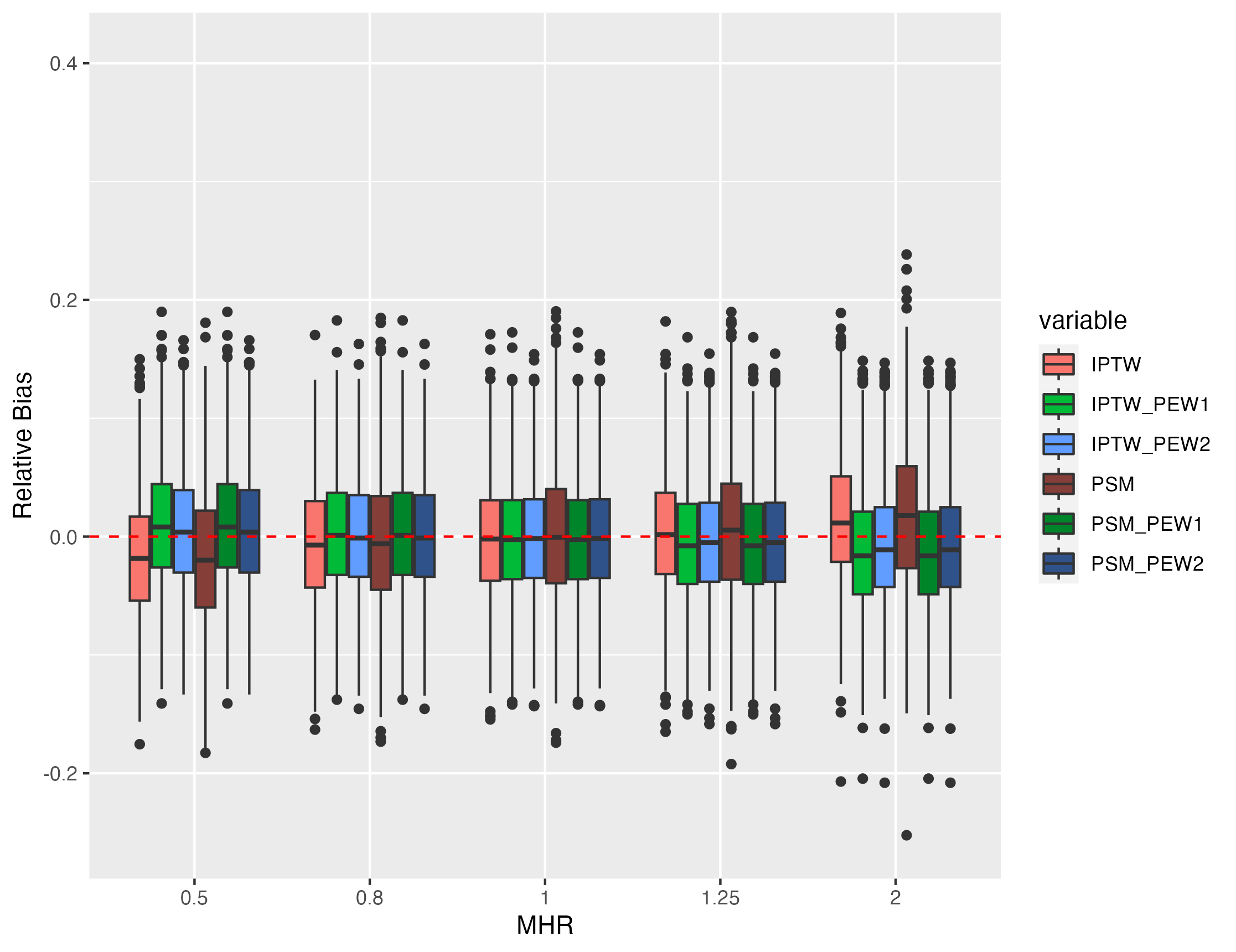}
\caption{Relative bias with censoring rate = 0.2}
\label{fig:n2000_notcounterfactual_censoring0.2}
\end{subfigure}

\medskip

\begin{subfigure}{0.49\columnwidth}
\centering
\includegraphics[width=\textwidth]{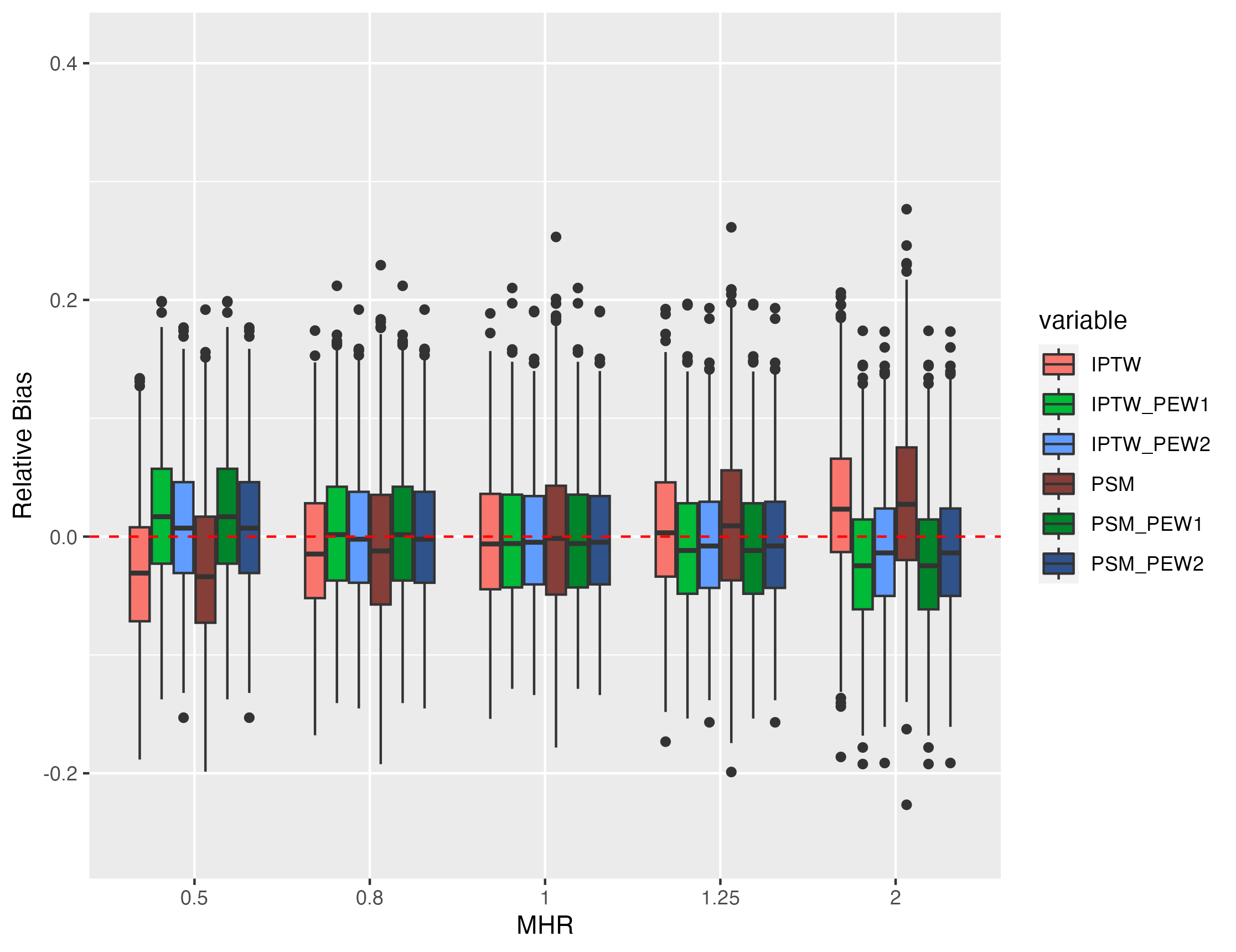}
\caption{Relative bias with censoring rate = 0.3}
\label{fig:n2000_notcounterfactual_censoring0.3}
\end{subfigure}\hfill
\begin{subfigure}{0.49\columnwidth}
\centering
\includegraphics[width=\textwidth]{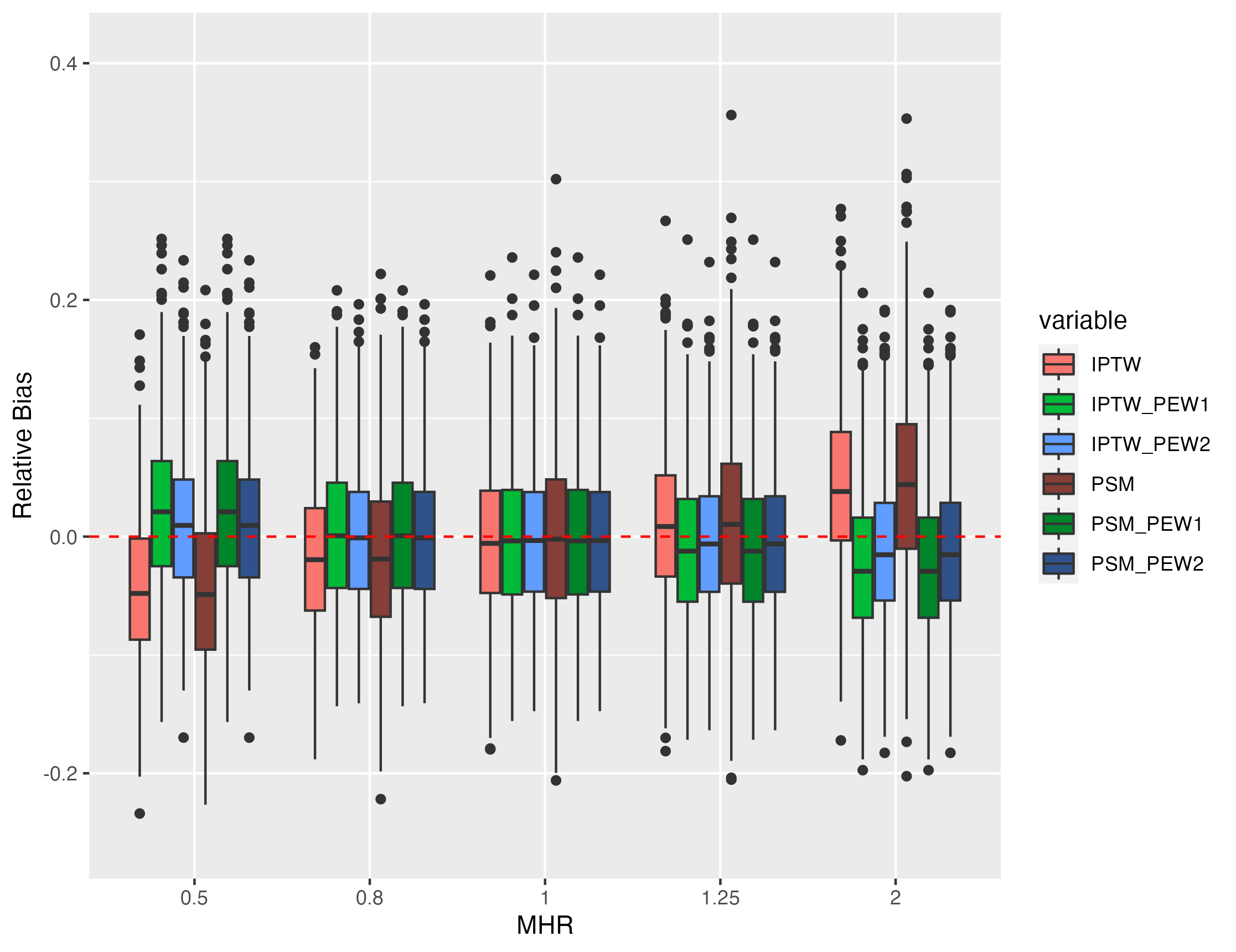}
\caption{Relative bias with censoring rate = 0.4}
\label{fig:n2000_notcounterfactual_censoring0.4}
\end{subfigure}

\medskip

\begin{subfigure}{0.49\columnwidth}
\centering
\includegraphics[width=\textwidth]{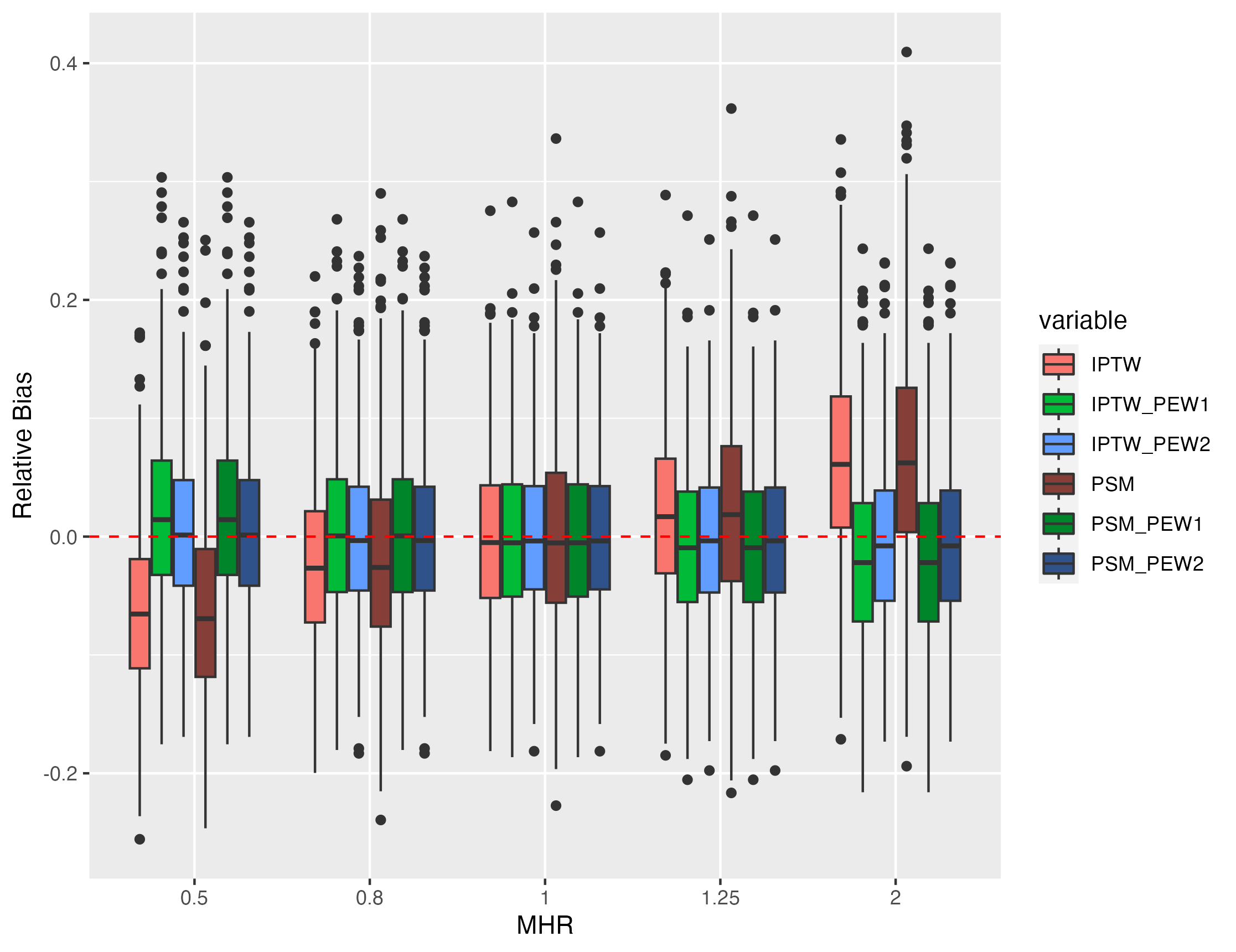}
\caption{Relative bias with censoring rate = 0.5}
\label{fig:n2000_notcounterfactual_censoring0.5}
\end{subfigure}

\caption{Observational setting, N = 2000, censoring rates = (0.1, 0.2, 0.3, 0.4, 0.5). Relative bias of estimation of MHR under several true values of MHR. Results based on 1000 simulation replicates.}
\label{fig:n2000_notcounterfactual_censoring_several_01_05}

\end{figure}

\begin{figure}[htp]
\centering

\begin{subfigure}{0.49\columnwidth}
\centering
\includegraphics[width=\textwidth]{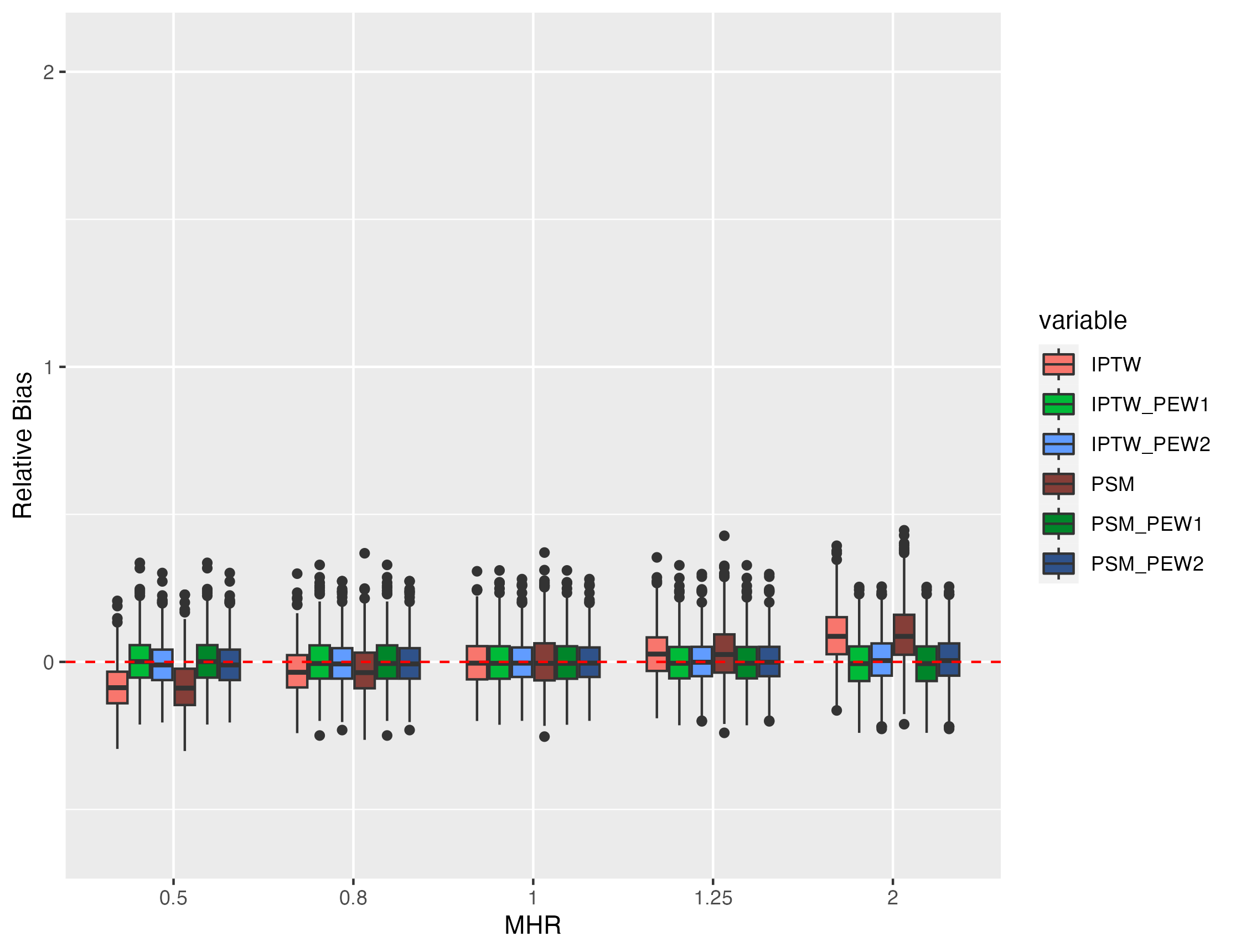}
\caption{Relative bias with censoring rate = 0.6}
\label{fig:n2000_notcounterfactual_censoring0.6}
\end{subfigure}\hfill
\begin{subfigure}{0.49\columnwidth}
\centering
\includegraphics[width=\textwidth]{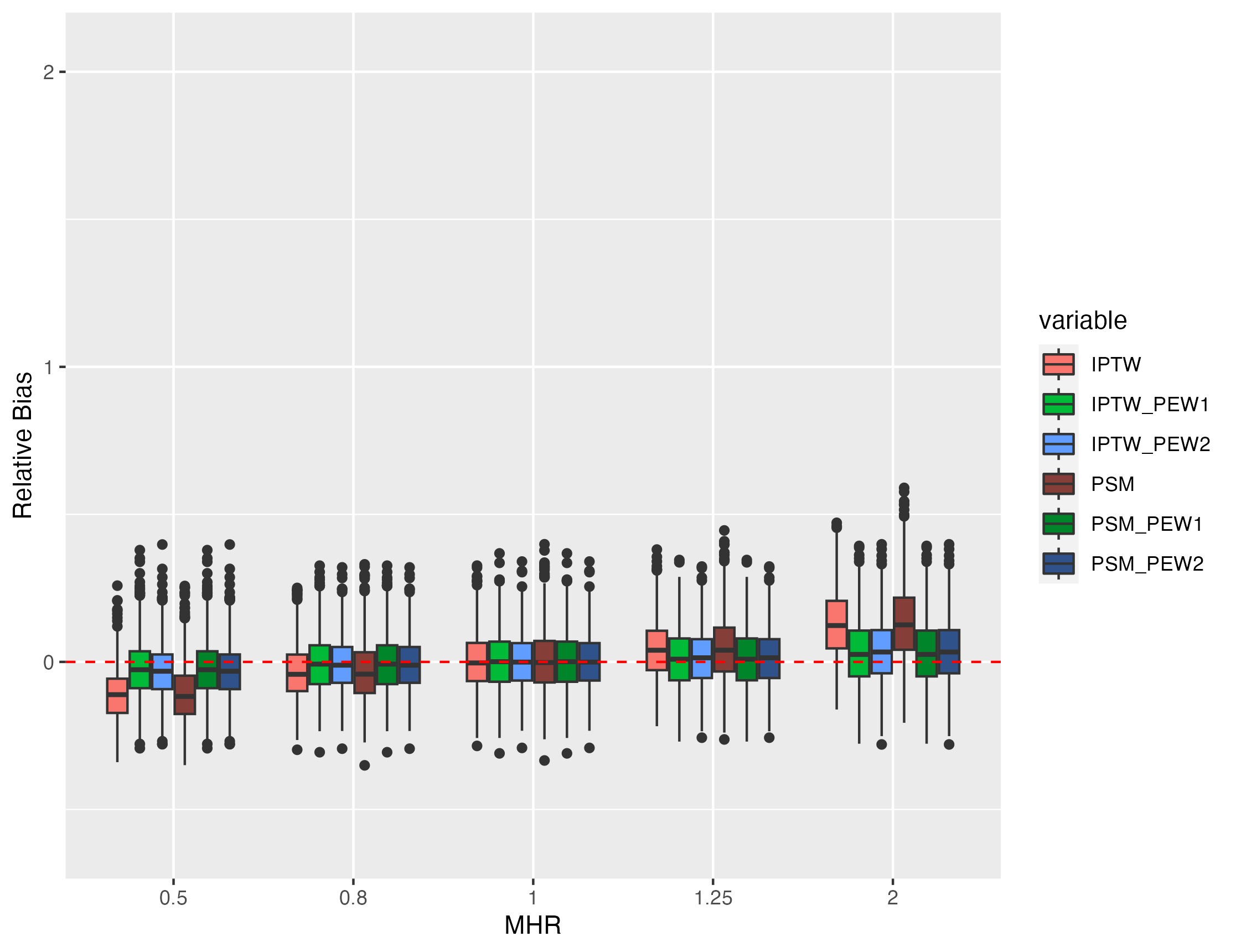}
\caption{Relative bias with censoring rate = 0.7}
\label{fig:n2000_notcounterfactual_censoring0.7}
\end{subfigure}

\medskip

\begin{subfigure}{0.49\columnwidth}
\centering
\includegraphics[width=\textwidth]{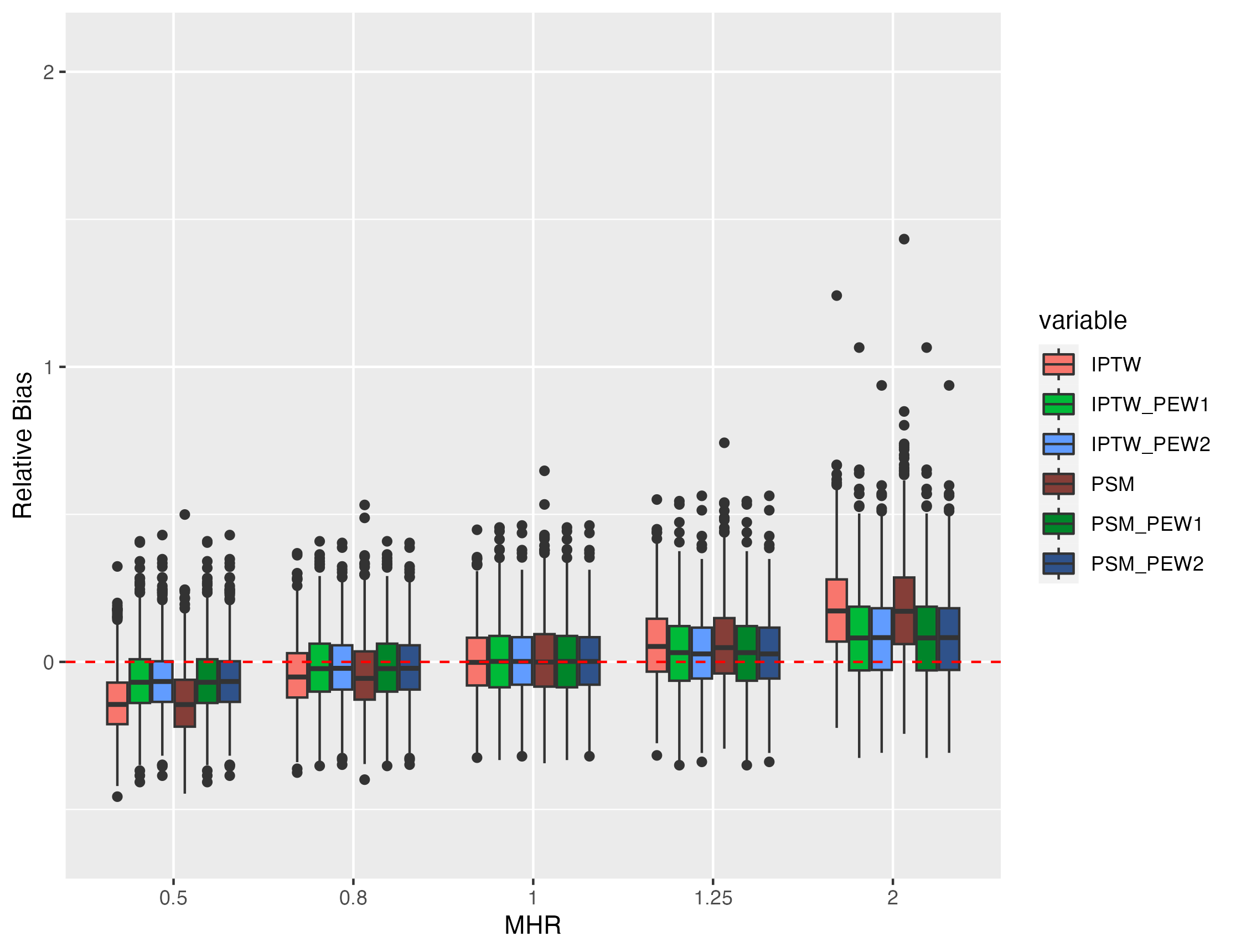}
\caption{Relative bias with censoring rate = 0.8}
\label{fig:n2000_notcounterfactual_censoring0.8}
\end{subfigure}\hfill
\begin{subfigure}{0.49\columnwidth}
\centering
\includegraphics[width=\textwidth]{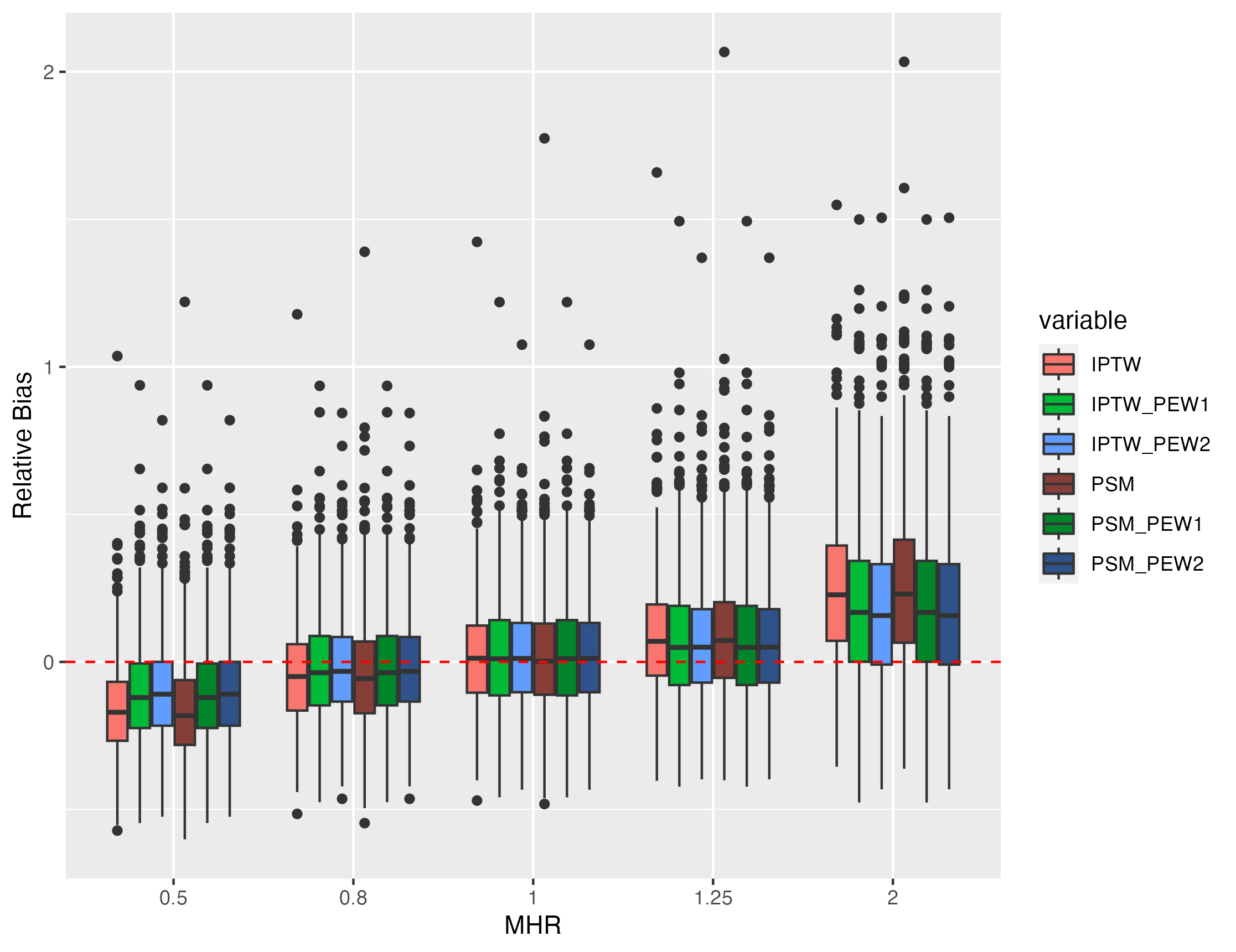}
\caption{Relative bias with censoring rate = 0.9}
\label{fig:n2000_notcounterfactual_censoring0.9}
\end{subfigure}

\caption{Observational setting, N = 2000, censoring rates = (0.6, 0.7, 0.8, 0.9). Relative bias of estimation of MHR under several true values of MHR. Results based on 1000 simulation replicates.}
\label{fig:n2000_notcounterfactual_censoring_several_06_09}

\end{figure}

\begin{figure}[htp]
\centering

\begin{subfigure}{0.49\columnwidth}
\centering
\includegraphics[width=\textwidth]{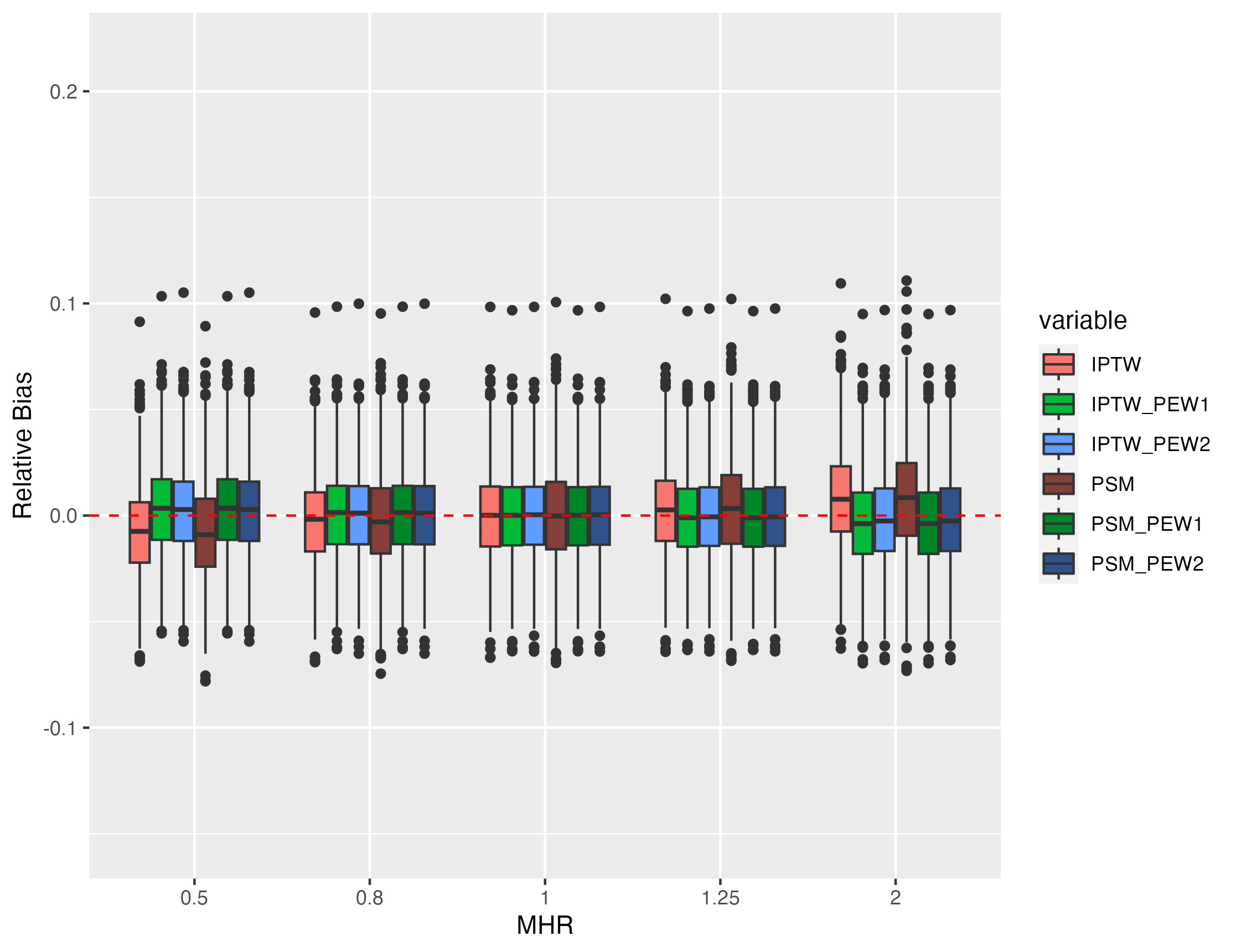}
\caption{Relative bias with censoring rate = 0.1}
\label{fig:n10000_notcounterfactual_censoring0.1}
\end{subfigure}\hfill
\begin{subfigure}{0.49\columnwidth}
\centering
\includegraphics[width=\textwidth]{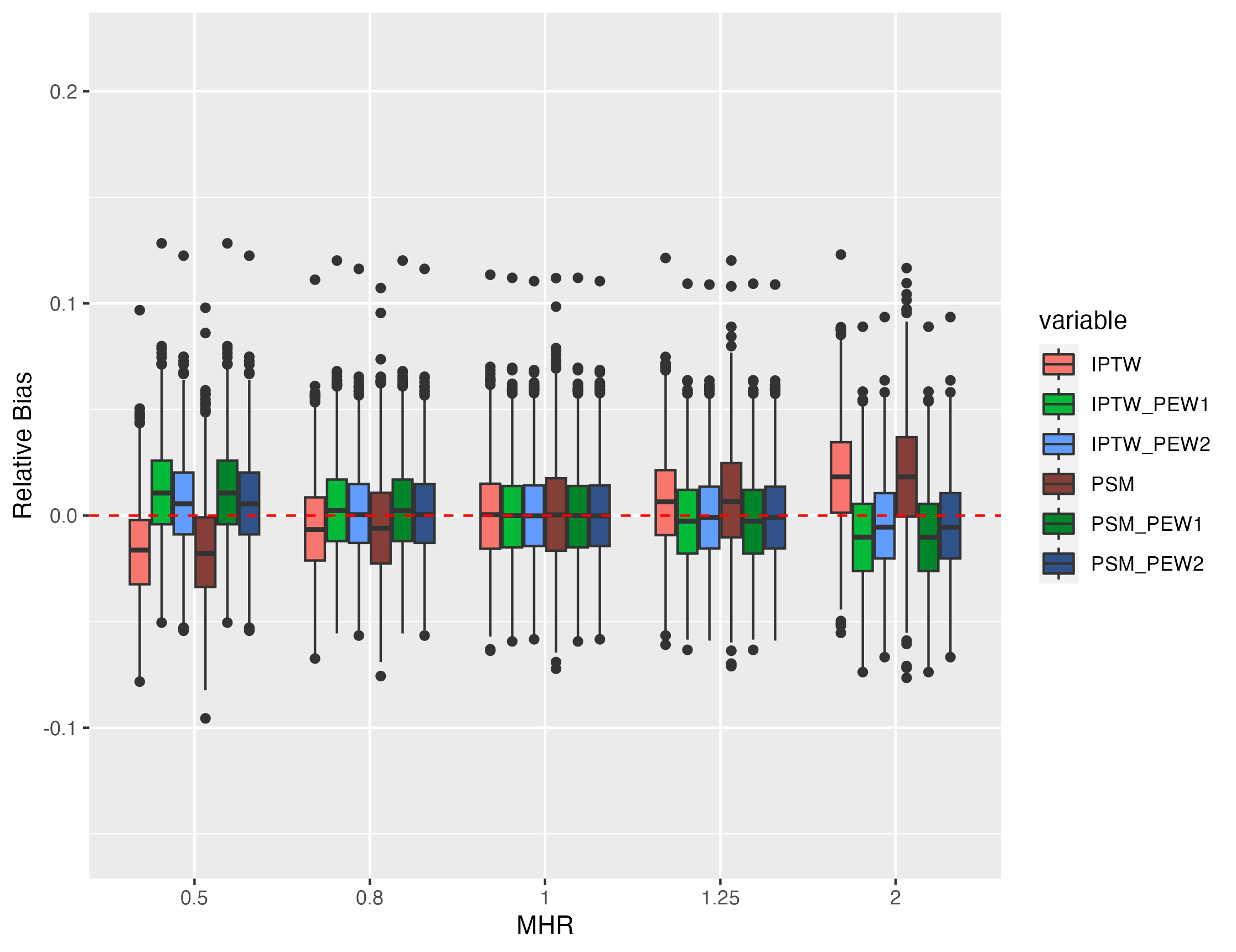}
\caption{Relative bias with censoring rate = 0.2}
\label{fig:n10000_notcounterfactual_censoring0.2}
\end{subfigure}

\medskip

\begin{subfigure}{0.49\columnwidth}
\centering
\includegraphics[width=\textwidth]{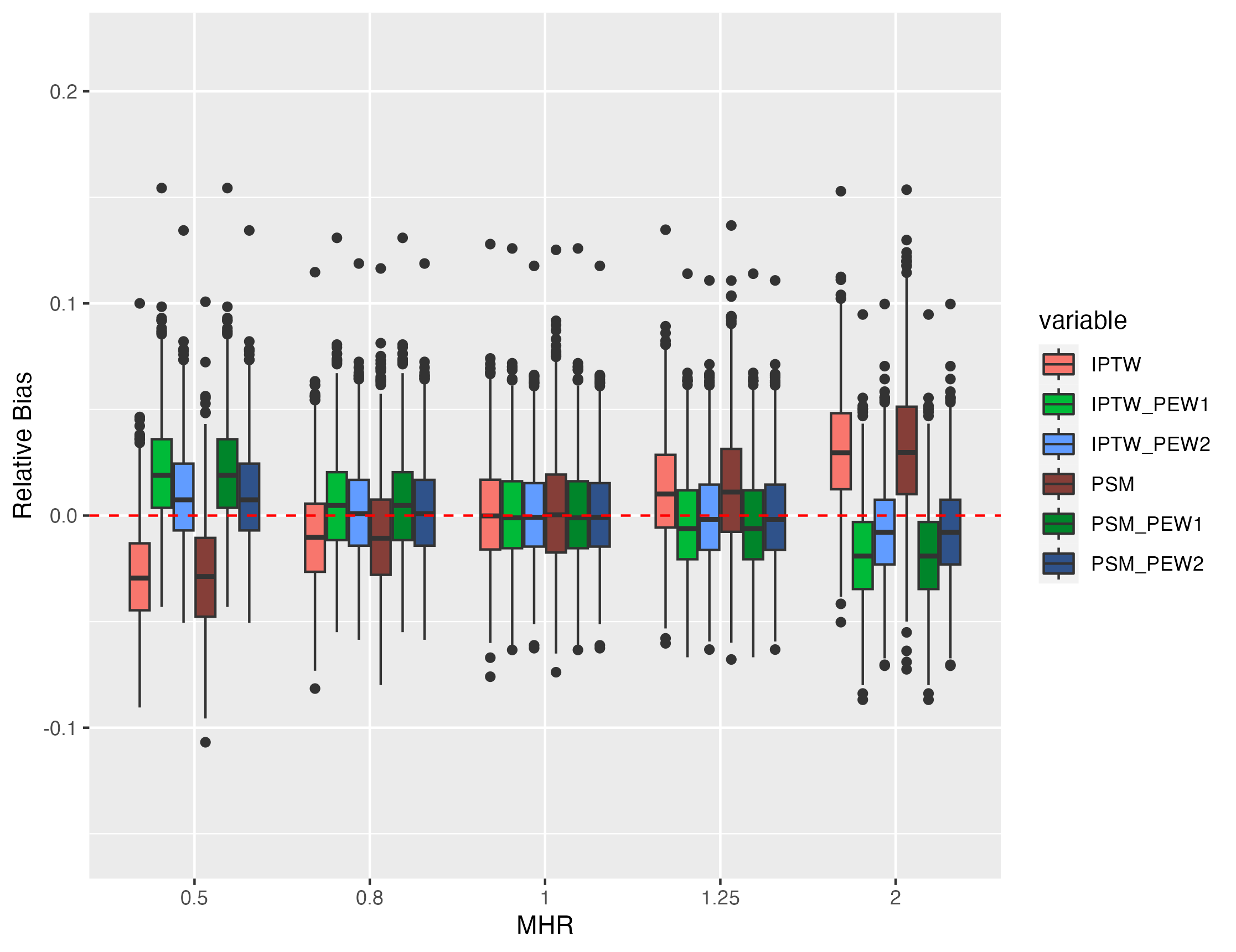}
\caption{Relative bias with censoring rate = 0.3}
\label{fig:n10000_notcounterfactual_censoring0.3}
\end{subfigure}\hfill
\begin{subfigure}{0.49\columnwidth}
\centering
\includegraphics[width=\textwidth]{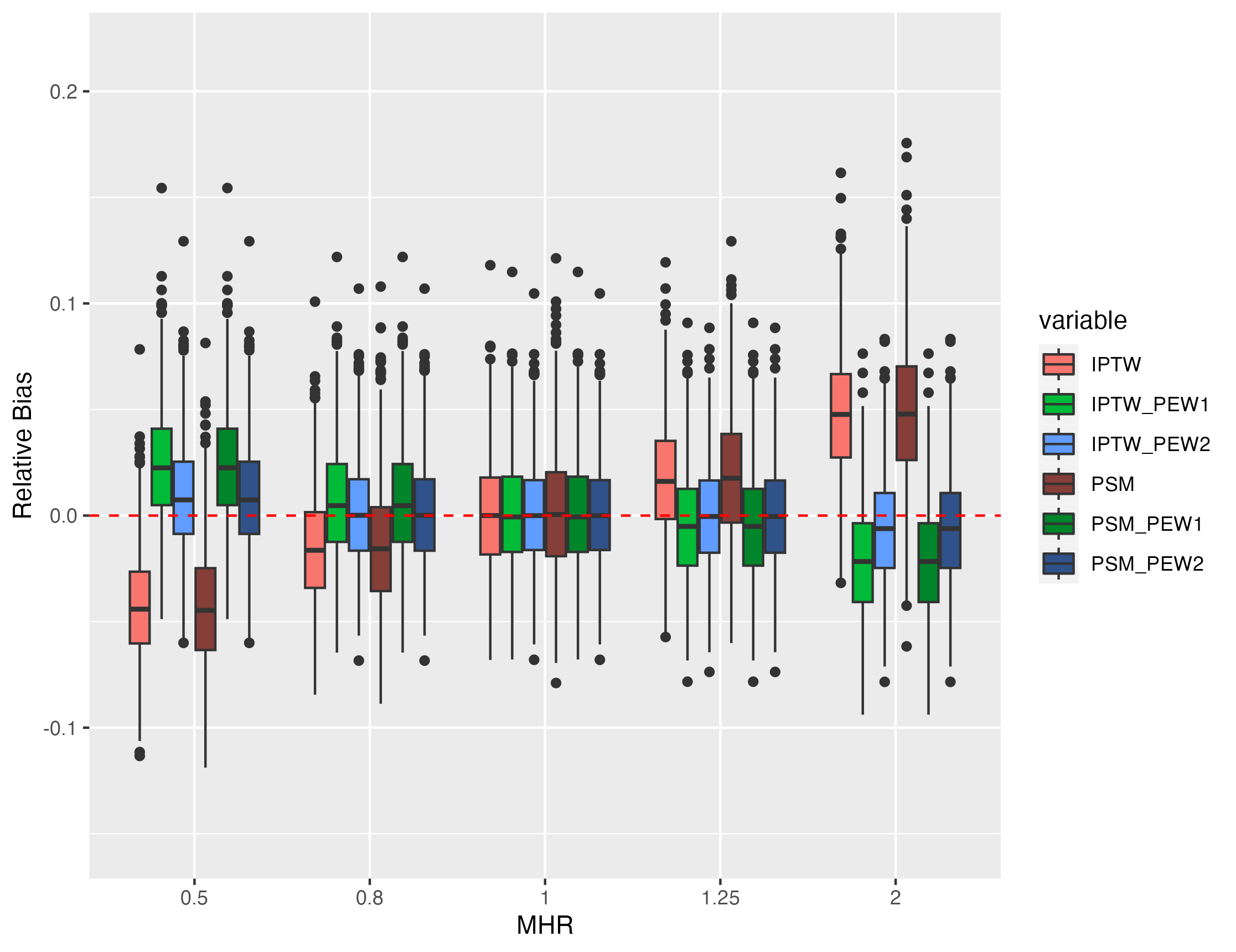}
\caption{Relative bias with censoring rate = 0.4}
\label{fig:n10000_notcounterfactual_censoring0.4}
\end{subfigure}

\medskip

\begin{subfigure}{0.49\columnwidth}
\centering
\includegraphics[width=\textwidth]{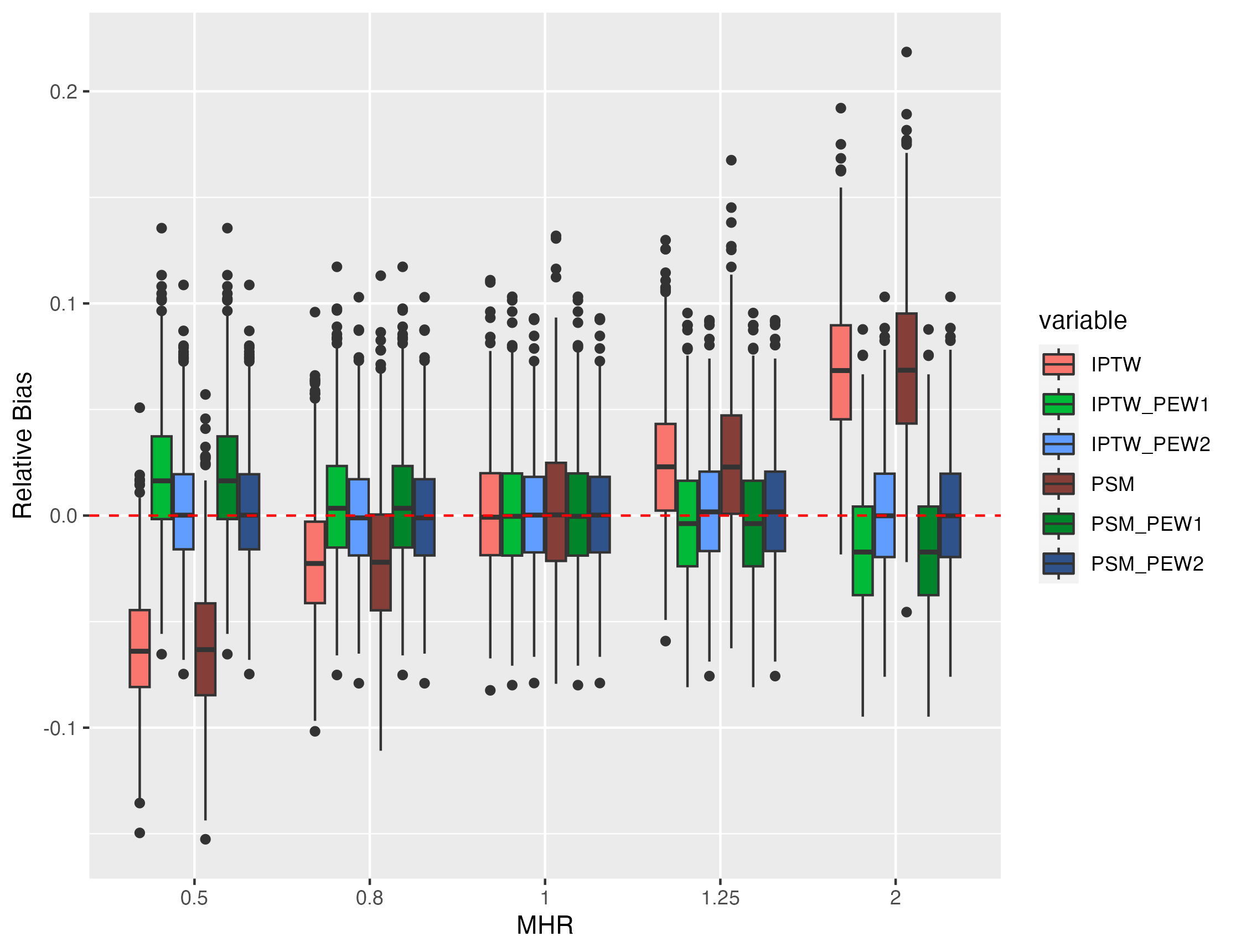}
\caption{Relative bias with censoring rate = 0.5}
\label{fig:n10000_notcounterfactual_censoring0.5}
\end{subfigure}

\caption{Observational setting, N = 10000, censoring rates = (0.1, 0.2, 0.3, 0.4, 0.5). Relative bias of estimation of MHR under several true values of MHR. Results based on 1000 simulation replicates.}
\label{fig:n10000_notcounterfactual_censoring_several_01_05}

\end{figure}

\begin{figure}[htp]
\centering

\begin{subfigure}{0.49\columnwidth}
\centering
\includegraphics[width=\textwidth]{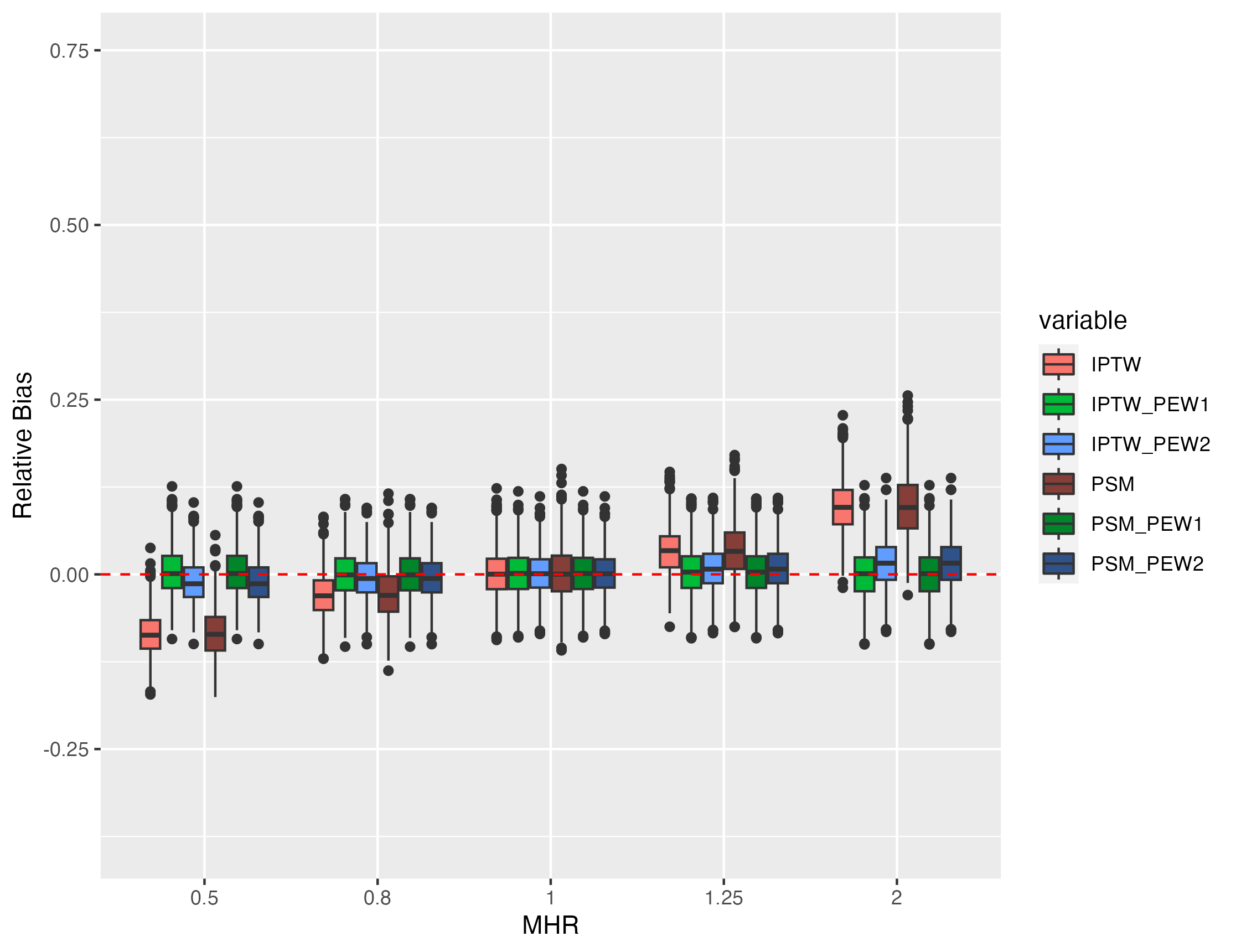}
\caption{Relative bias with censoring rate = 0.6}
\label{fig:n10000_notcounterfactual_censoring0.6}
\end{subfigure}\hfill
\begin{subfigure}{0.49\columnwidth}
\centering
\includegraphics[width=\textwidth]{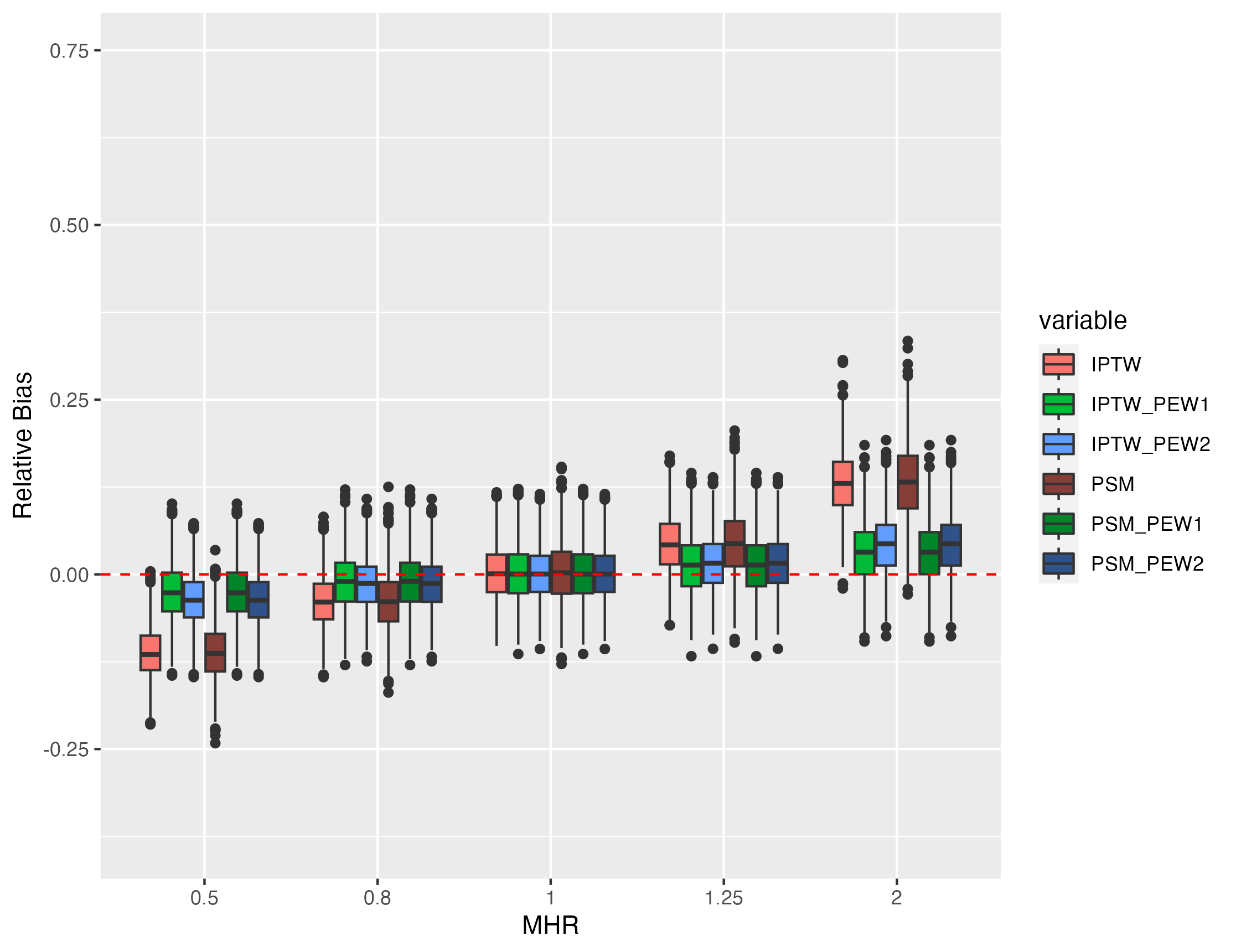}
\caption{Relative bias with censoring rate = 0.7}
\label{fig:n10000_notcounterfactual_censoring0.7}
\end{subfigure}

\medskip

\begin{subfigure}{0.49\columnwidth}
\centering
\includegraphics[width=\textwidth]{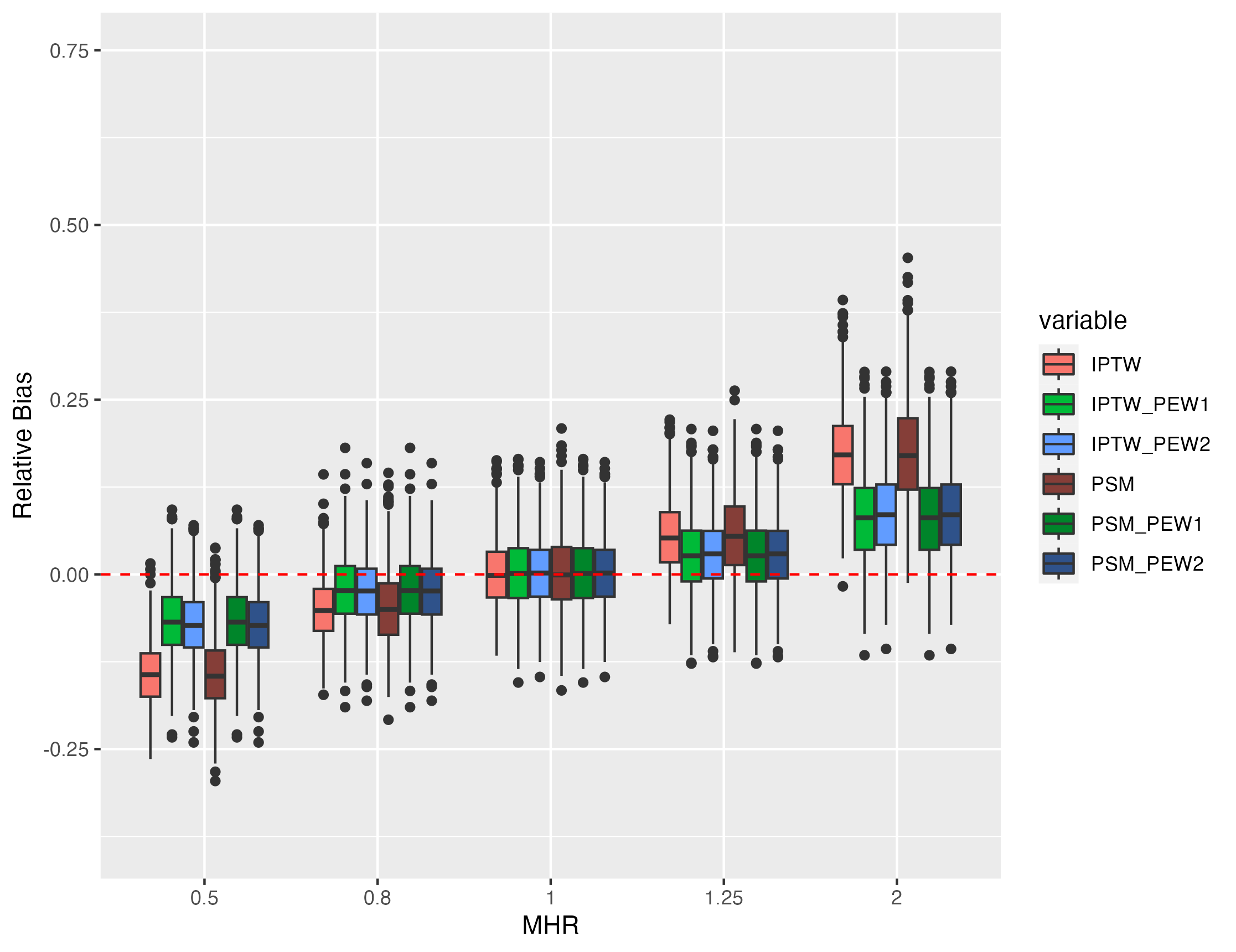}
\caption{Relative bias with censoring rate = 0.8}
\label{fig:n10000_notcounterfactual_censoring0.8}
\end{subfigure}\hfill
\begin{subfigure}{0.49\columnwidth}
\centering
\includegraphics[width=\textwidth]{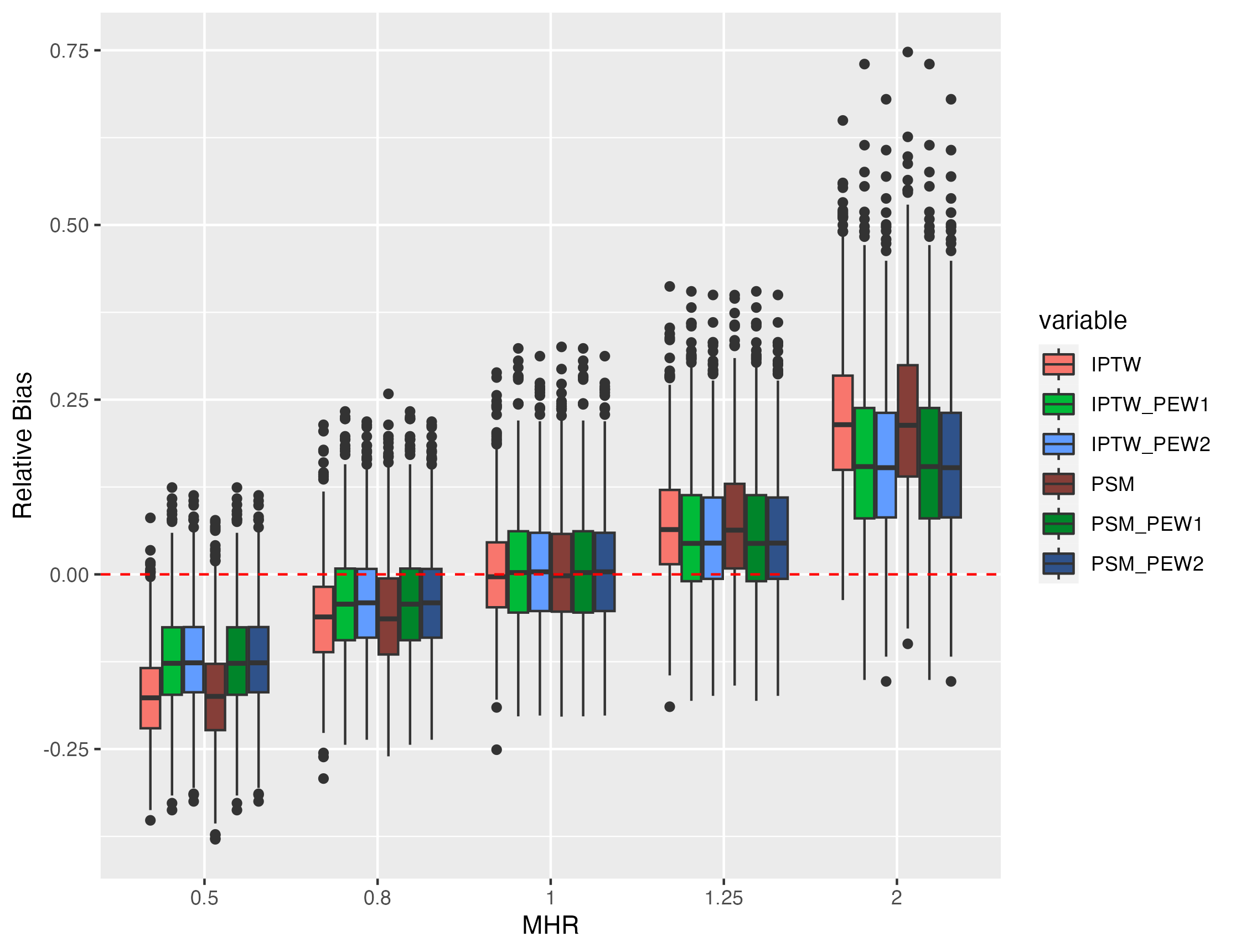}
\caption{Relative bias with censoring rate = 0.9}
\label{fig:n10000_notcounterfactual_censoring0.9}
\end{subfigure}

\caption{Observational setting, N = 10000, censoring rates = (0.6, 0.7, 0.8, 0.9). Relative bias of estimation of MHR under several true values of MHR. Results based on 1000 simulation replicates.}
\label{fig:n10000_notcounterfactual_censoring_several_06_09}

\end{figure}

\begin{table}
 \begin{center}
  \caption{Observational setup with population size $N = 2000$, true MHR = 0.5 and the censoring mechanism is uniform. Results of 1000 simulation replicates for estimators of MHR for IPTW, IPTW\_PEW1, IPTW\_PEW2, PSM, PSM\_PEW1 and PSM\_PEW2. Bias, Monte Carlo standard deviation (SD), root mean-squared error (RMSE), empirical coverage probability of 95\% confidence intervals (Coverage) and relative bias (Rel.Bias). }
 \begin{tabular}{@{}lcccccc@{}}
  \hline
  Method & \multicolumn{1}{c}{Censoring} &  \multicolumn{1}{c}{Bias} &  \multicolumn{1}{c}{SD} &  \multicolumn{1}{c}{RMSE} & Rel. Bias & Coverage\\
 \hline
  IPTW & \multirow{6}{*}{0.3} & -0.02 & 0.03 & 0.03 & -0.03 & 0.95 \\ 
  IPTW\_PEW1 & & 0.01 & 0.03 & 0.03 & 0.02 & 0.96 \\ 
  IPTW\_PEW2 & & 0.00 & 0.03 & 0.03 & 0.01 & 0.98 \\ 
  PSM & & -0.01 & 0.03 & 0.04 & -0.03 & 0.94 \\ 
  PSM\_PEW1 & & 0.01 & 0.03 & 0.03 & 0.02 & 0.96 \\ 
  PSM\_PEW2 & & 0.00 & 0.03 & 0.03 & 0.01 & 0.97 \\ 
  \hline
  IPTW & \multirow{6}{*}{0.4} & -0.02 & 0.03 & 0.04 & -0.05 & 0.91 \\ 
  IPTW\_PEW1 & & 0.01 & 0.03 & 0.03 & 0.02 & 0.96 \\ 
  IPTW\_PEW2 & & 0.00 & 0.03 & 0.03 & 0.01 & 0.97 \\ 
  PSM & & -0.02 & 0.03 & 0.04 & -0.04 & 0.92 \\ 
  PSM\_PEW1 & & 0.01 & 0.04 & 0.04 & 0.02 & 0.96 \\ 
  PSM\_PEW2 & & 0.00 & 0.03 & 0.03 & 0.01 & 0.97 \\ 
  \hline
  IPTW & \multirow{6}{*}{0.5} & -0.03 & 0.03 & 0.05 & -0.06 & 0.86 \\ 
  IPTW\_PEW1 & & 0.01 & 0.04 & 0.04 & 0.02 & 0.95 \\ 
  IPTW\_PEW2 & & 0.00 & 0.03 & 0.03 & 0.00 & 0.97 \\ 
  PSM & & -0.03 & 0.04 & 0.05 & -0.06 & 0.87 \\ 
  PSM\_PEW1 & & 0.01 & 0.04 & 0.04 & 0.02 & 0.96 \\ 
  PSM\_PEW2 & & 0.00 & 0.04 & 0.04 & 0.01 & 0.97 \\ 
  \hline
  IPTW & \multirow{6}{*}{0.6} & -0.04 & 0.04 & 0.06 & -0.09 & 0.81 \\ 
  IPTW\_PEW1 & & 0.00 & 0.04 & 0.04 & 0.00 & 0.96 \\ 
  IPTW\_PEW2 & & -0.00 & 0.04 & 0.04 & -0.01 & 0.97 \\ 
  PSM & & -0.04 & 0.04 & 0.06 & -0.08 & 0.82 \\ 
  PSM\_PEW1 & & 0.00 & 0.05 & 0.05 & 0.00 & 0.96 \\ 
  PSM\_PEW2 & & -0.00 & 0.04 & 0.04 & -0.01 & 0.97 \\ 
  \hline
  IPTW & \multirow{6}{*}{0.7} & -0.06 & 0.04 & 0.07 & -0.11 & 0.76 \\ 
  IPTW\_PEW1 & & -0.01 & 0.05 & 0.05 & -0.02 & 0.94 \\ 
  IPTW\_PEW2 & & -0.01 & 0.05 & 0.05 & -0.03 & 0.95 \\ 
  PSM & & -0.05 & 0.05 & 0.07 & -0.11 & 0.80 \\ 
  PSM\_PEW1 & & -0.01 & 0.05 & 0.06 & -0.02 & 0.94 \\ 
  PSM\_PEW2 & & -0.01 & 0.05 & 0.05 & -0.03 & 0.94 \\ 
  \hline
  IPTW & \multirow{6}{*}{0.8} & -0.07 & 0.05 & 0.09 & -0.14 & 0.77 \\ 
  IPTW\_PEW1 & & -0.03 & 0.06 & 0.07 & -0.06 & 0.91 \\ 
  IPTW\_PEW2 & & -0.03 & 0.06 & 0.06 & -0.06 & 0.93 \\ 
  PSM & & -0.07 & 0.06 & 0.09 & -0.14 & 0.79 \\ 
  PSM\_PEW1 & & -0.03 & 0.07 & 0.07 & -0.06 & 0.92 \\ 
  PSM\_PEW2 & & -0.03 & 0.06 & 0.07 & -0.06 & 0.93 \\ 
  \hline
  IPTW & \multirow{6}{*}{0.9} & -0.08 & 0.08 & 0.11 & -0.16 & 0.81 \\ 
  IPTW\_PEW1 & & -0.05 & 0.09 & 0.10 & -0.11 & 0.90 \\ 
  IPTW\_PEW2 & & -0.05 & 0.09 & 0.10 & -0.10 & 0.91 \\ 
  PSM & & -0.08 & 0.09 & 0.12 & -0.16 & 0.81 \\ 
  PSM\_PEW1 & & -0.05 & 0.10 & 0.11 & -0.11 & 0.90 \\ 
  PSM\_PEW2 & & -0.05 & 0.10 & 0.11 & -0.10 & 0.91 \\ 
      \midrule
\end{tabular}
\label{table:uniform_2000_MHR0.5}
\end{center}
\end{table}

 \begin{table}
 \begin{center}
  \caption{Observational setup with sample size $N = 2000$, true MHR = 2 and the censoring mechanism is uniform. Results of 1000 simulation replicates for estimators of MHR for IPTW, IPTW\_PEW1, IPTW\_PEW2, PSM, PSM\_PEW1 and PSM\_PEW2. Bias, Monte Carlo standard deviation (SD), root mean-squared error (RMSE), empirical coverage probability of 95\% confidence intervals (Coverage) and relative bias (Rel.Bias). }
 \begin{tabular}{@{}lcccccc@{}}
  \hline
  Method & \multicolumn{1}{c}{Censoring} &  \multicolumn{1}{c}{Bias} &  \multicolumn{1}{c}{SD} &  \multicolumn{1}{c}{RMSE} & Rel. Bias & Coverage\\
 \hline
  IPTW & \multirow{6}{*}{0.3} & 0.05 & 0.12 & 0.13 & 0.03 & 0.96 \\ 
  IPTW\_PEW1 & & -0.05 & 0.11 & 0.12 & -0.02 & 0.96 \\ 
  IPTW\_PEW2 & & -0.03 & 0.11 & 0.11 & -0.01 & 0.97 \\ 
  PSM & & 0.06 & 0.14 & 0.15 & 0.03 & 0.95 \\ 
  PSM\_PEW1 & & -0.04 & 0.13 & 0.14 & -0.02 & 0.96 \\ 
  PSM\_PEW2 & & -0.02 & 0.13 & 0.13 & -0.01 & 0.97 \\ 
  \hline
  IPTW & \multirow{6}{*}{0.4} & 0.08 & 0.14 & 0.16 & 0.04 & 0.94 \\ 
  IPTW\_PEW1 & & -0.05 & 0.12 & 0.14 & -0.03 & 0.95 \\ 
  IPTW\_PEW2 & & -0.03 & 0.12 & 0.12 & -0.01 & 0.97 \\ 
  PSM & & 0.09 & 0.16 & 0.18 & 0.04 & 0.94 \\ 
  PSM\_PEW1 & & -0.05 & 0.14 & 0.15 & -0.02 & 0.95 \\ 
  PSM\_PEW2 & & -0.02 & 0.14 & 0.14 & -0.01 & 0.96 \\ 
  \hline
  IPTW & \multirow{6}{*}{0.5} & 0.13 & 0.16 & 0.20 & 0.06 & 0.90 \\ 
  IPTW\_PEW1 & & -0.04 & 0.14 & 0.15 & -0.02 & 0.95 \\ 
  IPTW\_PEW2 & & -0.01 & 0.14 & 0.14 & -0.01 & 0.97 \\ 
  PSM & & 0.13 & 0.18 & 0.22 & 0.07 & 0.90 \\ 
  PSM\_PEW1 & & -0.04 & 0.16 & 0.17 & -0.02 & 0.95 \\ 
  PSM\_PEW2 & & -0.01 & 0.16 & 0.16 & -0.00 & 0.96 \\ 
  \hline
  IPTW & \multirow{6}{*}{0.6} & 0.18 & 0.18 & 0.26 & 0.09 & 0.85 \\ 
  IPTW\_PEW1 & & -0.01 & 0.17 & 0.17 & -0.00 & 0.95 \\ 
  IPTW\_PEW2 & & 0.02 & 0.16 & 0.16 & 0.01 & 0.97 \\ 
  PSM & & 0.19 & 0.21 & 0.28 & 0.09 & 0.87 \\ 
  PSM\_PEW1 & & -0.00 & 0.19 & 0.19 & -0.00 & 0.95 \\ 
  PSM\_PEW2 & & 0.02 & 0.18 & 0.18 & 0.01 & 0.96 \\ 
  \hline
  IPTW & \multirow{6}{*}{0.7} & 0.25 & 0.23 & 0.34 & 0.13 & 0.80 \\ 
  IPTW\_PEW1 & & 0.06 & 0.22 & 0.23 & 0.03 & 0.94 \\ 
  IPTW\_PEW2 & & 0.07 & 0.21 & 0.23 & 0.04 & 0.95 \\ 
  PSM & & 0.26 & 0.26 & 0.37 & 0.13 & 0.83 \\ 
  PSM\_PEW1 & & 0.06 & 0.25 & 0.26 & 0.03 & 0.94 \\ 
  PSM\_PEW2 & & 0.08 & 0.24 & 0.25 & 0.04 & 0.94 \\ 
  \hline
  IPTW & \multirow{6}{*}{0.8} & 0.36 & 0.32 & 0.48 & 0.18 & 0.77 \\ 
  IPTW\_PEW1 & & 0.17 & 0.32 & 0.36 & 0.09 & 0.92 \\ 
  IPTW\_PEW2 & & 0.17 & 0.30 & 0.35 & 0.09 & 0.93 \\ 
  PSM & & 0.36 & 0.35 & 0.51 & 0.18 & 0.82 \\ 
  PSM\_PEW1 & & 0.18 & 0.35 & 0.39 & 0.09 & 0.93 \\ 
  PSM\_PEW2 & & 0.18 & 0.34 & 0.38 & 0.09 & 0.93 \\ 
  \hline
  IPTW & \multirow{6}{*}{0.9} & 0.50 & 0.48 & 0.69 & 0.25 & 0.81 \\ 
  IPTW\_PEW1 & & 0.37 & 0.54 & 0.66 & 0.19 & 0.88 \\ 
  IPTW\_PEW2 & & 0.35 & 0.52 & 0.62 & 0.18 & 0.91 \\ 
  PSM & & 0.52 & 0.56 & 0.76 & 0.26 & 0.83 \\ 
  PSM\_PEW1 & & 0.40 & 0.62 & 0.74 & 0.20 & 0.87 \\ 
  PSM\_PEW2 & & 0.38 & 0.60 & 0.71 & 0.19 & 0.88 \\ 
    \midrule
\end{tabular}
\label{table:uniform_2000_MHR2}
\end{center}
\end{table}

\begin{table}
 \begin{center}
  \caption{Observational setup with sample size $N = 6000$, true MHR = 0.5 and the censoring mechanism is uniform. Results of 1000 simulation replicates for estimators of MHR for IPTW, IPTW\_PEW1, IPTW\_PEW2, PSM, PSM\_PEW1 and PSM\_PEW2. Bias, Monte Carlo standard deviation (SD), root mean-squared error (RMSE), empirical coverage probability of 95\% confidence intervals (Coverage) and relative bias (Rel.Bias). }
 \begin{tabular}{@{}lcccccc@{}}
  \hline
  Method & \multicolumn{1}{c}{Censoring} &  \multicolumn{1}{c}{Bias} &  \multicolumn{1}{c}{SD} &  \multicolumn{1}{c}{RMSE} & Rel. Bias & Coverage\\
 \hline
  IPTW & \multirow{6}{*}{0.3} & -0.01 & 0.02 & 0.02 & -0.03 & 0.87 \\ 
  IPTW\_PEW1 & & 0.01 & 0.02 & 0.02 & 0.02 & 0.93 \\ 
  IPTW\_PEW2 & & 0.00 & 0.02 & 0.02 & 0.01 & 0.97 \\ 
  PSM & & -0.02 & 0.02 & 0.02 & -0.03 & 0.90 \\ 
  PSM\_PEW1 & & 0.01 & 0.02 & 0.02 & 0.02 & 0.93 \\ 
  PSM\_PEW2 & & 0.00 & 0.02 & 0.02 & 0.01 & 0.96 \\ 
  \hline
  IPTW & \multirow{6}{*}{0.4} & -0.02 & 0.02 & 0.03 & -0.04 & 0.79 \\ 
  IPTW\_PEW1 & & 0.01 & 0.02 & 0.02 & 0.02 & 0.93 \\ 
  IPTW\_PEW2 & & 0.00 & 0.02 & 0.02 & 0.01 & 0.97 \\ 
  PSM & & -0.02 & 0.02 & 0.03 & -0.04 & 0.82 \\ 
  PSM\_PEW1 & & 0.01 & 0.02 & 0.02 & 0.02 & 0.93 \\ 
  PSM\_PEW2 & & 0.00 & 0.02 & 0.02 & 0.01 & 0.96 \\ 
  \hline
  IPTW & \multirow{6}{*}{0.5} & -0.03 & 0.02 & 0.04 & -0.06 & 0.68 \\ 
  IPTW\_PEW1 & & 0.01 & 0.02 & 0.02 & 0.02 & 0.94 \\ 
  IPTW\_PEW2 & & 0.00 & 0.02 & 0.02 & 0.00 & 0.98 \\ 
  PSM & & -0.03 & 0.02 & 0.04 & -0.06 & 0.73 \\ 
  PSM\_PEW1 & & 0.01 & 0.02 & 0.03 & 0.02 & 0.94 \\ 
  PSM\_PEW2 & & 0.00 & 0.02 & 0.02 & 0.00 & 0.96 \\ 
  \hline
  IPTW & \multirow{6}{*}{0.6} & -0.04 & 0.02 & 0.05 & -0.08 & 0.54 \\ 
  IPTW\_PEW1 & & 0.00 & 0.02 & 0.02 & 0.00 & 0.96 \\ 
  IPTW\_PEW2 & & -0.01 & 0.02 & 0.02 & -0.01 & 0.97 \\ 
  PSM & & -0.04 & 0.03 & 0.05 & -0.08 & 0.63 \\ 
  PSM\_PEW1 & & 0.00 & 0.03 & 0.03 & 0.00 & 0.95 \\ 
  PSM\_PEW2 & & -0.01 & 0.03 & 0.03 & -0.01 & 0.96 \\ 
  \hline
  IPTW & \multirow{6}{*}{0.7} & -0.06 & 0.03 & 0.06 & -0.11 & 0.45 \\ 
  IPTW\_PEW1 & & -0.01 & 0.03 & 0.03 & -0.02 & 0.92 \\ 
  IPTW\_PEW2 & & -0.02 & 0.03 & 0.03 & -0.03 & 0.91 \\ 
  PSM & & -0.05 & 0.03 & 0.06 & -0.11 & 0.54 \\ 
  PSM\_PEW1 & & -0.01 & 0.03 & 0.03 & -0.02 & 0.92 \\ 
  PSM\_PEW2 & & -0.02 & 0.03 & 0.03 & -0.03 & 0.91 \\ 
  \hline
  IPTW & \multirow{6}{*}{0.8} & -0.07 & 0.03 & 0.08 & -0.14 & 0.41 \\ 
  IPTW\_PEW1 & & -0.03 & 0.03 & 0.05 & -0.07 & 0.85 \\ 
  IPTW\_PEW2 & & -0.04 & 0.03 & 0.05 & -0.07 & 0.85 \\ 
  PSM & & -0.07 & 0.03 & 0.08 & -0.14 & 0.52 \\ 
  PSM\_PEW1 & & -0.03 & 0.04 & 0.05 & -0.06 & 0.87 \\ 
  PSM\_PEW2 & & -0.03 & 0.04 & 0.05 & -0.07 & 0.86 \\ 
  \hline
  IPTW & \multirow{6}{*}{0.9} & -0.09 & 0.04 & 0.10 & -0.18 & 0.50 \\ 
  IPTW\_PEW1 & & -0.06 & 0.05 & 0.08 & -0.13 & 0.76 \\ 
  IPTW\_PEW2 & & -0.06 & 0.04 & 0.08 & -0.13 & 0.78 \\ 
  PSM & & -0.09 & 0.05 & 0.10 & -0.17 & 0.60 \\ 
  PSM\_PEW1 & & -0.06 & 0.05 & 0.08 & -0.12 & 0.79 \\ 
  PSM\_PEW2 & & -0.06 & 0.05 & 0.08 & -0.12 & 0.81 \\ 
    \hline
 \end{tabular}
\label{table:uniform_6000_MHR0.5_notcounter}
\end{center}
\end{table}

\begin{table}
 \begin{center}
  \caption{Observational setup with sample size $N = 6000$, true MHR = 2 and the censoring mechanism is uniform. Results of 1000 simulation replicates for estimators of MHR for IPTW, IPTW\_PEW1, IPTW\_PEW2, PSM, PSM\_PEW1 and PSM\_PEW2. Bias, Monte Carlo standard deviation (SD), root mean-squared error (RMSE), empirical coverage probability of 95\% confidence intervals (Coverage) and relative bias (Rel.Bias). }
 \begin{tabular}{@{}lcccccc@{}}
  \hline
  Method & \multicolumn{1}{c}{Censoring} &  \multicolumn{1}{c}{Bias} &  \multicolumn{1}{c}{SD} &  \multicolumn{1}{c}{RMSE} & Rel. Bias & Coverage\\
 \hline
  IPTW & \multirow{6}{*}{0.3} & 0.06 & 0.07 & 0.10 & 0.03 & 0.89 \\ 
  IPTW\_PEW1 & & -0.04 & 0.07 & 0.08 & -0.02 & 0.94 \\ 
  IPTW\_PEW2 & & -0.02 & 0.07 & 0.07 & -0.01 & 0.97 \\ 
  PSM & & 0.06 & 0.08 & 0.10 & 0.03 & 0.89 \\ 
  PSM\_PEW1 & & -0.04 & 0.08 & 0.08 & -0.02 & 0.94 \\ 
  PSM\_PEW2 & & -0.01 & 0.07 & 0.08 & -0.01 & 0.97 \\ 
  \hline
  IPTW & \multirow{6}{*}{0.4} & 0.10 & 0.08 & 0.13 & 0.05 & 0.82 \\ 
  IPTW\_PEW1 & & -0.04 & 0.08 & 0.09 & -0.02 & 0.92 \\ 
  IPTW\_PEW2 & & -0.01 & 0.07 & 0.07 & -0.01 & 0.97 \\ 
  PSM & & 0.10 & 0.09 & 0.13 & 0.05 & 0.83 \\ 
  PSM\_PEW1 & & -0.04 & 0.08 & 0.09 & -0.02 & 0.93 \\ 
  PSM\_PEW2 & & -0.01 & 0.08 & 0.08 & -0.01 & 0.97 \\ 
  \hline
  IPTW & \multirow{6}{*}{0.5} & 0.14 & 0.10 & 0.17 & 0.07 & 0.69 \\ 
  IPTW\_PEW1 & & -0.03 & 0.09 & 0.09 & -0.02 & 0.94 \\ 
  IPTW\_PEW2 & & -0.00 & 0.08 & 0.08 & -0.00 & 0.97 \\ 
  PSM & & 0.14 & 0.11 & 0.18 & 0.07 & 0.74 \\ 
  PSM\_PEW1 & & -0.03 & 0.10 & 0.10 & -0.01 & 0.93 \\ 
  PSM\_PEW2 & & 0.00 & 0.09 & 0.09 & 0.00 & 0.96 \\ 
  \hline
  IPTW & \multirow{6}{*}{0.6} & 0.19 & 0.11 & 0.22 & 0.10 & 0.57 \\ 
  IPTW\_PEW1 & & -0.00 & 0.10 & 0.10 & -0.00 & 0.95 \\ 
  IPTW\_PEW2 & & 0.03 & 0.10 & 0.10 & 0.01 & 0.95 \\ 
  PSM & & 0.20 & 0.13 & 0.23 & 0.10 & 0.63 \\ 
  PSM\_PEW1 & & 0.00 & 0.11 & 0.11 & 0.00 & 0.95 \\ 
  PSM\_PEW2 & & 0.03 & 0.11 & 0.12 & 0.02 & 0.95 \\ 
  \hline
  IPTW & \multirow{6}{*}{0.7} & 0.26 & 0.14 & 0.29 & 0.13 & 0.49 \\ 
  IPTW\_PEW1 & & 0.06 & 0.13 & 0.14 & 0.03 & 0.93 \\ 
  IPTW\_PEW2 & & 0.08 & 0.12 & 0.15 & 0.04 & 0.92 \\ 
  PSM & & 0.27 & 0.15 & 0.31 & 0.13 & 0.55 \\ 
  PSM\_PEW1 & & 0.06 & 0.14 & 0.16 & 0.03 & 0.93 \\ 
  PSM\_PEW2 & & 0.09 & 0.14 & 0.16 & 0.04 & 0.92 \\ 
  \hline
  IPTW & \multirow{6}{*}{0.8} & 0.34 & 0.18 & 0.39 & 0.17 & 0.47 \\ 
  IPTW\_PEW1 & & 0.16 & 0.18 & 0.24 & 0.08 & 0.84 \\ 
  IPTW\_PEW2 & & 0.17 & 0.17 & 0.24 & 0.08 & 0.84 \\ 
  PSM & & 0.35 & 0.20 & 0.40 & 0.17 & 0.53 \\ 
  PSM\_PEW1 & & 0.17 & 0.20 & 0.26 & 0.08 & 0.86 \\ 
  PSM\_PEW2 & & 0.17 & 0.19 & 0.26 & 0.09 & 0.86 \\ 
  \hline
  IPTW & \multirow{6}{*}{0.9} & 0.45 & 0.27 & 0.53 & 0.23 & 0.56 \\ 
  IPTW\_PEW1 & & 0.33 & 0.29 & 0.44 & 0.16 & 0.80 \\ 
  IPTW\_PEW2 & & 0.32 & 0.29 & 0.43 & 0.16 & 0.82 \\ 
  PSM & & 0.47 & 0.30 & 0.56 & 0.23 & 0.63 \\ 
  PSM\_PEW1 & & 0.34 & 0.33 & 0.48 & 0.17 & 0.79 \\ 
  PSM\_PEW2 & & 0.34 & 0.33 & 0.47 & 0.17 & 0.80 \\ 
  \hline
 \end{tabular}
\label{table:uniform_6000_MHR2_notcounter}
\end{center}
\end{table}

\begin{table}
 \begin{center}
  \caption{Observational setup with sample size $N = 10000$, true MHR = 0.5 and the censoring mechanism is uniform. Results of 1000 simulation replicates for estimators of MHR for IPTW, IPTW\_PEW1, IPTW\_PEW2, PSM, PSM\_PEW1 and PSM\_PEW2. Bias, Monte Carlo standard deviation (SD), root mean-squared error (RMSE), empirical coverage probability of 95\% confidence intervals (Coverage) and relative bias (Rel.Bias). }
 \begin{tabular}{@{}lcccccc@{}}
 \hline
  Method & \multicolumn{1}{c}{Censoring} &  \multicolumn{1}{c}{Bias} &  \multicolumn{1}{c}{SD} &  \multicolumn{1}{c}{RMSE} & Rel. Bias & Coverage\\
 \hline
  IPTW & \multirow{6}{*}{0.3} & -0.01 & 0.01 & 0.02 & -0.03 & 0.87 \\ 
  IPTW\_PEW1 & & 0.01 & 0.01 & 0.02 & 0.02 & 0.91 \\ 
  IPTW\_PEW2 & & 0.00 & 0.01 & 0.01 & 0.01 & 0.97 \\ 
  PSM & & -0.01 & 0.01 & 0.02 & -0.03 & 0.88 \\ 
  PSM\_PEW1 & & 0.01 & 0.01 & 0.02 & 0.02 & 0.92 \\ 
  PSM\_PEW2 & & 0.00 & 0.01 & 0.01 & 0.01 & 0.97 \\ 
   \hline
  IPTW & \multirow{6}{*}{0.4} & -0.02 & 0.01 & 0.03 & -0.04 & 0.73 \\ 
  IPTW\_PEW1 & & 0.01 & 0.01 & 0.02 & 0.02 & 0.90 \\ 
  IPTW\_PEW2 & & 0.00 & 0.01 & 0.01 & 0.01 & 0.96 \\ 
  PSM & & -0.02 & 0.01 & 0.03 & -0.04 & 0.76 \\ 
  PSM\_PEW1 & & 0.01 & 0.02 & 0.02 & 0.02 & 0.91 \\ 
  PSM\_PEW2 & & 0.00 & 0.01 & 0.02 & 0.01 & 0.97 \\ 
   \hline
  IPTW & \multirow{6}{*}{0.5} & -0.03 & 0.01 & 0.03 & -0.06 & 0.50 \\ 
  IPTW\_PEW1 & & 0.01 & 0.01 & 0.02 & 0.02 & 0.93 \\ 
  IPTW\_PEW2 & & 0.00 & 0.01 & 0.01 & 0.00 & 0.98 \\ 
  PSM & & -0.03 & 0.02 & 0.03 & -0.06 & 0.61 \\ 
  PSM\_PEW1 & & 0.01 & 0.02 & 0.02 & 0.02 & 0.94 \\ 
  PSM\_PEW2 & & 0.00 & 0.02 & 0.02 & 0.00 & 0.98 \\ 
   \hline
  IPTW & \multirow{6}{*}{0.6} & -0.04 & 0.02 & 0.05 & -0.08 & 0.32 \\ 
  IPTW\_PEW1 & & 0.00 & 0.02 & 0.02 & 0.00 & 0.96 \\ 
  IPTW\_PEW2 & & -0.01 & 0.02 & 0.02 & -0.01 & 0.97 \\ 
  PSM & & -0.04 & 0.02 & 0.05 & -0.08 & 0.44 \\ 
  PSM\_PEW1 & & 0.00 & 0.02 & 0.02 & 0.00 & 0.96 \\ 
  PSM\_PEW2 & & -0.01 & 0.02 & 0.02 & -0.01 & 0.96 \\ 
   \hline
  IPTW & \multirow{6}{*}{0.7} & -0.06 & 0.02 & 0.06 & -0.11 & 0.22 \\ 
  IPTW\_PEW1 & & -0.01 & 0.02 & 0.02 & -0.02 & 0.92 \\ 
  IPTW\_PEW2 & & -0.02 & 0.02 & 0.03 & -0.04 & 0.90 \\ 
  PSM & & -0.06 & 0.02 & 0.06 & -0.11 & 0.31 \\ 
  PSM\_PEW1 & & -0.01 & 0.02 & 0.03 & -0.02 & 0.93 \\ 
  PSM\_PEW2 & & -0.02 & 0.02 & 0.03 & -0.03 & 0.92 \\ 
   \hline
  IPTW & \multirow{6}{*}{0.8} & -0.07 & 0.02 & 0.07 & -0.14 & 0.19 \\ 
  IPTW\_PEW1 & & -0.03 & 0.03 & 0.04 & -0.07 & 0.78 \\ 
  IPTW\_PEW2 & & -0.04 & 0.02 & 0.04 & -0.07 & 0.75 \\ 
  PSM & & -0.07 & 0.03 & 0.08 & -0.14 & 0.28 \\ 
  PSM\_PEW1 & & -0.03 & 0.03 & 0.04 & -0.07 & 0.80 \\ 
  PSM\_PEW2 & & -0.04 & 0.03 & 0.05 & -0.07 & 0.78 \\ 
   \hline
  IPTW & \multirow{6}{*}{0.9} & -0.09 & 0.03 & 0.09 & -0.17 & 0.31 \\ 
  IPTW\_PEW1 & & -0.06 & 0.04 & 0.07 & -0.12 & 0.66 \\ 
  IPTW\_PEW2 & & -0.06 & 0.04 & 0.07 & -0.12 & 0.67 \\ 
  PSM & & -0.09 & 0.04 & 0.09 & -0.17 & 0.40 \\ 
  PSM\_PEW1 & & -0.06 & 0.04 & 0.07 & -0.12 & 0.73 \\ 
  PSM\_PEW2 & & -0.06 & 0.04 & 0.07 & -0.12 & 0.74 \\  
  \midrule
\end{tabular}
\label{table:uniform_10000_MHR0.5_notcounterfactual}
\end{center}
\end{table}

\begin{table}
 \begin{center}
  \caption{Observational setup with sample size $N = 10000$, true MHR = 2 and the censoring mechanism is uniform. Results of 1000 simulation replicates for estimators of MHR for IPTW, IPTW\_PEW1, IPTW\_PEW2, PSM, PSM\_PEW1 and PSM\_PEW2. Bias, Monte Carlo standard deviation (SD), root mean-squared error (RMSE), empirical coverage probability of 95\% confidence intervals (Coverage) and relative bias (Rel.Bias). }
   \begin{tabular}{@{}lcccccc@{}}
   \hline
  Method & \multicolumn{1}{c}{Censoring} &  \multicolumn{1}{c}{Bias} &  \multicolumn{1}{c}{SD} &  \multicolumn{1}{c}{RMSE} & Rel. Bias & Coverage\\
 \hline
   IPTW & \multirow{6}{*}{0.3} & 0.06 & 0.05 & 0.08 & 0.03 & 0.86 \\ 
  IPTW\_PEW1 & & -0.04 & 0.05 & 0.06 & -0.02 & 0.94 \\ 
  IPTW\_PEW2 & & -0.01 & 0.05 & 0.05 & -0.01 & 0.98 \\ 
  PSM & & 0.06 & 0.06 & 0.09 & 0.03 & 0.88 \\ 
  PSM\_PEW1 & & -0.04 & 0.06 & 0.07 & -0.02 & 0.95 \\ 
  PSM\_PEW2 & & -0.01 & 0.06 & 0.06 & -0.01 & 0.98 \\ 
   \hline
  IPTW & \multirow{6}{*}{0.4} & 0.09 & 0.06 & 0.11 & 0.05 & 0.72 \\ 
  IPTW\_PEW1 & & -0.04 & 0.05 & 0.07 & -0.02 & 0.91 \\ 
  IPTW\_PEW2 & & -0.01 & 0.05 & 0.05 & -0.01 & 0.98 \\ 
  PSM & & 0.10 & 0.07 & 0.12 & 0.05 & 0.77 \\ 
  PSM\_PEW1 & & -0.04 & 0.06 & 0.07 & -0.02 & 0.92 \\ 
  PSM\_PEW2 & & -0.01 & 0.06 & 0.06 & -0.01 & 0.98 \\
  \hline
  IPTW & \multirow{6}{*}{0.5} & 0.14 & 0.07 & 0.15 & 0.07 & 0.54 \\ 
  IPTW\_PEW1 & & -0.03 & 0.06 & 0.07 & -0.02 & 0.94 \\ 
  IPTW\_PEW2 & & 0.00 & 0.06 & 0.06 & 0.00 & 0.98 \\ 
  PSM & & 0.14 & 0.08 & 0.16 & 0.07 & 0.62 \\ 
  PSM\_PEW1 & & -0.03 & 0.07 & 0.08 & -0.02 & 0.95 \\ 
  PSM\_PEW2 & & 0.00 & 0.07 & 0.07 & 0.00 & 0.97 \\ 
  \hline
  IPTW & \multirow{6}{*}{0.6} & 0.19 & 0.08 & 0.21 & 0.10 & 0.34 \\ 
  IPTW\_PEW1 & & 0.00 & 0.07 & 0.07 & 0.00 & 0.96 \\ 
  IPTW\_PEW2 & & 0.03 & 0.07 & 0.08 & 0.02 & 0.96 \\ 
  PSM & & 0.20 & 0.09 & 0.22 & 0.10 & 0.45 \\ 
  PSM\_PEW1 & & 0.00 & 0.08 & 0.08 & 0.00 & 0.96 \\ 
  PSM\_PEW2 & & 0.03 & 0.08 & 0.09 & 0.02 & 0.95 \\ 
  \hline
  IPTW & \multirow{6}{*}{0.7} & 0.26 & 0.09 & 0.28 & 0.13 & 0.23 \\ 
  IPTW\_PEW1 & & 0.06 & 0.09 & 0.11 & 0.03 & 0.93 \\ 
  IPTW\_PEW2 & & 0.08 & 0.09 & 0.12 & 0.04 & 0.90 \\ 
  PSM & & 0.26 & 0.11 & 0.29 & 0.13 & 0.32 \\ 
  PSM\_PEW1 & & 0.07 & 0.11 & 0.12 & 0.03 & 0.92 \\ 
  PSM\_PEW2 & & 0.09 & 0.10 & 0.14 & 0.04 & 0.90 \\ 
  \hline
  IPTW & \multirow{6}{*}{0.8} & 0.34 & 0.13 & 0.37 & 0.17 & 0.21 \\ 
  IPTW\_PEW1 & & 0.16 & 0.13 & 0.21 & 0.08 & 0.78 \\ 
  IPTW\_PEW2 & & 0.17 & 0.13 & 0.21 & 0.09 & 0.76 \\ 
  PSM & & 0.35 & 0.15 & 0.38 & 0.17 & 0.31 \\ 
  PSM\_PEW1 & & 0.17 & 0.15 & 0.22 & 0.08 & 0.81 \\ 
  PSM\_PEW2 & & 0.18 & 0.15 & 0.23 & 0.09 & 0.79 \\ 
  \hline
  IPTW & \multirow{6}{*}{0.9} & 0.45 & 0.21 & 0.49 & 0.22 & 0.36 \\ 
  IPTW\_PEW1 & & 0.33 & 0.24 & 0.40 & 0.16 & 0.69 \\ 
  IPTW\_PEW2 & & 0.32 & 0.23 & 0.40 & 0.16 & 0.70 \\ 
  PSM & & 0.45 & 0.24 & 0.51 & 0.23 & 0.47 \\ 
  PSM\_PEW1 & & 0.33 & 0.26 & 0.42 & 0.17 & 0.74 \\ 
  PSM\_PEW2 & & 0.33 & 0.26 & 0.42 & 0.16 & 0.74 \\ 
  \hline
\end{tabular}
\label{table:uniform_10000_MHR2_notcounterfactual}
\end{center}
\end{table}

\end{document}